\documentclass[ reprint, amsmath,amssymb, prb,floatfix]{revtex4-2}

\usepackage{graphicx}
\usepackage{color}

\usepackage{tikz}
\usetikzlibrary{shapes,arrows,positioning,calc,arrows.meta}
\usepackage{pgfplots} 

\usepackage{schemabloc}

\usepackage{xcolor}
\usepackage{hyperref}

\def\ket#1{|#1\rangle }
\def\bra#1{\langle#1 | }

\def\mat#1{\underline{\underline{{\bf #1 }}}}
\def\w{\omega}

\def\e{\epsilon}

\def\non{\nonumber\\ }
\definecolor{hotlead}{HTML}{F4684B}    
\definecolor{coldlead}{HTML}{3A78AF}   
\definecolor{interactcol}{HTML}{7B6FA8} 
\definecolor{arrowcol}{HTML}{4A3F35}   
\definecolor{chainsite}{HTML}{4A3A7A}   
\definecolor{qdcol}{HTML}{3D4F7C}       
\definecolor{reservoircol}{HTML}{C8BFE7}
\usepackage[nolist]{acronym}		

\def\acFull#1{\acl{#1} (\ac{#1})}

\begin{document}

\title{Open Wilson chain numerical renormalization group approach to \\steady-state 
nonequilibrium quantum transport}

\author{Anand Manaparambil}
\email{anand.manaparambil@tu-dortmund.de}
\author{Frithjof B. Anders}
\email{frithjof.anders@tu-dortmund.de}
\affiliation{Condensed Matter Theory, Department of Physik, TU Dortmund University,
Otto-Hahn-Str. 4, 44227 Dortmund,
Germany}

\date{\today}

\begin{abstract}
The numerical renormalization group (NRG)  approach was developed to identify and quantify different equilibrium regimes 
of  quantum impurity systems  with unprecedented  accuracy by a tailored finite-size representation. 
Out of equilibrium, the steady-state density operator is not of the Boltzmannian form but one that is determined by the imposed boundary conditions.
We extend the \ac{NRG} to the nonequilibrium setting by augmenting each Wilson site with a reservoir,
whose coupling functions are calculated via a continued fraction expansion in order to
recover the continuum limit exactly. The nonequilibrium parameters such as a finite bias  as well as a finite temperature gradient enter through the Bloch-Redfield tensor,
whose zero eigenvector gives the steady-state density operator.
We used the resulting open chain full density matrix  approach in combination with an effective description   to 
investigate the charge and spin transport through a \ac{QD} under finite bias and temperature gradient.
The influence of lead asymmetry and an external magnetic field on the transport properties are also studied. 
For completeness, we have also investigated local properties of the \ac{QD}, 
such as charge fluctuations and find excellent agreement with  \ac{RT-QMC} data.
The \ac{OC-FDM} approach was able to explore Kondo energy scales as low as 
$T_K/D \sim 10^{-8}$ in the nonequilibrium regime,
as well as show convergence with the established equilibrium benchmarks, such as a quantitative agreement with  \ac{FDM-NRG} 
and Fermi-liquid scaling at small bias and temperature.
Owing to the effective single lead description, a single  \ac{OC-FDM} data point takes orders of magnitude less time on a standard laptop, 
compared to other state-of-the-art numerical methods.
\end{abstract}

\maketitle

\begin{acronym}[time-dependent density matrix renormalization group]		
 \acro{TD}{time-dependent}
 \acro{NRG}{numerical renormalization group}
 \acro{TD-NRG}{time-dependent  numerical renormalization group}
 \acro{ED}{exact diagonalization}
\acro{FDM}{full density matrix}
\acro{FDM-NRG}{full density matrix numerical renormalization group}
 \acro{DMRG}{density matrix renormalization group}
 \acro{TD-DMRG}{time-dependent density matrix renormalization group}
 \acro{QIS}{quantum impurity system}
\acro{OC}{open chain}
 \acro{NEQ}{nonequilibrium}
 \acro{GF}{Green's function}
 \acro{QI}{quantum impurity}
 \acro{DOF}{degrees of freedom}
 \acro{OC-FDM}{open-chain full-density matrix}
 \acro{NCA}{noncrossing approximation}
 \acro{HWHM}{half-width-half-maximum}
 \acro{SNRG}{scattering states NRG}
 \acro{RT-QMC}{real-time quantum Monte Carlo}
 \acro{QMC}{quantum Monte Carlo}
 \acro{RG}{renormalization group}
 \acro{QPC}{quantum point contact}
\acro{DMFT}{dynamical mean field theory}
 \acro{SIAM}{single impurity Anderson model}
 \acro{BR}{Bloch-Redfield}
 \acro{BRT}{Bloch-Redfield tensor}
 \acro{QD}{quantum dot}

\end{acronym}

\section{Introduction}

Current and heat transport through interacting junctions
are examples of boundary-driven transport problems. 
The  difficulty in calculating the exact charge or heat current at an arbitrary finite DC bias  is related to the Kondo effect \cite{Wilson75} at very low temperatures.  The Kondo effect opens an additional transport channel and lifts the Coulomb blockade below the Kondo temperature.
The relevance for the quantum transport through a nano-device was predicted theoretically \cite{HershfieldDaviesWilkins1991,meirTransportSIAM93} and
later observed in single electron transistors \cite{goldhaberSET98,NatureGoldhaberGordon1998} and 
in the spectroscopy of adatoms on a metallic surface \cite{AdatomKondo1998}.
The recently observed resistive switching in transition metal oxides \cite{SAWA2008,LeeSung2025} 
was analyzed \cite{DiazHan2023} in terms of an  interacting  orbital out of equilibrium embedded into
a lattice\cite{AronWeberKotliar2013} using the \ac{DMFT}.
The local entropy of a  \acFull{QD}
was related to temperature dependent change of the \ac{QD}
occupancy via a Maxwell relation \cite{PhysRevLett.123.147702,Hartman:2018hw}. However, a 
discrepancy between the theoretical entropy of the \ac{QD} predicted by the \acFull{NRG}  and the experimental
entropy measurement protocol  \cite{Child-QPC-Entropy2022} was found in the Kondo regime which might be related to a back-action effect
of the quantum point contact used to detect the local occupation \cite{MaMeirQPC2023}. Solving this problem requires a reliable and numerically efficient algorithm to address this nonequilibrium back-action.

An exact expression for the current through an interacting region has been derived by Meir and Wingreen \cite{MeirWingreen1992} 
requiring the knowledge of the exact nonequilibrium many-body Green's functions for the interacting region.
Only a few approaches can correctly address the crossover from  local moment physics into
the strong coupling regime \cite{BullaCostiPruschke2008,GulletAl2011} in equilibrium.
Approximate solutions to finite bias are typically based on  Keldysh perturbation theory in the coupling
or the interaction, which either underestimate the charge fluctuations  \cite{MeirWingreen1994}
in \ac{NCA} \cite{Grewe83, Kuramoto83,Anders1995} based approaches 
or overestimates the charge screening \cite{spataruGwMillis2009,thygesenNonEqGW07}.
Also nonequilibrium extensions \cite{Eckern_2020,ChargeHeatTransportCorrelatedHopping2021}
 to the equation-of-motion decoupling schemes developed over 40 years ago, as well as the slave boson mean-field approaches  \cite{SlaveBosonQuantumTransport2017} have been used 
to investigate thermal transport in the strong coupling limit including an extension to correlated hopping problem \cite{ChargeHeatTransportCorrelatedHopping2021} on an approximate level.
Substantial advances have been reported using \ac{RT-QMC} approaches \cite{GuyGullMillis2014}
which allows to access the nonequilibrium  spectral function for a \ac{QD} coupled to two leads using a so-called inchworm \ac{QMC} approach \cite{CohenGullQMC2023}.
Charge fluctuations were investigated by Dirks et al. \cite{DirksSchmitt2013}
comparing \ac{RT-QMC}  and Matsubara voltage \ac{QMC} approaches with the \ac{SNRG} \cite{ AndersSSnrg2008}
and Keldysh perturbation theory.

In this paper we present a novel hybrid approach to the steady state quantum transport problem. It combines the virtue of Wilson's \ac{NRG} \cite{Wilson75,BullaCostiPruschke2008}
for the accurate  description of the Kondo problem by a discretized system 
with coupling of the Wilson chain to a set of reservoirs  enforcing the correct boundary conditions
of the leads  \cite{BoekerAnders2020}. The exact reservoir coupling functions are determined by a continued fraction expansion \cite{OpenChains2017,BoekerAnders2020}
ensuring that the  continuum limit is  exactly  recovered \cite{OpenChains2017,BoekerAnders2020}.
 Since the \ac{NRG} has been proven to describe the equilibrium physics with unprecedented low-energy and low-temperature resolution without these additional reservoirs, we treat the coupling to the reservoirs as
perturbation which generate temperature and bias-dependent rates for a \ac{BR} master equation \cite{BoekerAnders2020}. The steady-state
nonequilibrium density matrix is given by the eigenvector to the zero eigenvalue of the \ac{BRT}. 
In equilibrium, it was shown analytically, that the Boltzmannian form of the density operator
used by the \ac{NRG} is recovered \cite{BoekerAnders2020}. We use the nonequilibrium density matrix
obtained at finite bias or temperature gradient  
to calculate the  retarded spectral function  at finite bias as well as for a finite temperature gradient between the leads
which enters the Meir-Wingreen transport integral \cite{MeirWingreen1992}.

The \ac{TD-DMRG} uses  finite-size tight-binding chains as a discretized representation of the lead continua
\cite{Schmitteckert2004,silvaDagotto2008,Schmitteckert2013,Heidrich-Meisner-SIAM-transport2009} and drives a current through
the interacting region by a charge imbalance of the leads. Due to the finite-size of the system and the lack of dissipation in the leads, 
a steady-state cannot be reached \cite{Schmitteckert2013} 
and the quasi-stationary current must be read off before the charge package can reach the chain end. Recent advances
propose a hybrid \ac{NRG}/\ac{TD-DMRG} approach \cite{GuettgeAndersSchiller2013} to access the charge transport in the strong coupling regime.
Schwarz et al.\ \cite{SchwarzPRL-TransportQD2018}  constructed an effective low energy Hamiltonian
using the \ac{NRG}  until the bias window is reached and solve the residual problem using a two-lead  \ac{TD-DMRG} thermofield approach. This approach can access the strong-coupling regime for $|eV|\ll T_K$ and extract the current from the real-time evolution of the quantum state but uses a finite-size
representation.
The auxiliary master equation approach \cite{DoraArrigioni2015, LindbladContinuumVonDelft2016} maps the two-lead transport problem onto a discretized  auxiliary Lindblad problem and recovers the original continuum and the boundary conditions by fitting the Lindblad rates to
the exact lead Keldysh Green's functions.

A different approach to the steady-state current was proposed by Hershfield \cite{Hershfield1993} using Lippmann-Schwinger scattering states for an infinitely large system. He derived a Boltzmannian form of the steady state density operator.
While the \ac{SNRG} \cite{AndersSSnrg2008} using \ac{TD-NRG} to derive the density operator of the interacting system from the analytically known for the  non-interacting system \cite{Hershfield1993,SchillerHershfield96,Oguri2007},
May \cite{MayPhD2020} and Han \cite{han2026} independently proposed  to use the \ac{NRG} eigenstates of the interacting Hamiltonian to directly   evaluate Hershfield's  expressions \cite{Hershfield1993}. It is, however, not apparent how
to connect the discretized representation in the \ac{NRG} with the continuum limit \cite{MayPhD2020,han2026}.
May \cite{MayPhD2020} found that the current calculated directly by the approach tends to be suppressed by a finite Coulomb interaction \cite{MayPhD2020}.
Han \cite{han2026} circumvented the problem, by using the density matrix only to calculate the retarded Green's function out of equilibrium and resort to the Meir-Wingreen formula \cite{MeirWingreen1992} to obtain their  current representation, but still observes a current suppression at intermediate bias
in the strong coupling regime.

The paper is organized as follows: In Sec. \ref{sec:II}, we give an overview on the theory of quantum transport.
We present our novel hybrid \ac{OC-FDM} approach to boundary driven quantum systems in Sec.\ \ref{sec:III}. 
In Sec.\ \ref{Sec:IV}, this general approach is tailored to a quantum transport through 
single-orbital quantum dot, and the results for bias and temperature driven charge currents are presented in Sec.\ \ref{sec:results}. 
It has been realized  that the influence of the several leads onto the dynamics of single-orbital quantum dot can be modeled by an effective problem \cite{GlazmanRaikh1988} 
where the different chemical potentials \cite{MeirWingreen1992,Oguri2007}  enter via linear superposition of lesser and greater \acs{GF}.  
We conclude the paper with a short summary.

The general transport problem is defined in Sec.\ \ref{sec:II-intro}. 
We briefly address some drawbacks of using a closed system representation in Sec.\ \ref{sec:II-closed-systems},  remind the reader of the central exact formula for the steady-state charge transport by Meir and Wingreen in Sec. \ref{sec:meir-wingreen}, and
review the Bloch-Redfield and Lindblad approaches in Sec.\ \ref{BR-Lindblad}. 
We present our new approach to boundary driven quantum systems in Sec.\ \ref{sec:III}. We set the stage by reviewing 
the combination of the \ac{NRG} with the \ac{BR} approach \cite{BoekerAnders2020} in Sec.\ \ref{sec:NRG+BRT}: the coupling 
of the Wilson chain sites to the individual reservoirs\cite{OpenChains2017,BoekerAnders2020} leads to
a \ac{BR} master equation. A simplified master-equation is presented in Sec.\ \ref{sec:III-B}, from which a numerically very efficient \ac{FDM} approach  \cite{WeichselbaumDelft2007} is derived.
By using the mapping onto an effective single-lead problem \cite{GlazmanRaikh1988,Oguri2007},
the  general approach is applied to a quantum transport through 
single-orbital quantum dot  in Sec.\ \ref{sec:mapping-single-lead}. The algorithm is summarized in Sec.\ \ref{sec:algorithm-single-lead}. 

The results are presented in Sec.\ \ref{sec:results}. We start with a benchmark of our \ac{OC-FDM} spectral function in equilibrium	 Sec.\ \ref{sec:equilibrum}.
We then investigate the finite bias evolution of the retarded spectral
function in Sec.\ \ref{sec:neq-spectral-functions} and calculate local observables such as charge fluctuations on the \ac{QD} in Sec.~\ref{sec:charge-fluctuation}.
The charge transport through a \ac{QD} in the presence of a finite bias is presented in Sec.\ \ref{sec:charge-currents}.
We extend the discussion to charge and spin currents in the presence of 
an external magnetic field in Sec.\ \ref{sec:magneto-transport} and wrap up the results section
with a discussion of our calculations for the  thermoelectric 
charge and spin currents driven by a finite temperature difference in the leads at zero bias 
in Sec.\ \ref{sec:thermo-current}.

\section{Quantum transport through a nano-constriction}
\label{sec:II}

\subsection{Introduction to the problem: Expression for steady state currents}
\label{sec:II-intro}

\begin{figure}[b]
\begin{center}

\begin{tikzpicture}
 \node (l0) at (0,0) [fill=hotlead, text=white,rounded corners=5pt,inner sep =5.5]{Left lead: $T_L,\mu_L$};
 \node (r0) at (60mm,0) [fill=coldlead, text=white, rounded corners=5pt,inner sep =5.5]{Right lead: $T_R,\mu_R$};
 \node (QD) at (29.5mm,0) [fill=interactcol, text=white, rounded corners=10pt,inner sep =5]{Interacting region};

\draw[-{Triangle[length=8pt, width=5pt]}, line width=1.5pt, arrowcol,
      bend left=25]
  (l0) to node[above, yshift=2pt, font=\small\itshape] {$V_L$} (QD);

\draw[-{Triangle[length=8pt, width=5pt]}, line width=1.5pt, arrowcol,
      bend left=25]
  (QD) to node[above, yshift=2pt, font=\small\itshape] {$V_R$} (r0);
 \end{tikzpicture}
\caption{Schematic setup for transport: Two leads ($L,R$) kept at different chemical potentials $\mu_L, \mu_R$, and temperatures $T_L, T_R$ are coupled to an interacting region. A nonequilibrium steady state current flows through the interacting region driven by potential bias $eV=\mu_L-\mu_R$, or temperature gradient $T_L \ne T_R$, or both.}
\label{fig:1}
\end{center}
\end{figure}
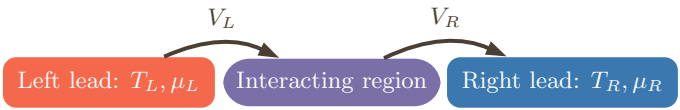

Quantum transport through a finite-size nano-constriction is schematically depicted in Fig.\ \ref{fig:1}. A current is driven
between the leads through the interacting region by an imbalance of the chemical potentials $\mu_L$ and $\mu_R$, or by a temperature gradient, i. e., $T_L\not=T_R$, defining this as a boundary-driven transport problem \cite{JauhoWingreenMeir1994}.

The dynamics of the coupled problem is governed by the total Hamiltonian,
\begin{eqnarray}
H = H_S + \sum_{\alpha=R,L} H_\alpha^\text{lead} + H_T,
\end{eqnarray}
where the finite-size system Hamiltonian $H_S$ describes the interacting region, $H_\alpha^\text{lead}$ denotes the Hamiltonian of the non-interacting leads and $H_T$ the coupling between the two subsystems.
The lead Hamiltonian takes the form,
\begin{equation}
H_\alpha^\text{lead}=\sum_{k\nu} \epsilon_{k\alpha\nu} c^{\dagger}_{k\alpha\nu}c_{k\alpha\nu},
\end{equation}
where $c^\dagger_{k\alpha\nu}$ creates  an electron in lead $\alpha=L,R$ and  channel $\nu$ with energy $\epsilon_{k\nu\alpha}$.
$H_T$ is given by the bilinear hopping (or tunneling) term,
\begin{equation}
\label{eq:1}
H_{T}=  \sum_{\alpha}  \sum_{n,k\nu} \left[V_{k\alpha \nu,n} c^\dagger_{k\alpha\nu} d_{n} + H.c. \right],
\end{equation}
where $d_{n} $   annihilates an electron in the orbital $n$ of the  interacting region ($n$ is a shorthand for
the orbital, spatial and the spin degree of freedom) .
The channel index $\nu$ contains the spin and other quantum number relevant for the lead $\alpha$. 
We define the corresponding coupling function matrix elements as,
\begin{eqnarray}
\label{eq:4}
\Gamma^\alpha_{m,n}(\omega)  = \pi \sum_{k\nu} V_{k\alpha \nu,m}^* V_{k\alpha \nu,n}  \delta(\omega-\epsilon_{k\alpha\nu}),
\end{eqnarray}
assuming the leads are diagonal in the quantum numbers $k,\nu$.

This type of problem belongs to the class of \acp{QIS} where 
a small subsystem S of interest with a finite
size Hilbert space is coupled to  fermionic leads
characterized by an energy continuum.  The energy dependent coupling functions
to the system in combination with the boundary conditions 
fully specify the effect of these baths onto the dynamics of the interacting region.

The current operator can be derived from change of the charge in the  interacting region. 
Let $Q_S=eN_S$ be the total charge operator of interacting region.
Charge conservation relates the change of the charge 
\begin{equation}
\frac{d}{dt} Q_S(t) = i\frac{e}{\hbar} [H, N_S] =  i\frac{e}{\hbar} [H_T, N_S]  = \hat I_L +  \hat I_R,
\end{equation}
to the current operators $I_\alpha$ which contain all  contributions from the lead $\alpha$. In the steady state $ \langle \dot Q_S\rangle =0$ and the two average currents are exactly equal and opposite to each other, i.e., $I_R=-I_L$. Taking the leads as non-interacting, the steady state
current is given by the expression \cite{MeirWingreen1992},
\begin{eqnarray}
\label{eq:3}
I_\alpha &=& \frac{2e}{\hbar}
\sum_{n,k\nu} \,{\rm Im}  \left[V_{k \alpha \nu,n} \langle c^\dagger_{k \alpha\nu} d_n\rangle
\right]
\\
\label{eqn:2}
 &=&  \frac{2e}{\hbar} \,{\rm Im}\left[\sum_{n,k\nu}
V_{k \alpha \nu,n}  \sum_m \rho_m \langle m| c^\dagger_{k \alpha\nu} d_n |m\rangle 
\right],
\end{eqnarray}
where we use the spectral decomposition in the joint eigenbases of $H$ and $\rho_{\rm steady}$ in the second line since 
$[\rho_{\rm steady},H]=0$. Since $0\le \rho_m\le 1$ and real, the expectation values  $\langle m| c^\dagger_{k \alpha\nu} d_n |m\rangle$
must be complex for a finite steady state current. This implies a complex superposition of states of the number operator basis forming the eigenbasis   $\{|m\rangle\}$.

\subsection{Closed systems and discretized representation}
\label{sec:II-closed-systems}

In a discretized representation, the lead continuum is replaced by a bilinear Hamiltonian comprising of finite number of non-interacting orbitals that are connected to the interacting region. A popular choice are tight binding chains for both leads, whose parameters can be derived by a continued fraction expansion \cite{Schmitteckert2004,Heidrich-Meisner-SIAM-transport2009} from the original continuum. Tailored to the Kondo problem  logarithmic Wilson chains have  been employed \cite{ AndersSSnrg2008, LotemWeichselbaum2020}.

Using the \ac{TD-DMRG} \cite{Schollwoeck-2005,Schollwoeck2011} approach, the current transport through an interaction region as depicted in Fig.\ \ref{fig:1}
has been investigated after   a quench of the system \cite{Schmitteckert2004,Heidrich-Meisner-SIAM-transport2009}, 
by evaluating the time dependent density operator, $\rho(t)= \exp(-iH t)\rho_0\exp(iH t)$. A finite time transport current $I_\alpha(t)$ is observed, 
and a representation of the actual steady-state current is read off once a metastable state has been reached  \cite{Schmitteckert2004,Heidrich-Meisner-SIAM-transport2009,conductanceDMRG2010}.  
Since the complex expectation values in  Eq.~\eqref {eqn:2} stems from the unitary time evolution of $\rho(t)$ in that approach and not from the complex wave function, a steady current carrying state can never be reached in a closed system. 
It turns out that the propagating charges driving the current are reflected at the end of the chain  after the time t exceeds the transit time $T_R$
in the finite-size chain yielding a reversed current as reported by  Branschaedel et al.~\cite{conductanceDMRG2010}.  Hence the numerical simulations are usually stopped before $T_R$ is reached. The  long time averaged current
\begin{eqnarray}
\bar I_\alpha = \lim_{T\to \infty} \frac{1}{T} \int_0^T dt I_\alpha(t)\,,
\end{eqnarray}
vanishes  since closed systems cannot carry a steady state current.

\subsection{Open system approaches}

\subsubsection{Meir-Wingreen current expression}
\label{sec:meir-wingreen}

In a seminal paper, \cite{MeirWingreen1992} Meir and Wingreen derived an exact expression for the transport current through an interacting
nano-device,
\begin{align}
\label{eqn:15}
I_\alpha = &i\frac{2e}{h}  \sum_{nm} \int_{-\infty}^\infty d\omega \Gamma^\alpha_{m,n}(\omega) 
\left[f_\alpha(\omega)(G^r_{n,m}(\omega) - G^a_{n,m}(\omega)) 
\right. \nonumber \\
 &\left. \hspace{15mm}
+ G^<_{n,m}(\omega)
\right],
\end{align}
using Keldysh \acp{GF}
where $ \Gamma^\alpha_{m,n}(\omega) $ is defined in Eq.~\eqref{eq:4}.
 The boundary condition driving the current is explicitly entering via the
Fermi function $f_\alpha(\omega)= [\exp(\beta_\alpha(\omega-\mu_\alpha))+1]^{-1}$ including the chemical potential
$\mu_\alpha$ and the inverse temperature $\beta_\alpha$ of the lead $\alpha$ as well as implicitly through
the full retarded [advanced] \ac{GF}, $G^r_{n,m}(\omega)$ [$G^a_{n,m}(\omega)$] and the lesser \ac{GF},
$G^<_{n,m}(t) = i\langle d^\dagger_m d_n(t) \rangle$ of the interacting region.
In case of  $\Gamma^L_{m,n}(\omega) =R\Gamma^R_{m,n}(\omega)$, the averaged  current 
is given by,
\begin{eqnarray}
I  &=&  \frac{R}{1+R} I_R - \frac{1}{1+R} I_L 
\nonumber \\
&=&
\label{eq:meir-wingreen-Qdot}
-\frac{R}{(1+R)^2} \frac{4e}{h} \int_{-\infty}^\infty d\omega 
 \left(f_L(\omega) -f_R(\omega) \right)\nonumber \\
 && \hspace{20mm} \times\, {\rm Im} \left\{ \text{Tr}\left[\mat{\Gamma}(\omega)\mat{G}^r (\omega)\right]\right\} ,
 \end{eqnarray}
by using a matrix representation in the orbital space $(n,m)$.
The current is positive and flows from left to right when $\mu_L>\mu_R$.
These formulas are exact but require the determination of the exact full many-body retarded 
\ac{GF} matrix of the interacting region. While this is possible
for a non-interacting problem, for a fully interacting problem some approximations are usually applied.

The bias voltage is defined via the difference in the chemical potentials $eV=\mu_L-\mu_R$. In an experiment, a finite bias $eV$ is applied,
and the chemical potential adjust accordingly. Let us assume, the interacting region is only coupled to the right lead, and $\Gamma_L=0$, i.e., $R=0$. In this case the orbital occupation of the  interacting region is completely determined by $\mu_R$ and $\Gamma_R$, so that we can take $\mu_R=0$ as the reference point and assign $\mu_L=eV$. For a symmetric junction, $R=1$, we expect that $\mu_R=-\mu_L =-eV/2$. We can interpolate between this two limits by
$\mu_L= 1/(1+R) eV$ and $\mu_R=-R/(1+R) eV$. This is a consequence that the ratio of the resistivity of the link between left lead/\ac{QD} and the link between right lead/\ac{QD} is given by $R$, so that the total voltage drops in steps over the junction, as in a serial resistor.

From Eq.\ \eqref{eq:meir-wingreen-Qdot}, we  define the unit of conductance 
\begin{equation}
G_0= \frac{e^2}{h} \frac{4R}{(1+R)^2}\, ,
\end{equation}
which includes the asymmetry of the
junction.  Obviously $G_0(R)= G_0(1/R)$ so that only $1\le R$ need to be investigated.
For a symmetric junction ($R=1$) we obtain $G_0=e^2/h$, the conductance quantum per channel for perfect transmission, while for large
$R$  where $G_0(R)\approx 4e^2/(h R)$ for $1\ll R$ describes the tunnel regime.
For a symmetric junction, $R=1$, the universal conductance quanta per transport channel is $e^2/h$.

\subsubsection{Bloch-Redfield  master-equation and Lindblad approach}
\label{BR-Lindblad}

Another popular approach  is based on an exact diagonalization of the interacting region and
the introduction of dissipators $D_\nu$ which accounts for the effect of the reservoir $\nu$ onto the eigenstates of $H_S$.
In the Schrödinger picture, the reduced density operator for the interacting region obeys
 
\begin{eqnarray}
\dot \rho_S = -i [ H_S,\rho_S ] + \sum_\nu D_\nu(\rho_S) \, .
\end{eqnarray}
One choice for such dissipators is the superposition of Lindblad terms \cite{PhysRevE.76.031115,Dzhioev_2012,DoraArrigioni2015,LindbladContinuumVonDelft2016,LotemWeichselbaum2020}
which have well-defined rates associated with the Kraus operators. 

In the Bloch-Redfield approach \cite{MayKuehn2000,BoekerAnders2020,BoekerPhD2021,RevModPhys.NEQ-QS-2022} 
the dissipators are directly calculated  from the coupling functions. Applying the  secular approximation \cite{MayKuehn2000}
the resulting master equations are very similar to the Lindblad approach: the master equation for the diagonal and off-diagonal matrix elements become independent.  While the relaxation rates  are free parameters in the  Lindblad approach, they are derived from $H_T$  in the Bloch-Redfield approach. 

Focusing on the steady state, only the diagonal matrix elements are relevant which obey the rate equation
\begin{eqnarray}
\dot \rho_{aa} = \sum_b \left[\sum_{\alpha\nu} W^{\alpha\nu}_{ab} \rho_{bb} - \sum_{\alpha\nu} W^{\alpha\nu}_{ba} \rho_{aa} \right] \, .
\end{eqnarray}
The boundary conditions enter  $W_{ab}$  via the Fermi functions  \cite{BoekerAnders2020,BoekerPhD2021}--see also below.

From its derivation, it becomes clear that calculating the current 
is only possible if the coupling to the reservoirs is the smallest energy scale in the problem. 
If links inside of the system S are much smaller than the coupling to
the reservoirs, then the system is  partitioned 
into two or more rigid parts, and the current is controlled by the internal link properties \cite{PhysRevE.76.031115,RevModPhys.NEQ-QS-2022} 
rather than the explicit dissipators.
 As a consequence, an adapted local master-equation approach is required to access the current. An extensive discussion can be found in the review \cite{RevModPhys.NEQ-QS-2022}.  Therefore such approaches are only applicable to the Coulomb blockade regime excluding the Kondo effect.

\section{Open Wilson approach to Quantum transport}
\label{sec:III}

This section contains the main body of our work.  
Below, we present  a general approach to the steady state quantum transport that can be easily adapted to different junction geometries such as double \acp{QD} or short interacting chains.
In this work, however,  we apply the approach only to a
 \ac{SIAM} coupled to two leads at different chemical potential or at temperature gradient.
For solving the nonequilibrium steady state we will employ Wilson's \ac{NRG} \cite{Wilson75}  to access the Kondo effect and the strong coupling limit.

While the equilibrium density operator and the dynamics is fully specified by the Hamiltonian, the 
characterization of the nonequilibrium steady state 
requires the nonequilibrium density operator $\rho_{\text{steady}}$ as well.

Since $[\rho_{\text{steady}},H]=0$, there exist a joint eigenbasis of $H$ and $\rho_{\text{steady}}$, Therefore, 
we propose to use the \ac{NRG} to construct the accurate eigenstates of the \ac{NRG} Hamiltonian which serves as $H_S$ for \ac{BR} approach to determine the full-density operator $\rho_{\text{steady}}$ under  the two lead-boundary conditions. Combining
these two pieces of information is sufficient to determine the nonequilibrium retarded \ac{GF} 
 of the problem using the standard \ac{NRG}  approach
to spectral functions \cite{PetersPruschkeAnders2006,WeichselbaumDelft2007}.
This spectral function  enters the Meir-Wingreen formula, Eq.\ \eqref{eq:meir-wingreen-Qdot}, for determining the transport current in the case of a \ac{QD}. 

For more complicated interacting regions, such as a system of several coupled sites, 
 one has to calculate the lesser \ac{GF} as well, and resort to a more general
transport equation Eq.\ \eqref{eqn:15}  \cite{MeirWingreen1992}.
One example is the interacting resonant level model, where the effective local lead orbital interacting with
the quantum dot via a density-density term must be included in the interacting region. Another example
is a double quantum dot geometry where the each quantum dot only couples to one lead.

By combining the \ac{NRG}, the \ac{BR} approach in combination with the construction of the accurate spectral function, the generic hierarchy problem \cite{PhysRevE.76.031115,RevModPhys.NEQ-QS-2022} 
of the current calculation within the  \ac{BR} approach 
is circumvented.

\subsection{Combining the \ac{NRG} with the Bloch-Redfield approach}
\label{sec:NRG+BRT}

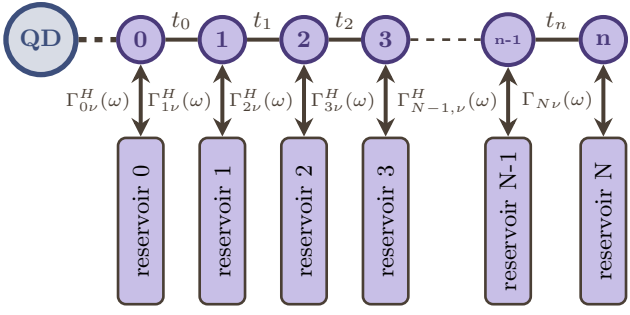
\begin{figure}[t]
\begin{center}
\begin{tikzpicture}[scale=0.6]

\node (A) at (-4mm,0)
  [circle, minimum size=8mm, draw=qdcol, fill=qdcol!20,
   line width=2pt, font=\bfseries, text=qdcol]
  {QD};

\node (f0) at (18mm,0)
  [circle, minimum size=6mm, draw=chainsite, fill=reservoircol,
   line width=1.5pt, font=\small\bfseries, text=chainsite] {0};

\node (f1) at (36mm,0)
  [circle, minimum size=6mm, draw=chainsite, fill=reservoircol,
   line width=1.5pt, font=\small\bfseries, text=chainsite] {1};

\node (f2) at (54mm,0)
  [circle, minimum size=6mm, draw=chainsite, fill=reservoircol,
   line width=1.5pt, font=\small\bfseries, text=chainsite] {2};

\node (f3) at (72mm,0)
  [circle, minimum size=6mm, draw=chainsite, fill=reservoircol,
   line width=1.5pt, font=\small\bfseries, text=chainsite] {3};

\node (fnm) at (99mm,0)
  [circle, minimum size=6mm, draw=chainsite, fill=reservoircol,
   line width=1.5pt, font=\tiny\bfseries, text=chainsite] {n-1};

\node (fn) at (120mm,0)
  [circle, minimum size=6mm, draw=chainsite, fill=reservoircol,
   line width=1.5pt, font=\small\bfseries, text=chainsite] {n};

\draw[dashed, line width=2pt, arrowcol] (A) -- (f0);

\draw[ dashed, line width=1pt, arrowcol] (f3) -- (fnm);

\draw[arrowcol, line width=1.5pt]
  (f0) to node[above, font=\small\itshape] {$t_0$} (f1);

\draw[arrowcol, line width=1.5pt]
  (f1) to node[above, font=\small\itshape] {$t_1$} (f2);

\draw[arrowcol, line width=1.5pt]
  (f2) to node[above, font=\small\itshape] {$t_2$} (f3);

\draw[arrowcol, line width=1.5pt]
  (fnm) to node[above, font=\small\itshape] {$t_{n-1}$} (fn);

\node (r0)  at (18mm,-40mm)
  [fill=reservoircol, draw=arrowcol, line width=1pt,
   rounded corners=3pt, rotate=90, minimum width=2.2cm, minimum height=0.6cm] {reservoir 0};

\node (r1)  at (36mm,-40mm)
  [fill=reservoircol, draw=arrowcol, line width=1pt,
   rounded corners=3pt, rotate=90, minimum width=2.2cm, minimum height=0.6cm] {reservoir 1};

\node (r2)  at (54mm,-40mm)
  [fill=reservoircol, draw=arrowcol, line width=1pt,
   rounded corners=3pt, rotate=90, minimum width=2.2cm, minimum height=0.6cm] {reservoir 2};

\node (r3)  at (72mm,-40mm)
  [fill=reservoircol, draw=arrowcol, line width=1pt,
   rounded corners=3pt, rotate=90, minimum width=2.2cm, minimum height=0.6cm] {reservoir 3};

\node (rnm) at (99mm,-40mm)
  [fill=reservoircol, draw=arrowcol, line width=1pt,
   rounded corners=3pt, rotate=90, minimum width=2.2cm, minimum height=0.6cm] {reservoir N-1};

\node (rn)  at (120mm,-40mm)
  [fill=reservoircol, draw=arrowcol, line width=1pt,
   rounded corners=3pt, rotate=90, minimum width=2.2cm, minimum height=0.6cm] {reservoir N};

\draw[{Stealth[length=7pt,width=7pt]}-{Stealth[length=7pt,width=7pt]},
      line width=1.5pt, arrowcol]
  (f0) to node[left, font=\scriptsize\itshape]
  {$\Gamma^H_{0\nu}(\omega)$} (r0);

\draw[{Stealth[length=7pt,width=7pt]}-{Stealth[length=7pt,width=7pt]},
      line width=1.5pt, arrowcol]
  (f1) to node[left, font=\scriptsize\itshape]
  {$\Gamma^H_{1\nu}(\omega)$} (r1);

\draw[{Stealth[length=7pt,width=7pt]}-{Stealth[length=7pt,width=7pt]},
      line width=1.5pt, arrowcol]
  (f2) to node[left, font=\scriptsize\itshape]
  {$\Gamma^H_{2\nu}(\omega)$} (r2);

\draw[{Stealth[length=7pt,width=7pt]}-{Stealth[length=7pt,width=7pt]},
      line width=1.5pt, arrowcol]
  (f3) to node[left, font=\scriptsize\itshape]
  {$\Gamma^H_{3\nu}(\omega)$} (r3);

\draw[{Stealth[length=7pt,width=7pt]}-{Stealth[length=7pt,width=7pt]},
      line width=1.5pt, arrowcol]
  (fnm) to node[left, font=\scriptsize\itshape]
  {$\Gamma_{N-1,\nu}^H(\omega)$} (rnm);

\draw[{Stealth[length=7pt,width=7pt]}-{Stealth[length=7pt,width=7pt]},
      line width=1.5pt, arrowcol]
  (fn) to node[left, font=\scriptsize\itshape]
  {$\Gamma_{N\nu}(\omega)$} (rn);

\end{tikzpicture}

\caption{Schematic representation of the extension of the Wilson chain by coupling high-energy reservoirs to each Wilson site by a flavor $\nu$-dependent coupling function $\Gamma_{m\nu}^H(\omega)$ as introduced in Refs.\ \cite{OpenChains2017,BoekerAnders2020}
}
\label{fig:2} 
\end{center}
\end{figure}

Recently, Böker and Anders  proposed and implemented an \ac{OC} extension to the time-dependent \ac{NRG} \cite{BoekerAnders2020}. 
In this section, we provide only a short summary of Sec II.C of Ref.~\cite{BoekerAnders2020}. Starting from the exact coupling function
$\Gamma_{\nu}(\omega)$  where  $\nu$  denotes the flavor  index,
it was proven  \cite{OpenChains2017,BoekerAnders2020}  that every Wilson chain of length 
N~\footnote{Actually, the Wilson chain has $N+1$ sites since the first is traditionally labeled by $m=0$ and the last by $N$ \cite{BullaCostiPruschke2008}.} can be augmented by a coupling of each chain site $m$ to a reservoir continuum $\Gamma_{m\nu}^H(\omega)$ to reproduce the exact initial coupling function $\Gamma_\nu(\omega)$ determined by a continued fraction expansion. The
coupling functions $\Gamma_{m\nu}^H(\omega)$ compensate for the exponential decay of the hopping matrix element $t_{m\nu}\propto \Lambda^{-m/2}$ and vanish in the  limit of the \ac{NRG} discretization parameter $\Lambda \to 1^+$
where the semi-infinite chain becomes an exact representation of the continuum as shown in Refs.\ \cite{KrishWilWilson80a,BullaCostiPruschke2008}.
This is schematically depicted in Fig.\ \ref{fig:2}. For $\Lambda \to 1^+$, only a single reservoir at the end of the chain remains present \cite{DMRGwithDMFT2004}.

The conventional  \ac{NRG} \cite{BullaCostiPruschke2008} can be interpreted as an approximation where all coupling functions $\Gamma_{m\nu}^H(\omega)$ are set to zero.
Due to the tremendous  success of the  \ac{NRG} in the last five decades, 
it is well justified to view the $\Gamma_{m\nu}^H(\omega)$ as a weak perturbation, since their spectral integrals vanish for  $\Lambda \to 1$. 

By combining the \ac{NRG} with a \ac{BR} approach  in secular approximation  \cite{MayKuehn2000,RevModPhys.NEQ-QS-2022},
the time evolution of  the reduced density matrices in the interaction representation 
was calculated \cite{BoekerAnders2020}   and used to obtain the nonequilibrium time evolution of local operators \cite{AndersSchiller2005,AndersSchiller2006} 
or the spectral functions \cite{BoekerPhD2021,BoekerGf2022}. 

While the  reduced density matrices are stationary in the \ac{TD-NRG} approach \cite{AndersSchiller2005,AndersSchiller2006},
the coupling to the reservoirs generate a weak time dependence on $\rho$.
The rates in this hybrid approach are governed by the set $\{\Gamma_{m\nu}^H(\omega)\}$
as well as the boundary conditions. It can be shown analytically \cite{BoekerAnders2020,BoekerPhD2021}  that the steady-state density matrix  after a quench corresponds to the \ac{FDM} at infinite time for equilibrium boundary conditions: the system relaxes to
thermal equilibrium.

In the secular approximation the master equations for the diagonal and the off-diagonal matrix elements of the density matrix decouple  \cite{MayKuehn2000,BoekerAnders2020,BoekerPhD2021,RevModPhys.NEQ-QS-2022}. The off-diagonal matrix elements contain information about
the dephasing, the Lamb shift, coherent oscillations and the approach of the steady state from
an arbitrary initial condition. Due to the decoupling, we can focus on the master equation for
diagonal matrix elements.

Using the  complete \ac{NRG} basis \cite{AndersSchiller2005,AndersSchiller2006},
$\{|l,e;m\rangle\}$, where $l$ denotes the approximate energy eigenstate present in the \ac{NRG} at iteration $m$ and discarded after the iteration, i.e., $H_m^{\rm NRG} |l,e;m\rangle = E_l^{m} |l,e;m\rangle$, and $e$ accounts for all \acl{DOF} of the remaining parts of the Wilson chain $m+1,\cdots,N$, 
the master equation for the diagonal matrix elements,
\begin{widetext}
\begin{subequations}
\label{eq:master-equation-block}
\begin{eqnarray}
\label{eq:master-equation}
\dot \rho_{l_1,l_1}(m_1;t) &=& \sum_{m_2} \sum_{l_2} 
(
\Gamma_{l_2,l_1}(m_2,m_1)    \rho_{l_2,l_2}(m_2;t)  
-\Gamma_{l_1,l_2}(m_1,m_2)    \rho_{l_1,l_1}(m_1;t)  
)
\end{eqnarray}
was derived in Ref.\ \cite{BoekerAnders2020} where the rates are given by
\begin{eqnarray}
\label{eqn:91}
\Gamma_{l_1,l_2}(m_1,m_2) &=& 
\frac{2}{d^{N-m_1}} \sum_{\alpha }
\left(W^{(m_1,m_2)}_{l_1,l_2}(\alpha)[1 - f_\alpha(\w_{l_1,l_2})]
 +W^{(m_2,m_1)}_{l_2,l_1}(\alpha)f_\alpha(\w_{l_2,l_1})
\right),
\\
\label{equ-Wl1l2-m1m2}
W^{(m_1,m_2)}_{l_1,l_2}(\alpha)  &=& 
\sum_{\tilde m}^{N}  \sum_{ \nu} \Gamma_{ \tilde m\alpha\nu}(\w_{l_1,l_2})
X^{\tilde m\alpha\nu}_{l_1,l_2}(m_1,m_2),
\\
X^{\tilde m\alpha\nu}_{l_1,l_2}(m_1,m_2) &=& \sum_{e_1,e_2} 
\bra{l_1,e_1;m_1} f^\dagger_{\tilde m \alpha\nu } \ket{l_2,e_2;m_2}
\bra{l_2,e_2;m_2} f_{\tilde m\alpha\nu } \ket{l_1,e_1;m_1} . \label{eq:14d}
\end{eqnarray}
\end{subequations}
\end{widetext}
where $d$ denotes the  number of local states when adding one additional Wilson site to the chain ($d=4$ for a single spin degenerate lead.)
The index $\tilde m$ runs over all reservoirs of each flavor $\nu$ and lead $\alpha$. Each lead $\alpha$ can have its individual chemical potential
typical for nonequilibrium boundary-driven quantum systems as well as different
temperatures. Both aspects are encoded in the Fermi function $f_\alpha(\omega)= [\exp(\beta_\alpha (\omega-\mu_\alpha))+1]^{-1}$
where $\mu_\alpha$ encodes the potential different chemical potential of each lead. $\w_{l_2,l_1}$ 
denotes the energy differences $\w_{l_2,l_1}
=E^{m_2}_{l_2}-E_{l_1}^{m_1}$.

Böker and Anders \cite{BoekerAnders2020} were interested in
the modification of the local time evolution after a quench as predicted by the \ac{TD-NRG}.
In this case, not only the full time dependency of the diagonal density matrix elements was needed
but also the real-time solution of the off-diagonal matrix elements.
In this paper, however, we focus on the boundary-driven steady-state solution of Eq.\ \eqref{eq:master-equation}. 
Eq.\ \eqref{eq:master-equation} can be casted into an eigenvalue equation with
a non-symmetric \ac{BRT}. The steady-state density matrix is given by the normalized 
eigenvector of the zero eigenvalue. For the equilibrium, it was shown  \cite{BoekerAnders2020}
that this corresponds to the thermal \ac{FDM}.

\subsection{Simplified master-equation approach}
\label{sec:III-B}

It was analytically shown \cite{BoekerAnders2020} that in thermal equilibrium, i.e., for $\beta_\alpha=\beta,\mu_\alpha=\mu=0$,
 the steady state solution of the master equation is given by the thermal density matrix  \cite{WeichselbaumDelft2007}: Independent of the magnitude of the decay rates $W^{(m_1,m_2)}_{l_1,l_2}(\alpha)$, the density operator
 $\rho$ approaches the Boltzmann form in the steady state provided the reservoirs
 are in thermal equilibrium.
 Essentially the closed quantum system defined by the finite-size Hamiltonian $H_{\text{NRG}}$ thermalizes in the presence of the external reservoirs  via exchange of particles and energy consistent with  textbook  statistical mechanics  \cite{Feynman72}.

The major disadvantage of the full master equation, Eq.\ \eqref{eq:master-equation}, is the coupling of the reduced matrix elements over all NRG 
energy shell $m_1$ and $m_2$ which makes this approach numerically very expensive \cite{BoekerAnders2020}. 
While an efficient implementation of the \ac{NRG} heavily relies on the partitioning the 
Hamiltonian into smaller  sub-blocks, defined by the conserved quantum numbers such as charge
and total spin, the \ac{BRT} generated by the rates $\Gamma_{l_1,l_2}(m_1,m_2) $ in Eq.\ \eqref{eq:master-equation}
couples the density matrix elements of all these different  sub-blocks due the exchange of electrons with the reservoirs
characteristic for an open quantum systems.
In a full two-lead setup, with typically $N_s=$4000-6000 kept state, and a chain length of $N=50$, this would imply finding the steady state of the density matrix  which is of the order $N \times N_s(d-1)\approx 3\times 10^{6}$ elements. 

The challenge is to find a good single shell
approximation by making use of the energy hierarchy for a numerically efficient and accurate algorithm.

In the NRG, the discarded high-energy states at iteration $m$ remain unaltered in the following later iterations $m'>m$. Only the kept states are refined in the next iterations.
Therefore, we propose to  solve the master equation  in Eq.\ \eqref{eq:master-equation-block} only within a Wilson shell. For that purpose we calculate the transition rates $\Gamma_{r,s}(m,m)$  for all states present at iteration $m$ by pretending that this is the last iteration. This implies that (i) the degeneracy prefactor in Eq.\ \eqref{eqn:91} is absent, (ii) we use the full coupling
function $\Gamma_m$ for the iteration $m$ and $\Gamma^H_{\tilde m}(\omega)$ for $\tilde m<m$.
The reservoirs $\tilde m' >m$ can be neglected \cite{BoekerAnders2020}.

\subsubsection{Open-chain full-density matrix approach }
 
Using the assumptions outlined in the previous section, we propose an algorithm which utilizes the \ac{NRG} hierarchy of eigenstates and is based on the resulting
hierarchy of  master-equations each only involve states present at any given \ac{NRG} iteration. This   \ac{OC-FDM} approach
provides an approximate solution for the steady state density matrix and is numerically  efficient.

Let us consider the \ac{NRG} iteration $m_\text{min}$ after which the first time NRG states are discarded. 
As outlined above, we consider all states as discarded for the master equation, focus on the diagonal matrix elements
and determine the stationary density matrix from  the transition rates $\Gamma_{l,l'}(m_\text{min},m_\text{min})$ introduced in Eq.\
\eqref{eq:master-equation-block}.

Since the non-symmetric \ac{BRT}  of the states even on a single Wilson shell
is usually too large for exact diagonalization due to the lack of locally conserved quantum numbers, we use a bi-orthogonal Lanczos algorithm \cite{SaadSparseLinearSystemsBook2003} with a typical Krylov space dimension  $N_k=100$ and the thermal density matrix as starting vector 
$\rho^0_{rr}$ to obtain an approximated representation 
of the eigenstates of the \ac{BRT} \cite{BoekerAnders2020} which is then diagonalized exactly.

We end up with a normalized right eigenvector of the \ac{BRT},
\begin{eqnarray}
\sum_{r} \rho_{rr}(m_\text{min}) = 1 ,
\end{eqnarray}
for the eigenvalue zero, representing the steady-state solution of the master equation. 
The index $r$ runs over all  \ac{NRG} eigenstates present at iteration $m$.

The Lanczos algorithm is known for representing the smallest and the largest eigenvalue very accurately while the representation of the intermediate spectrum is usually poor.  
To find the steady state, only the single eigenvector with the eigenvalue zero  needs to be determined.
Consequently, the Lanczos algorithm is optimally suitable for this problem. Using thermal density matrix as starting vector 
ensure fast convergence since the Lanczos algorithm stops already after a single step in equilibrium. This even holds at finite
bias as long as the Wilson chain energy scale $\omega_m\gg eV$.

The boundary temperatures $T_\alpha$ and the chemical potentials $\mu_\alpha$ of lead $\alpha$
enter the Fermi functions in definition of the rates, Eq.\ \eqref{eqn:91} as well as the finite bias $\mu_\alpha$. In case of the equilibrium, we find $\rho_{rr}(m) = \exp(-\beta E_r)/Z_m, Z_m= \sum_r \exp(-\beta E_r)$.

After each iteration, we partition the states into discarded states $l$ and kept states $k$ refined in the next \ac{NRG} iteration. The preliminary representation of the density operator in the \ac{NRG} at iteration $m_\text{min}$
is therefore given by the two parts
\begin{eqnarray}
\hat \rho &=&   
\frac{1}{d^{N-m_\text{min}}} \sum_{le} \rho_{ll}^{m_\text{min}} |le;m_{min}\rangle\langle le;m_\text{min}|
\nonumber
\\
&& + \frac{1}{d^{N-m_\text{min}}} \sum_{ke} \rho_{kk}^{m_\text{min}} |ke;m_{min}\rangle\langle ke;m_\text{min}| .
\end{eqnarray}
The prefactor $d^{m_\text{min}-N}$ cancels the degeneracy factor $d^{N-m_\text{min}}$ of the states when tracing over 
the environment variable $e$ for a given Wilson chain length of $N$ and ensures that $\text{Tr}[\hat \rho]=1$.

The total spectral weight contributions of the density operator  to each sector are given by
\begin{eqnarray}
\rho^D(m) &=& \sum_l  \rho_{ll}^{m} \,,\\
\rho^K(m) &=&1- \rho^D(m) =  \sum_k  \rho_{kk}^{m},
\end{eqnarray}
where the superscript $D(K)$ refer to the discarded (kept) sector.

Our hierarchy algorithm assumes that the  density matrix elements $\rho_{ll}^{m}$ of the discarded states are frozen, and the \ac{NRG}
algorithm just refines the kept states by adding another Wilson site to the Hamiltonian. Applying the same argument as above, we obtain
a new steady state solution of the master equation at iteration $m_\text{min} +1$. 
We recall, however, that the refined total spectral weight is $\rho^K(m)$, 
and we arrive at the representation
\begin{eqnarray}
\hat \rho(m_\text{min}+1) =
\frac{1}{d^{N-m_\text{min}}} \sum_{le} \rho_{ll}^{m_\text{min}} |le;m_\text{min}\rangle\langle le;m_\text{min}|
\nonumber \\
+ \frac{d \rho^K(m_\text{min})}{d^{N-m_\text{min}}}  \sum_{re'} \rho_{rr}^{m_\text{min}+1} |re';m_{min}+1\rangle\langle re';m_\text{min}+1| \,,
\nonumber \\
\end{eqnarray}
where $r$ runs over all states present at the NRG iteration $m_\text{min}+1$.

This defines a recursion: Essentially, the spectral weight of the coarse-grained representation of $\hat \rho(m_{\rm min})$ is refined in $\hat \rho(m_{\rm min}+1)$ by splitting the
kept states of iteration $m_{\rm min}$ into a finer low energy resolution \cite{BullaCostiPruschke2008}.
This step is repeated in all following iterations. 

To this end we arrive at the final result for the approximate \ac{OC}-\ac{FDM} density operator,
\begin{eqnarray}
\hat \rho(N) &=&\sum_{m=m_{\rm min}}^N \frac{\bar \rho^K_{m-1} }{d^{N-m} }
\sum_{l e}  \rho_{ll}^{m}   |le; m  \rangle \langle le; m|,
\label{eq:rho-final}
\end{eqnarray}
where we label all states at iteration $N$ as discarded \cite{AndersSchiller2005,AndersSchiller2006}, define the product
\begin{eqnarray}
\bar \rho^K_{m-1} = \prod_{\bar m=m_{\rm min}}^{m-1} \rho^K(\bar m)
\end{eqnarray}
and set $\rho^K(m_{\rm min}-1)=1$ since no truncation has occurred prior to $m_{\rm min}$. It is straight forward to show that 
$\text{Tr}[\hat \rho(N)]=1$ by construction.

The density operator can be brought into the form of a \ac{FDM} \cite{WeichselbaumDelft2007},
\begin{eqnarray}
\label{eq:32}
\hat \rho(N) &=&\sum_{m=m_{\rm min}}^N w_m \hat \rho_{dd}^{(m)},
\end{eqnarray}
where  the trace of $\hat \rho_{dd}^{(m)} $ acting only on the Fockspace spanned by the discarded states
at iteration $m$,
\begin{eqnarray}
\hat \rho_{dd}^{(m)} = \frac{1}{d^{N-m} \rho^D(m)} \sum_{l e}  \rho_{ll}^{m}   |le; m  \rangle \langle le; m|,
\end{eqnarray}
is normalized. i.e., $\text{Tr}[\hat \rho_{dd}^{(m)}]=1$, and the weight factors $w_m$ are given by
\begin{eqnarray}
\label{eq:w-m-OC-FDM}
 w_m = \bar \rho^K_{m-1} \rho^D(m).
\end{eqnarray}

The factor $\bar \rho^K_{m-1}$ determines the relative weight which the normalized eigenvector of the
\ac{BRT} contribute to the density operator at the iteration $m$. In the first iterations, $\bar \rho^K_{m-1}\approx 1$, since the discarded states do not contribute to the reduced density matrix.
Once $\omega_m< T$, we reach equal population in equilibrium, and the contribution of the discarded states starts to increase significantly. Consequently $\rho^K_{m-1}$ decreases while $\rho^D(m)$ increases. Hence $w_m$ increases, goes through a maximum and decreases rapidly.
Eventually, the RG procedure does not make any sense any more and must be stopped. In  Wilson's
\ac{NRG} the iteration is stopped once $\omega_m\approx  T$. Since we are applying the
\ac{OC}-\ac{FDM} approach to nonequilibrium, we use $\rho^K_{m-1}$ as a cutoff criteria.
We define a minimum value $\rho_c$ and stop the renormalization group flow if $ \rho^K_{m-1}<\rho_c$.

Let us consider the equilibrium. For $m\ll N$, $\rho^D(m)=0$. When setting all the discarded matrix elements $\rho_{ll}(m<N)=0$, and define $T\approx \omega_N$
we recover the Wilson representation \cite{Wilson75} of the density matrix at the final iteration $N$.

\subsubsection{Infinite temperature limit}

The question arises, how accurate is this approach representing the correct density matrix. 
One extreme case would be the infinite temperature limit $(\beta=0)$
where all states are equally populated. For $\bar N_s= N_s d$ states present at iteration $m_\text{min}$, we have $\rho_r^{m_\text{min}}= 1/N_s d$. After the iteration, we only keep $N_s$ states, which are refined in the next step. Then
$\rho^D(m_\text{min})= N_s(d-1)/(N_s d)= (d-1)/d$. Since we have again $N_s d$ states present in the next iteration, the factors $\rho_r^m=1/N_s d$ are independent of m and we finally arrive at
\begin{eqnarray}
\hat \rho(N) &=&
\frac{1}{N_s d^{N+1-m_{\rm min}}}
\sum_{m=m_{\rm min}}^N \sum_{l e}  |le; m  \rangle \langle le; m|
\\
&=& \frac{1}{N_s d^{N+1-m_{\rm min}}}\hat 1,
\end{eqnarray}
which proves that our algorithm correctly reproduce an equal spectral weight of all states.

\subsubsection{Equilibrium \ac{FDM} representation}

In the \ac{FDM} representation introduced by Weichselbaum and von Delft \cite{WeichselbaumDelft2007}, all \ac{NRG} energy shells contribute, and the thermal density operator is given by the expression
\begin{eqnarray}
\label{eq:29}
\hat \rho(N) &=&
\sum_{m=m_{\rm min}}^N \sum_{l e}  \frac{e^{-\beta E_l^m}}{Z}|le; m  \rangle \langle le; m| 
\nonumber \\
&=&
 \sum_{m=m_{\rm min}}^N w_m \hat \rho_{dd}^{(m)},
\end{eqnarray}
with the partition function
\begin{eqnarray}
Z_{\rm{FDM}} = \sum_{m=m_{\rm min}}^N  d^{N-m} \sum_{l}  e^{-\beta E_l^m} 
= \sum_{m=m_{\rm min}}^N\tilde Z^D_m.
\label{eq:35}
\end{eqnarray}
The factor $d^{N-m}$ accounts for the state degeneracies in the partition function,
$w_m=  \tilde Z^D_m/Z_{\text{FDM}}$, and $Z^D_m= \sum_{l}  e^{-\beta E_l^m}= \tilde Z^D_m/d^{N-m}$
contains the contribution of the discarded states.  

When operating the NRG in a limit when
the chain length N is much larger then the iteration $M$ at which the temperature is of the order of the
energy scale $\omega_M\propto \Lambda^{-\frac{N-1}{2}}$ of Wilson shell, i.e., $\omega_M \approx T$,
the \ac{FDM} approach differs from the  hierarchy algorithm outlined above. 
However, we will see below that the \ac{OC-FDM} and the \ac{FDM} approach yields almost identical spectral functions and weight factors $w_m$. The reason is the following. In the \ac{FDM} approach, the ground state energy shift must be taken into account such that
\begin{eqnarray}
Z^D_m &=& e^{\beta \Delta E_g(m)}  \sum_{l}  e^{-\beta E_l^m({\rm NRG})}
\nonumber 
\\
&=&  e^{\beta \Delta E_g(m)} Z^D_m({\rm NRG})
\end{eqnarray}
where $E_l^m({\rm NRG})$ is the NRG spectrum used in the standard NRG run
as well as the single-shell master equation,
and $\Delta E_g(m)$ is the relative ground state energy shift at iteration $m$ with respect to the absolute ground state energy $E_g(N)$.
$Z^D_m({\rm NRG})$ enters the $\rho^m_{ll}$ of the \ac{OC-FDM} approach.  The condition
$Z^D_m < Z^D_m({\rm NRG})$  holds: The \ac{OC-FDM} ignores the refinement of the kept states in the later iterations such that the contribution $\rho^K(m)$ is overestimated  at the iteration $m$. As a consequence, 
a reduction of the numerator in $\tilde Z^D_m$  is partially canceled by the reduction of $Z_{\text{FDM}}$ in the weight factors $w_m$ .

\section{Boundary -riven current through a single orbital quantum dot}
\label{Sec:IV}

We consider the prototypical problem \cite{KrishWilWilson80a,MeirWingreen1994,goldhaberSET98,Kouwenhoven2000} where
the interaction region depicted in Fig.\ \ref{fig:1} is a
\ac{QD} represented by
 a single spin-degenerate orbital subject,
\begin{eqnarray}
H_S= \sum_\sigma \epsilon_{d\sigma} d^\dagger_\sigma d_\sigma + U n^d_\uparrow n_\downarrow^d,
\end{eqnarray}
where $d^\dagger_\sigma$ creates an electron with spin $\sigma$ and energy $ \epsilon_{d\sigma} $ in the \ac{QD}
orbital and $U$ denotes the Coulomb repulsion between two localized electrons with opposite spin. An external magnetic field $B_0$
enters via Zeeman splitting of the local orbitals $ \epsilon_{d\sigma} =\epsilon_{d} + \sigma g\mu_b B_0
= \epsilon_{d} + \sigma h_0, h_0=g\mu_b B_0$. When adding the two leads, we obtain a \ac{SIAM}
which contains the Kondo physics in equilibrium.

Throughout the paper, we neglect weak logarithmic
correction stemming from the lead polarization in a finite magnetic field due to the wide band limit \cite{KondoModelComparisonGebhard2020}, and  the  
details of the lead spectral function. We use a constant
$\Gamma_\alpha(\omega)=\Gamma_\alpha$ on the support $I=[-D:D]$ where $D/\Gamma=10^{3}$ unless specified otherwise, and define $\Gamma=\Gamma_L +\Gamma_R$.

The challenge of this problem is the treatment of the strong coupling regime 
out of equilibrium. In equilibrium, the relevant crossover energy scale $T_K$
is exponentially suppressed with an increasing 
Coulomb interaction $U$ \cite{KrishWilWilson80a,KrishWilWilson80b}.

\subsection{Effective description of the quantum dot problem}

\label{sec:mapping-single-lead}

In  a two-lead (or multiple lead) \ac{QD} problem, only the bonding combination 
\begin{eqnarray}
\tilde V_\sigma  c^\dagger_{0\sigma} =  \sum_{k\alpha} V_{k\alpha \sigma} c^\dagger_{k\alpha\sigma}
\end{eqnarray}
couples to the local orbital of the \ac{QD}
where the effective coupling strength $\tilde V$ is obtained from the anticommutator
of the fermion operators,
\begin{eqnarray}
|\tilde V_\sigma  |^2= \sum_{k\alpha} |V_{k\alpha\sigma} |^2,
\end{eqnarray}
where we identified the flavor $\nu$ with the spin.
The influence of the leads onto to
local orbital dynamics is fully determined \cite{GlazmanRaikh1988} by the coupling function
$\Gamma_\sigma (\omega) = \Gamma^L_\sigma (\omega) +\Gamma^R_\sigma (\omega)$
and the  non-interacting part of lesser self-energy,
\begin{eqnarray}
\Sigma_{\sigma,0}^<(\omega) &=& 2if_L(\omega)  \Gamma^L_\sigma(\omega) + 2i f_R(\omega)  \Gamma^R_\sigma(\omega).
\end{eqnarray}
where the different chemical potentials of the two leads enter through the different Fermi functions.
This allows the definition of an
effective occupation function
\begin{eqnarray}
f_{\text{eff}}\,(\omega) &=& \frac{\Sigma_{\sigma,0}^<(\omega)}{i2\,\Gamma_\sigma (\omega)}
\end{eqnarray}
imposing the nonequilibrium boundary condition on the   \ac{QD}. 

Assuming that the Meir-Wingreen condition holds, i.e., $ \Gamma^L_\sigma(\omega) =R  \,\Gamma^R_\sigma(\omega)$,  we arrive at
\begin{eqnarray}
\label{eq:f-eff}
f_{\text{eff}}\,(\omega) &=&  \left[
\frac{R}{1+R}\, f_L(\omega) + \frac{1}{1+R}\, f_R(\omega) 
\right]
\end{eqnarray}
which coincides with the Fermi-Dirac distribution for $\mu_L=\mu_R$ and $\beta_L=\beta_R$ in equilibrium 
but accounts for a two step function at finite bias.
Note that $f_\alpha(\omega)= [\exp(\beta_\alpha(\omega -\mu_\alpha))+1]^{-1}$  includes not only lead dependent chemical potentials $\mu_\alpha$
but also different lead temperatures $T_\alpha=1/\beta_\alpha$ relevant for thermotransport currents.

In the effective description of the problem, $f_{\text{eff}}(\omega)$ accounts for both of 
the lead or flavor dependent Fermi functions in Eq.\ \eqref{eqn:91}. 
Since the current transport is determined by Eq.\ \eqref{eq:meir-wingreen-Qdot}, it is sufficient to calculate the 
nonequilibrium retarded \ac{GF} in an effective model by substituting $f_{\text{eff}}(\omega)$
in Eq.\ \eqref{eqn:91} for the rates entering the master equation.

\subsection{Effective open-chain \ac{NRG} nonequilibrium approach}
\label{sec:algorithm-single-lead}

A Wilson chain \cite{BullaCostiPruschke2008} of length $N$ is constructed 
which is then augmented by the appropriate $N$ reservoir coupling functions $\Gamma^H_\nu(\omega)$  
obtained by a continued fraction expansion as derived in Refs.\ \cite{OpenChains2017,BoekerAnders2020}
using the total coupling function $\Gamma_\sigma (\omega)=(1+R)\,\Gamma^R(\omega)$ of both leads.
The effective embedding of the Wilson chain in the set of reservoir is depicted in Fig.\ \ref{fig:2}. 
The standard \ac{NRG} \cite{BullaCostiPruschke2008} determines the approximate eigenstates and eigenbasis of the \ac{NRG} chain Hamiltonian. 
This is done for different z-values ($0<z\le 1$) to allow for z-averaging \cite{YoshidaWithakerOliveira1990,Oliveira1994,AndersSchiller2005,AndersSchiller2006}.
It was shown \cite{YoshidaWithakerOliveira1990,Oliveira1994} that by selecting the first \ac{NRG} discretization interval
$I_0=D[\Lambda^{-z},1]$ and varying  $z$ from zero to unity is sweeping over the conduction state continuum. For practical
purpose, a finite set of $z_i\in (0:1]$ values are selected and the results averaged over all $z$ values to eliminate some discretization
artifacts \footnote{For more details see Refs.~\cite{YoshidaWithakerOliveira1990,Oliveira1994} and the NRG review \cite{BullaCostiPruschke2008}.}
Unless otherwise stated, we used $N_z=2$ and $z=0.25,0.75$, $\Lambda=1.8$ and average the results as  proposed by  Oliveira and Oliveira \cite{Oliveira1994}.

After each  iteration $m$,
the simplified master-equation approach is employed. The 
transition rates $\Gamma_{r,s}(m,m)$  are obtained from Eq.\ \eqref{eqn:91}, where we extended the rates to all states $r,l$ present at the iteration.
We store the density matrix elements of the discarded states which contribute to the \ac{OC-FDM} approach  
while matrix elements of the kept states  are refined in the next iteration.
The boundary condition of a finite bias and a finite temperature gradient enters
via  effective nonequilibrium bath distribution function $f_{\text{eff}}(\omega)$ 
the transition rates $\Gamma_{r,s}(m,m)$. 

The effective nonequilibrium bath distribution function $f_{\text{eff}}(\omega)$ 
has a very interesting property for the symmetric junction, $R\approx 1$:
It agrees with the Fermi-Dirac distribution for excitation energies $|\omega|\gg |eV=\mu_L-\mu_R|$ 
while $f_{\text{eff}}(\omega) \to 1/2$ for  $|\omega|< |eV|$ located
in the bias window. In the latter range the distribution function approaches the infinite temperature limit \cite{Oguri2007,LotemWeichselbaum2020}.
As a consequence the many-body eigenstates become equally
populated once the low-energy scale $\omega_m \propto \Lambda^{-(m-1)/2}$ lies within the bias window, i.e., $\omega_m< \text{min}(|\mu_L|,|\mu_R|)$. In the Wilsonian NRG, the iteration (RG-flow) stops at this point, while in the \ac{FDM} approaches additional iteration lead to a refined
 \cite{WeichselbaumDelft2007} spectrum in this window.

The quantity $\bar \rho^K_{M-1}$ denotes the total contribution to the density operator 
of all states discarded all further iterations $m$, $M\le m$. 
For  $\omega_m< \text{min}(|\mu_L|,|\mu_R|)$, $\bar \rho^K_{m-1}$ is rapidly
decaying similar to the equilibrium FDM approach when $\omega_m\ll T$. We define a cutoff $\rho_{c}$ such that we stop the NRG at iteration  $M$, once   if $\bar \rho^K_{M}<\rho_{c} $.
This condition  substitutes NRG equilibrium criteria where the iterations are typically stopped at  $\omega_m \approx  T$ \cite{Wilson75,KrishWilWilson80a,KrishWilWilson80b}. We used typical values $\rho_{c}=0.01-0.02$
which corresponds a stop of the RG flow  after about 10 additional iterations once  $\omega_m \approx  T$ is reached which corresponds to a reasonably application of the \ac{FDM} approach \cite{WeichselbaumDelft2007}.
  
 \begin{figure}[t]
\begin{center}
\includegraphics[width=0.9\columnwidth, trim=0 0 0 0, clip]{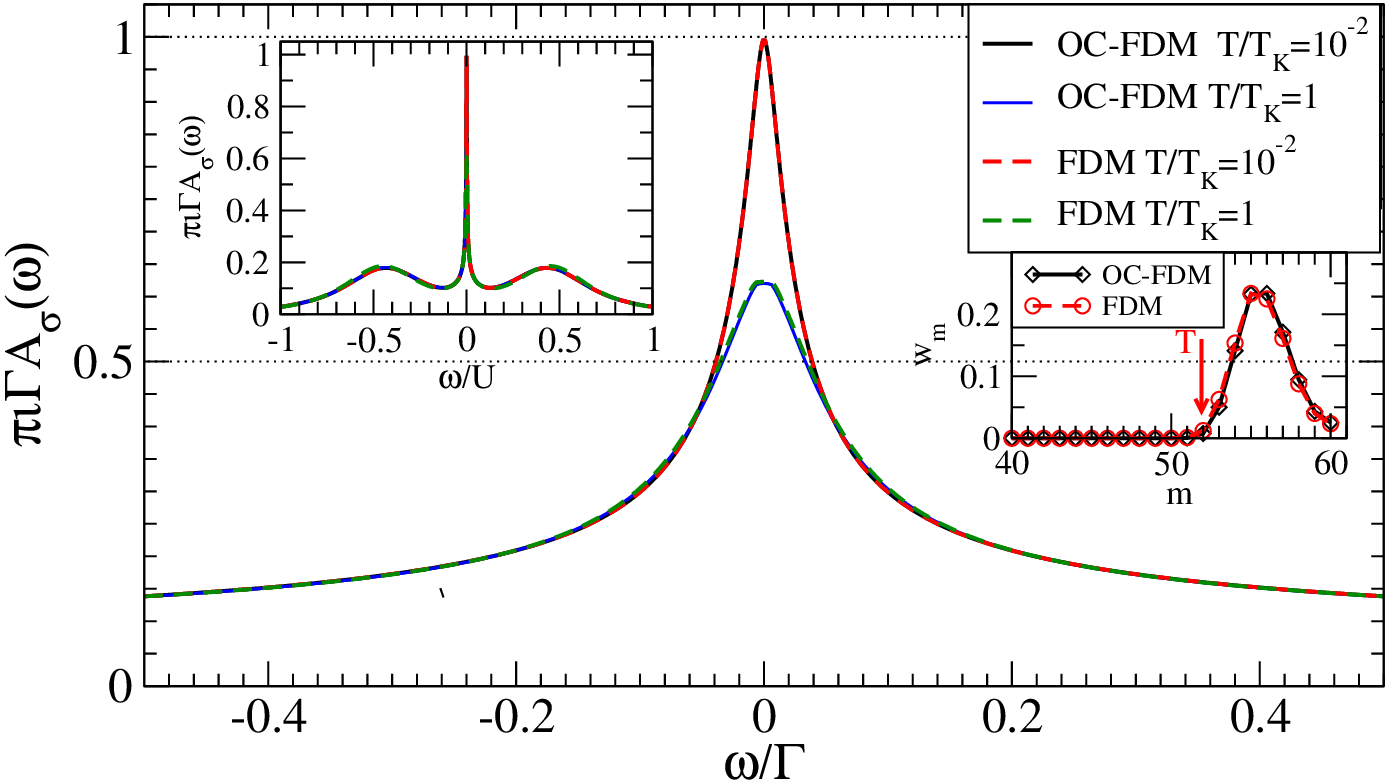}
\caption{
Spectral functions $A_\sigma(\omega)= -{\rm Im}(G_\sigma(\omega+i\delta)/\pi)$  obtained by the \ac{FDM}-algorithm \cite{WeichselbaumDelft2007}
and  the \ac{OC-FDM} outlined above  for $T=T_K,0.01\,T_K$ and
 $U/\Gamma=12,\epsilon_d/\Gamma=-6$ where  $T_K/\Gamma=0.0242$ was obtained by the condition $G(T_K)=1$.
The inset on the left-hand side  shows the same spectral function but over a larger frequency range. The inset on the right-hand side the weight factors $w_m$ 
defined in Eq.\ \eqref{eq:32}, OC-FDM in black diamonds and the FDM weight in red circles,
as functions of the iteration $m$  at  $T=0.01\,T_K$ with $N_\text{iter}=60$ NRG iterations.
The red arrow indicated the iteration $m$ at which $\omega_m\approx T$.
NRG parameter: $\Lambda=1.8,N_s=1000,D=10^3\,\Gamma, N_{\text{iter}}=70,b=0.5,\omega_0=0.1\,T,N_z=2$.}

\label{fig:3}
\end{center}
\end{figure}

To this end the \ac{OC-FDM} is used to calculate the equilibrium and nonequilibrium retarded Green's function whose spectrum enters the current integral. The broadening of the spectral function must be adapted to  accommodate the regime of equal occupation.   We adapted a  transition from log-Gaussian to a Gaussian broadening of the Lehmann representation  \cite{WeichselbaumDelft2007}  which occurs on the Wilson shell close to maximum of the  weight factors $w_m$ making the approach similar to the \ac{FDM} approach. We also performed $z$-averaging with mostly $N_z=2$ z-values as suggested in Ref.\ \cite{Oliveira1994}. 
In all calculations the hybridization is corrected  by the factor $A_\Lambda$ defined in Eq.\ (5.20) in the seminal  \ac{NRG} paper by Krishnamurthy et al \cite{KrishWilWilson80a} to
connect the results with a continuum theory.

\section{Results}
\label{sec:results}

\subsection{Equilibrium spectral function}
\label{sec:equilibrum}

We compare the spectral functions $A_\sigma(\omega)$
for the \ac{QD} orbital obtained from the \ac{FDM}-algorithm,\cite{WeichselbaumDelft2007},
and the  \ac{OC-FDM}  for two temperatures $T=0.01\,T_K$ and $T=T_K$ in Fig.\ \ref{fig:3}
for $U/\Gamma=12, \e_d=-U/2$.

We determined the transport Kondo temperature \cite{goldhaberSET98}
$T_K^{\text{trans}}$ from the zero-bias conductance 
\begin{eqnarray}
G(T)=\left .\frac{\partial I(T,V)}{\partial V}\right|_{V=0}
\end{eqnarray}
by setting $G(T_K^{\text{trans}})/G(0)=1/2$ \footnote{The absolute value of $T_K$ depends slightly on the number of kept NRG state $N_s$
and whether the correction factor $A(\Lambda)$ -- see Eq (5.20) in the seminal \ac{NRG} paper of Krishna-Murthy et al \cite{KrishWilWilson80a} --
is taken into account which we did for all our calculations}. 
The Kondo temperatures $T_K({\rm NRG})$ determined by the Wilson criteria \cite{KrishWilWilson80a} given by ${\chi_{loc}}(T_K)\, T_K=0.7$
are about a factor $0.45$ smaller than the $T_K^{\text{trans}}$, i.e., $T_K({\rm NRG})/\Gamma= 4.578 \times 10^{-2}$ for $U/\Gamma=8$
and $T_K({\rm NRG})/\Gamma= 1.09  \times 10^{-2}$ for $U/\Gamma=12$. Throughout the paper, we use $T_K=T_K^{\text{trans}}$ which allows to make contact to experiments \cite{goldhaberSET98}:
The crossover scale $T_K$ is only defined up to some constant of $O(1)$.
We found $T_K^{\text{trans}}/\Gamma = 2.42 \times 10^{-2}$ for $U/\Gamma=12,\e_d=-U/2$
from our algorithm which is very close to the value reported in
Ref.\  \cite{SchwarzPRL-TransportQD2018}. 

\begin{figure}[b]
\centering

\includegraphics[width=0.45\textwidth,trim = 0 0 0 0, clip]{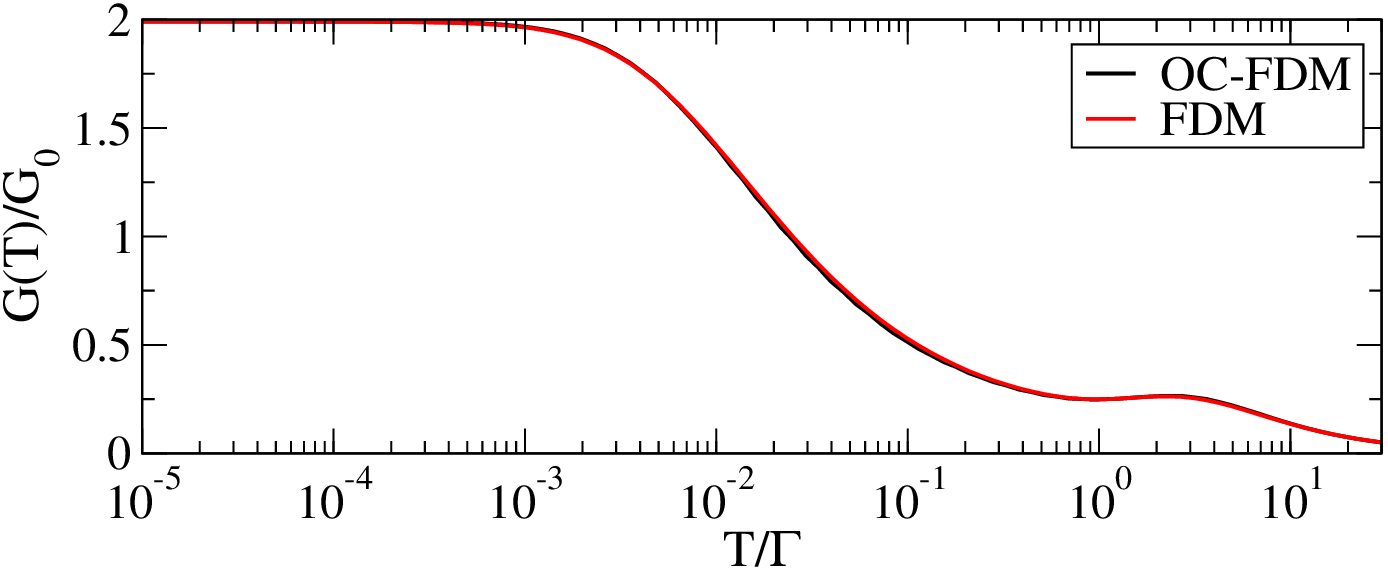}

\caption{Comparison between the zero-bias conductance, calculated with the \ac{OC}-\ac{FDM} spectral function 
and obtained by the \ac{FDM}-algorithm \cite{WeichselbaumDelft2007} combined with the 
equation of motion approach \cite{BullaHewsonPruschke98} for the same NRG parameters
and $U/\Gamma=12,\epsilon_d/\Gamma=-6$. 
}

\label{fig:4-g-vs-t}
\end{figure}

To have a direct comparison between both approaches, we used the identical Wilson chain lengths, $N=60$ for $T/T_K=0.01$ and $N=44$ for $T/T_K=1$,
roughly identical broadening parameter \cite{WeichselbaumDelft2007} for both approaches
and $\rho_c=0$. This would correspond to a cutoff $\rho_c\approx 0.04$ in the \ac{OC}-\ac{FDM} approach
when setting $N\ll 60$. 
In this case we observe an excellent agreement between the approaches which are of slightly different nature:
The \ac{NRG} spectrum is identical and the only difference comes from the different setup of the full-density matrix. Since the weight factors $w_m$ 
differ only slightly--see the right-hand side inset in Fig.\ \ref{fig:3}--the resulting \ac{FDM} are almost identical. In this case, difference in the spectral function are mainly due to the choice of broadening functions and parameter \cite{BullaCostiPruschke2008}. 

Our  implementation of the \ac{FDM} approach \cite{WeichselbaumDelft2007}  takes into account the shift of the ground state energies in each iteration when evaluation the Boltzmann factors $\exp(-\beta E^m_l)$ in Eq.\ \eqref{eq:29}:  $E^m_l$ is not the NRG eigenenergy obtain at the  iteration $m$
but the eigenenergy relative to the ground state energy at the final iteration $N$. In the \ac{OC}-\ac{FDM} approach, however, the matrix elements $\rho_{ll}^m$ are related to $\exp[-\beta E^m_l(\text{NRG})]$ in equilibrium, since the final ground state energy shift is not known.  At the crossover iteration, when the density matrix weights $w_m$ become nonvanishing, the  occupation of the kept states are overemphasized in the \ac{OC}-\ac{FDM} approach since these states are refined at a later iteration. In the  \ac{FDM} approach the ground state energy shift  reduces the weights of the discarded states which apparently has the same effect, and the weight factors are almost identical in both approaches as demonstrated by the right-hand side inset of Fig.\ \ref{fig:3}.

To set the stage for the nonequilibrium quantum transport, we present a comparison of the zero-bias conductance $G(T)$ calculated  by the \ac{OC}-\ac{FDM} equilibrium spectral function  and spectra obtained by the \ac{FDM}-algorithm \cite{WeichselbaumDelft2007}.
 $G(T)$ is nearly identical for both approaches  and the difference are hard to detect as shown in Fig.\ \ref{fig:4-g-vs-t}. $G(T)/G_0\to 2$ for $T\to 0$
 as expected for two transport channels. Defining $G(T_K)=1$, yield $T_K(\text{OC-FDM})/\Gamma= 2.42 \times 10^{-2}$ and $T_K(\text{FDM})/\Gamma=2.51\times 10^{-2}$ for $U/\Gamma=12,\epsilon_d= -U/2$ which was already used as $T_K$ above.

\subsection{Nonequilibrium spectral functions at finite bias}
\label{sec:neq-spectral-functions}

\begin{figure}[tb]
\begin{center}

\includegraphics[width=0.45\textwidth, trim = 0 0 0 0, clip]{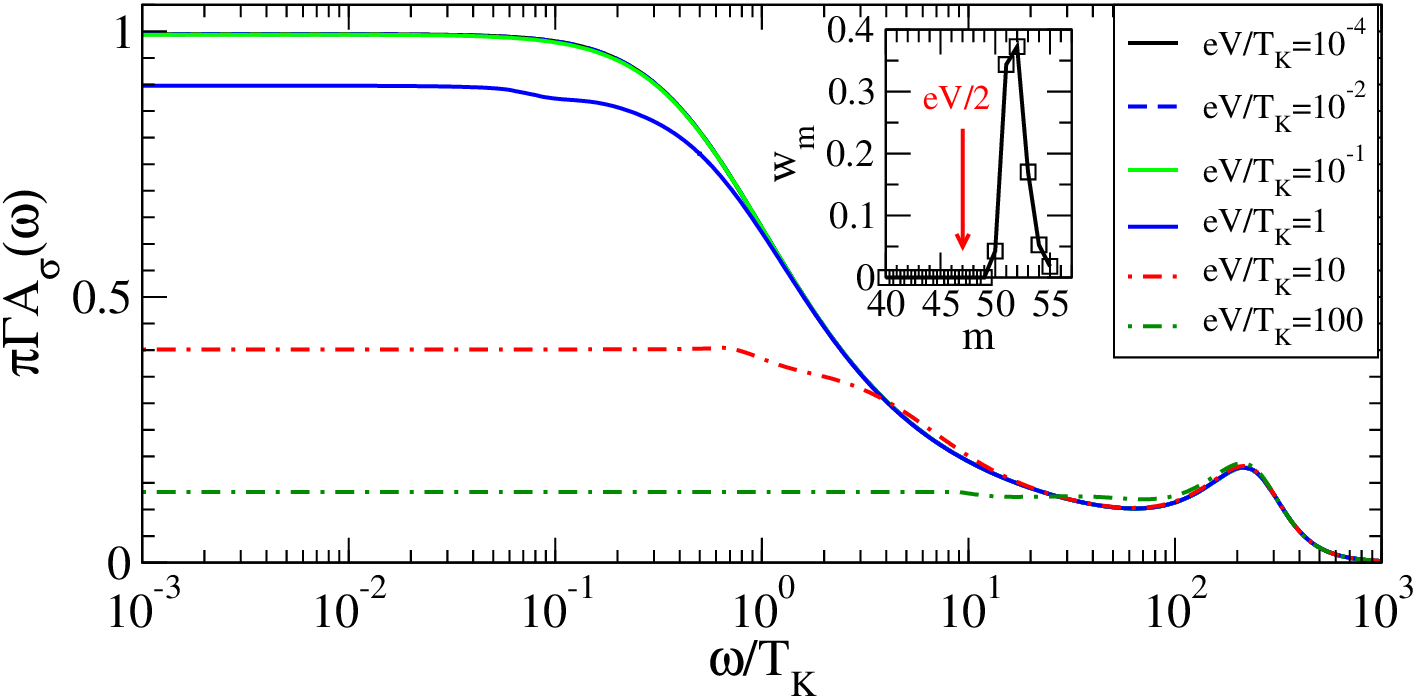}

\caption{nonequilibrium spectral function $\pi\Gamma A_\sigma(\omega)$ vs $\omega$
for different values of the bias voltage $eV/T_K$ and $T=10^{-6}\Gamma\to 0$ for 
$U/\Gamma=12,\epsilon_d/\Gamma=-6$. 
The inset shows the distribution of the weights $w_m$ for $eV/T_K=0.1$ calculated for $z=1$.
The Wilson shell corresponding to $eV$ is indicated by the red arrow.
NRG parameters: $\Lambda=1.8,N_s=1000,D=10^{3}\Gamma,\rho_c=0.02,T_K^{\text{trans}}/\Gamma = 2.42 \times 10^{-2}$, $N_z=2$.
}

\label{fig:rho-NEQ}
\end{center}
\end{figure}

To exemplify the evolution of the spectral function with the bias voltage we
present \ac{OC}-\ac{FDM} results Fig.\ \ref{fig:rho-NEQ} for a particle-hole symmetric case
using  $U/\Gamma=12, \epsilon_d=-U/2,T\to 0$ and $D/\Gamma=10^{3}$.  The results for the spectra are depicted 
for $V/T_K=10^{-4},10^{-2},10^{-1},10^{0},10^{1},10^{2}$.

We also added the distribution of the statistical weight factors $w_m$ in the \ac{OC-FDM}  defined in 
Eq.\ \eqref{eq:w-m-OC-FDM}
as inset to   Fig.\ \ref{fig:rho-NEQ} for the bias $eV/T_K=0.1$. We indicated the Wilson shell corresponding to the same energy as the bias energy, i.e., $|\mu_\alpha|=eV/2\approx \omega_m$, with
a red arrow. The distribution of the statistical weight factors $w_m$ peaks at much later iteration. We stopped the NRG iteration when the total spectral weight of the remaining kept states
 $\bar\rho_K(m) < \rho_c=0.02$: the states become equally populated deep in the bias window and the contribution to the spectral function decays rapidly.
While the standard NRG used the temperature T as a cutoff scale \cite{Wilson75,KrishWilWilson80a,KrishWilWilson80b},
the energy scale  is replaced by a density matrix criteria  $\rho_c$
in the \ac{OC}-\ac{FDM} approach. The choice $\rho^K_M < \rho_c$ therefore provides an a-priori bound on the truncation error of $\rho$, i.e., at most a fraction $\rho_c$ of the total density-matrix weight is discarded by stopping the RG flow at iteration $M$.
Typically we include 8-12 NRG additional iterations after the chemical potential reaches 
the  Wilson shell energy scale; this is  comparable with the \ac{FDM} approach.

The Kondo resonance remains unaltered to the equilibrium result for
$eV<10^{-1} T_K$, then deviations start to show
up below $|\omega|< eV$. The Kondo peak in the spectral function is
continuously reduced with increasing bias $eV$ due to the enhancement of the charge
fluctuations  inside the bias window which leads to a finite current.

In several  perturbative approaches, the spectral function shows a splitting of the Kondo
resonance at finite bias at low temperatures $T\ll T_K$. The first example is the seminal paper
by Meir et al \cite{meirTransportSIAM93} uses an equation-of-motion method and the \ac{NCA}.
Both methods are expansions with respect to the hybridization with the leads: they only include the leading order charge fluctuations and tend to overestimate peaks that follow the chemical potential steps by construction.  
In the perturbative \ac{RG} approach \cite{RoschKrohaWoelfe2001,RoschPaaskeKrohaWoelfe2003} charge fluctuations are eliminated 
by starting from a Kondo model (local moment fixed point.) 
Including only the leading order processes in a poor man's approach yields to a shallow double peak structure for the energy dependent coupling functions already around $eV\approx T_K$.

In the \ac{SNRG} \cite{AndersSSnrg2008} no clear splitting of the Kondo resonance was reported. Essentially the Kondo resonance is destroyed, and a shallow splitting was reported for $eV\gg T_K$.  A similar feature was also obtained using auxiliary master equation approach
\cite{PhysRevB.89.165105}.  A recent  \ac{RT-QMC} study
\cite{PhysRevB.100.201104} using massively parallel implementation of the
inchworm  \ac{QMC} solver tracks the evolution of spectral function as function of time
after a coupling quench. The authors also reported a shallow splitting for $U/\Gamma=8$ ($T_K/\Gamma \approx 0.1$)  and very large bias $|eV|>\Gamma\approx 10T_K$ similar to the \ac{SNRG} approach.

The absence of a Kondo peak splitting in our spectral function might be related to the effective
\ac{NRG} approach pursued here for maximum numerical efficiency. 
In our effective  \ac{NRG} approach a conventional Wilson discretization is used which implies
that the energy resolution around the chemical potentials $\pm eV/2$ become rather coarse-grained, such that very shallow structures are difficult to resolve.
Specifically, the energy resolution at an energy $\omega$ can be approximated from the energy scales of the neighboring Wilson shells, $\omega_n \sim \Lambda^{-n}$ and $\omega_{n+1} \sim \Lambda^{-(n+1)}$, that covers the energy regime such that 
$\omega_n>\omega>\omega_{n+1}$. The Wilson mesh at frequency $\omega_n$ has
 then a spacing $\Delta\omega\, (\omega_n) = \omega_n\, (1-\Lambda^{-1})$.
 For our value of $\Lambda = 1.8$, the resolution at the energy scale of applied bias, $\Delta\omega\,(eV/2) \approx 0.22\, eV$, so spectral 
 features of intrinsic width below this scale are not resolved by the single-lead discretization.

A two-lead \ac{NRG} calculation would allow for zoom into the low-energy excitations around the chemical
potential  of each lead separately but is numerically very expensive due to a lack of flavor conservation
and the much larger number of kept states. 
 A partial and much cheaper mitigation is z-averaging \cite{YoshidaWithakerOliveira1990,Oliveira1994}. 
Since the $N_z$ runs as discussed in the starting of Sec. \ref{sec:algorithm-single-lead} are independent, the cost scales only linearly, in contrast to the two-lead extension. 
 However, z-averaging increases the density of the mesh uniformly in $\log|\omega|$ and does not place additional points selectively at 
 $\pm eV/2$. It can therefore reduce, but not eliminate, the resolution limitation discussed above. Resolving any genuine sub-$\Delta\omega(eV/2)$ Kondo splitting at moderate bias ultimately requires the two-lead construction, which we leave to future work.

While the latest
inchworm  \ac{QMC} approach \cite{CohenGullQMC2023} to nonequilibrium steady state
requires about 10000 core hours CPU time on a supercomputer, our effective single lead approach requires  less than two minutes on a laptop with an ARM M1 processor.
Therefore, we leave this subtle question for a future investigation using a two-lead approach, and
focus here on exploring this very fast approach to nonequilibrium quantum transport.

\subsection{Charge fluctuations}
\label{sec:charge-fluctuation}

\begin{figure}[tb]
\begin{center}

\includegraphics[width=0.45\textwidth, trim = 0 0 0 0, clip]{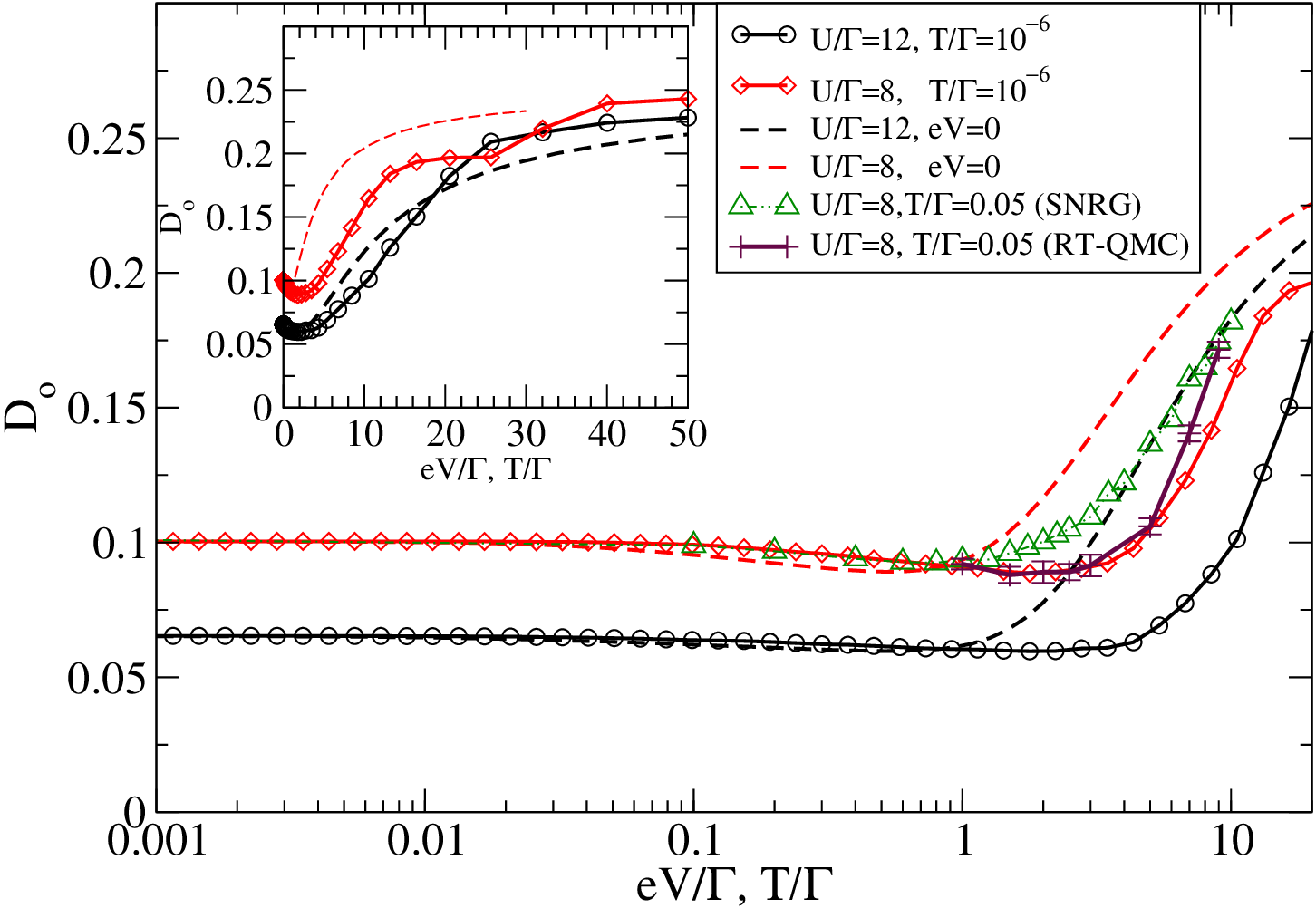}

\caption{Double occupancy as function of voltage at fixed temperature $T=10^{-6}\Gamma$
or at zero voltage and function of temperature 
for $U/\Gamma=8$ (red data) and  for $U/\Gamma=12$  (black data) 
obtained via \ac{OC}-\ac{FDM} approach. The finite bias results are marked by symbols
while the equilibrium data is shown as dashed line in the same color.
We added two data sets from Fig.\  3 of Ref.\ \cite{DirksSchmitt2013}:
the green triangles were obtained  by the scattering states \ac{NRG} 
\cite{AndersSSnrg2008}
 while the  magenta crossed were obtained by a real-time quantum Monte Carlo approach (RT-QMC) \cite{Werner09}
 both at a finite temperature $\beta=20\Gamma^{-1}$.
 The inset shows the \ac{OC}-\ac{FDM} on a linear scale and larger parameter range.
NRG parameters are the same as in Fig.\ \ref{fig:rho-NEQ}.
}

\label{fig-6}
\end{center}
\end{figure}

The charge fluctuation on the \ac{QD}
 is related to the double occupancy $D_o=\langle n_\uparrow n_\downarrow \rangle$,
\begin{eqnarray}
\Delta Q^2 = \langle (\hat n_d -\bar n_d)^2 \rangle = \bar n_d(1-\bar n_d) +2 D_o \,.
\end{eqnarray}
At half filling,  i.e., $\bar n_d=1$, the first term vanished, and $D_o=\Delta Q^2/2$.  
$D_o$ is a local property of the \ac{QD} and can be directly calculated  as local
expectation value from the \ac{OC}-\ac{FDM} approach. 

For an isolated  \ac{QD}, $\Gamma=0$, $D_o=1/4$ at infinitely high temperatures and crossed over to $D_o\to 0$ when $\beta U/2> 1$. A finite coupling to the leads induces virtual charge fluctuations 
so that $D_o>0$: the larger the ratio $U/\Gamma$, the smaller the zero temperature limit $D_o(0)$.

In nonequilibrium, $D_o=D_o(T,V)$ is a function of temperature and voltage. Since the charge can freely fluctuate when $V\to\pm\infty$,
$D_o(T,V\to\infty)=1/4$ as pointed out in Ref.\ \cite{DirksSchmitt2013}. 

The \ac{OC}-\ac{FDM} results obtained for $D_o(T,V=0)$ and $D_o(T\to 0,V)$ are shown in Fig. \ref{fig-6} for $U/\Gamma=8$ (red curves and symbols) and $U/\Gamma=12$ (black curves and symbols)
for a particle-hole symmetric junction $\e_d=-U/2$ and $R=1$. 
The finite bias results are marked by symbols
while the equilibrium data is shown as dashed line in the same color.  The inset shows the \ac{OC}-\ac{FDM} results on a linear 
temperature and voltage scale over a larger parameter range. 

We clearly observe an excellent agreement between the equilibrium data and the finite bias results for $T,eV\ll T_K$.
$D_o$ is stronger suppressed for larger $U$ as expected.  
Since the strong coupling fixed point describes a strong entanglement between the \ac{QD} orbital
and the lead, and finite temperature or finite bias will shift the system away from the strong coupling fixed point
and therefore causes a small reduction of $D_o$. As expected from a Fermi liquid expansion  \cite{OguriFermiLiquid2005} the  $D_o(T,0)$ shows this behavior earlier than $D_o(0,V)$. After a shallow minimum located  clearly above $T_K$, $D_o$ approaches its asymptotic value of $1/4$ rapidly once $T> U/2$ or $eV>U$. However, $D_o(T,0)$ and $D_o(0,V)$ cannot be mapped on top of each other by some scaling factor $a$, $eV=a T$ demonstrating that voltage and temperature are not interchangeable.

To make contact to the literature, we added two original 
data sets from Fig.\  3 of Ref.\ \cite{DirksSchmitt2013} to Fig.\ \ref{fig-6}:
The green triangles were obtained  by the \ac{SNRG}
\cite{AndersSSnrg2008}  while the  magenta crossed were obtained by a real-time quantum Monte Carlo approach \cite{Werner09}  both at a finite temperature $\beta=20\Gamma^{-1}$.
 We note that the  \ac{SNRG} perfectly agrees with the \ac{OC}-\ac{FDM} for small voltages $eV<T_K$
 although slightly different  \ac{NRG} parameters and a two-lead approach was used.  It indicates that $D_o$ essentially depends on the ratio $U/\Gamma$ and
 is barely influenced by a change of the band width. At larger voltages, $eV>\Gamma$, we observe deviations from the 
 \ac{OC}-\ac{FDM} result. Interestingly, the RT-\ac{QMC} data \cite{DirksSchmitt2013,Werner09} 
 calculated by Ph.\ Werner agrees very nicely
 with the \ac{OC}-\ac{FDM} result up to around $eV\approx 5\Gamma$. The deviation might be due to the finite temperature
 of the  RT-\ac{QMC}  approach: the crossover to the asymptotic value of $D_o\to 1/4$ occurs at a smaller voltage in the  RT-\ac{QMC} that in our  \ac{OC}-\ac{FDM} approach where $T/\Gamma=10^{-6}$ was used
 which cannot be accessed by the RT-\ac{QMC}.

\subsection{Charge transport through a quantum dot}
\label{sec:charge-currents}

\subsubsection{Symmetric junction}

\begin{figure}[tbh]
\begin{center}

\includegraphics[width=0.45\textwidth]{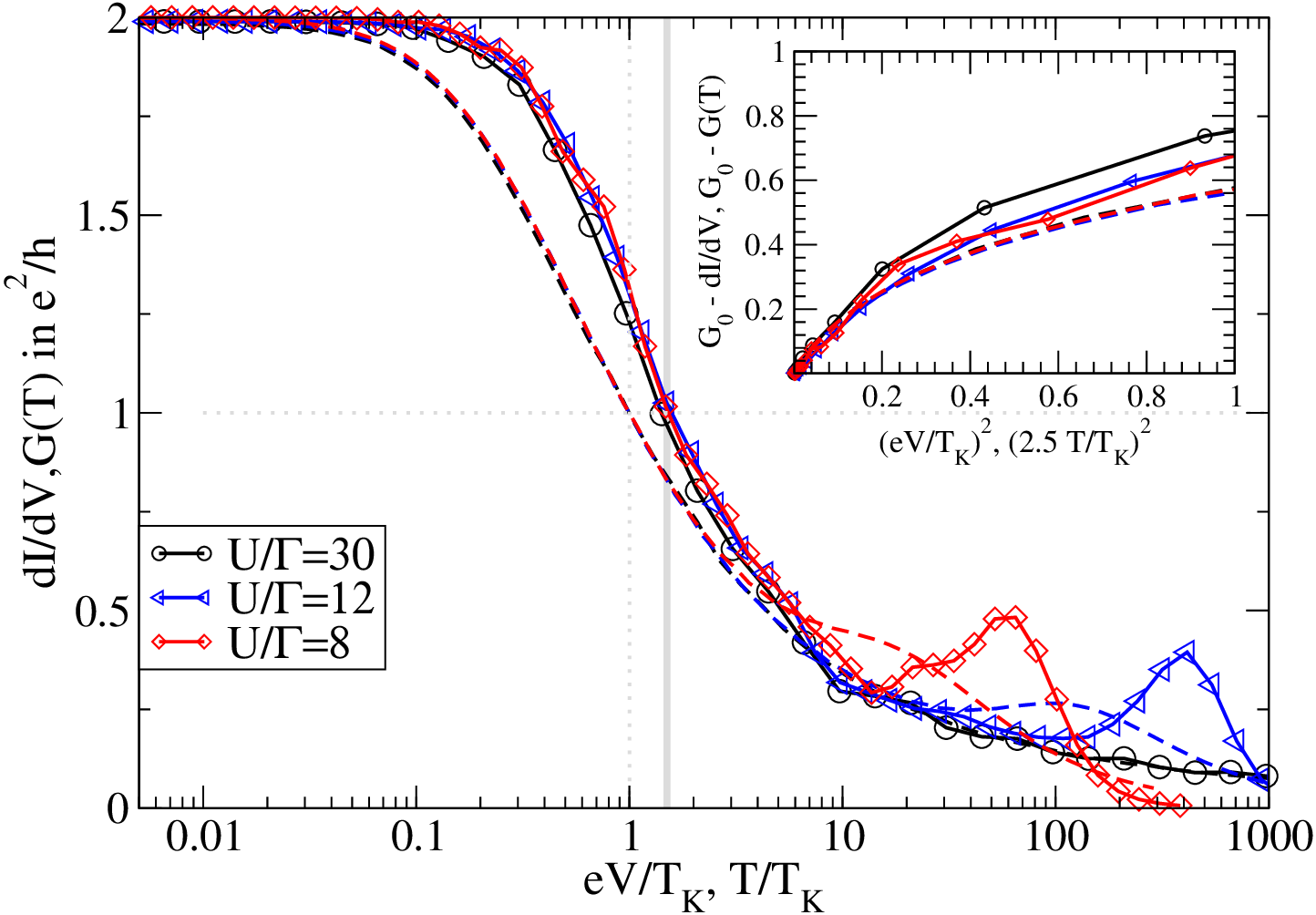}

\caption{Zero-bias conductance $G(T)$ vs $T/T_K$ (dashed color-coded curves) and
the nonequilibrium differential conductance $dI(V)/dV$  (solid color-coded curves) 
at $T/\Gamma=10^{-6}$ vs $eV/T_K$, for $\epsilon_d=-U/2$ with $U/\Gamma=30$ (black circle), $U/\Gamma=12$ (blue triangle), and  $U/\Gamma=8$ (red diamond)
for a symmetric coupling.  The values of $T_K$ for $U/\Gamma=8,12,30$
are $T_K/\Gamma=1.04 \times 10^{-1}, 2.42\times 10^{-2}, 3.3 \times 10^{-5}$.
The vertical gray line indicates the half-maxima in bias, $V=1.6\,T_K \approx V_K$.
The inset shows $G(0)-dI(V)/dV$ and $G(0)-G(T)$ but in the nonequilibrium 
Fermi-liquid regime as a function of $(eV/T_K)^2$ and a rescaled $(2.5\,T/T_K)^2$.
NRG parameters are the same as in Fig.\ \ref{fig:rho-NEQ}.
}

\label{fig-7}
\end{center}
\end{figure}

Figure~\ref{fig-7} shows a comparison between the zero-bias conductance $G(T)$ and the differential conductance $dI(V)/dV$
for $T\to 0$.  While the  zero-bias conductance $G(T)$  is obtained using the equilibrium spectral function and the analytic
derivative of the Fermi function, the differential conductance is calculated numerically. 
The results are qualitatively consistent with Fig.\ 3(a)  in Ref.\  \cite{SchwarzPRL-TransportQD2018},
with $G(T)$ and $dI/dV$ decaying monotonically with increasing temperature and applied bias respectively. 

To demonstrate the virtue of our approach, we present the data for $U/\Gamma=30$, with an equilibrium Kondo temperature
$T_K/D=3.31 \times 10^{-8}$, which is beyond the accessibility of most other approaches.  Our 
transport calculation is able to bridge nine decades in energy scales, owing to the RG approach employed here.
It demonstrates universality up to $eV/T_K,T/T_K \sim 1$.

For small temperatures and voltages the differential conductance for a symmetric junction 
is given by the Fermi-liquid expansion \cite{OguriFermiLiquid2005},
\begin{eqnarray}
\frac{dI}{dV} = \frac{2e^2}{h}\left[
1 - c_T \left(\frac{\pi T}{T_K}\right)^2 -  c_V \left(\frac{V}{T_K}\right)^2  + \cdots
\right],
\end{eqnarray}
so that $G(T)=dI/dV(T,0)$ and $dI/dV=dI(0,V)/dV$ should be universal  
when setting 
\begin{eqnarray}
V =  \pi \sqrt{\frac{c_T}{c_V}} T
\end{eqnarray}
within the range of the validity of the expansion.

In the literature, various values for the $c_T$ and $c_V$ are reported using different methods which are summarized in   the review section  of Ref.\  \cite{OguriFermiLiquid2005}. Oguri used Ward identities
to show that $c_V=(3/2)\, c_T$ for the strong coupling limit \cite{OguriFermiLiquid2005}, which yields $eV\approx 2.565\,T$
for mapping $G(T)$ onto $dI/dV$.  
The inset of Fig.\  \ref{fig-7} demonstrates the convergence in the Fermi-liquid regime after rescaling $G(T)$  with a prefactor of $2.5$ in the $T$-axis, for a direct comparison with $dI/dV$.
The prefactor $2.5$ corresponds to the Fermi-liquid ratio $c_V/c_T= 1.58$,
which is very close to Oguri's analytic result in the strong coupling limit, considering that the $dI/dV$ has been obtained by numerical differentiation.

Our results show that deviations from Fermi-liquid approximation start to show up at voltages as low as $eV \approx 0.3\,T_K$ [cf. Fig. \ref{fig-7}].
Moreover, the results using \ac{NRG}-{TD-DMRG} from Refs.~\cite{SchwarzPRL-TransportQD2018, Manaparambil2022} and
real-time renormalization group (RTRG) from Ref.~\cite{Pletyukhov2012} places the bias-to-temperature ratio based on nonequilibrium conductance $V_K/T_K \approx 1.6-1.8$. Where $V_K$ and $T_K$ are defined as, $\frac{dI}{dV} (T\to0, V=V_K) = G (T_K,V\to 0) = \frac{1}{2}$.
Our \ac{OC}-\ac{FDM} calculations in Fig. \ref{fig-7} yields $V_K/T_K \approx 1.6$, showing excellent agreement with the literature.

\subsubsection{Asymmetric junction}

\begin{figure}[tb]
\begin{center}
 \includegraphics[width=0.45\textwidth]{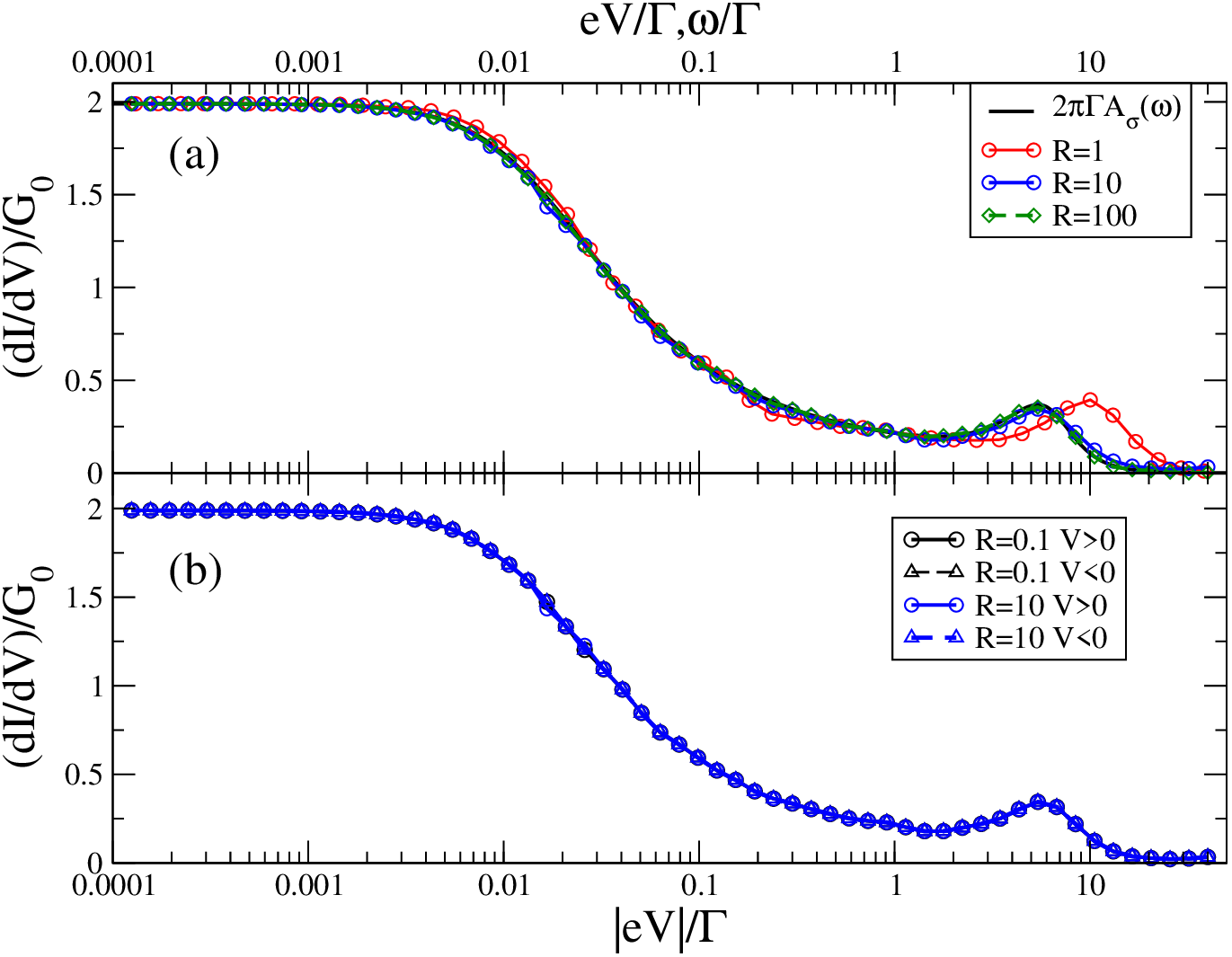}
\caption{
(a): Differential conductance $dI/dV$   for asymmetric junctions $R=1,10,100$ at $T\to 0$ 
and particle-hole symmetric case with $U/\Gamma=12, \e_d=-U/2$ for
$V > 0$ augmented by the equilibrium spectral function.
(b): $R=0.1,10$ for positive and negative voltages showing equivalence with simultaneous reversals of bias and lead asymmetry.
}

\label{fig-8}
\end{center}
\end{figure}

While for $R=1$, the spectral functions are strongly bias voltage dependent, they become independent for $R\to \infty$ since $f_{\text{eff}}(\omega)$
approaches $f_L(\omega)$: an equilibrium spectral function with the chemical potential $\mu=\mu_L$ characterizes this regime and the right lead
is only a probe of the current.  

In all calculations, we keep $\Gamma=\Gamma_L+\Gamma_R=const$
and only change $R$ ensuring that the equilibrium Kondo scale remains invariant of $R$. Therefore the figures in this section focus on the effect of the asymmetry factor $R$ 
on the dI/dV curves: The trivial reduction of the Kondo temperature as well as the change of $G_0$ 
is eliminated when quenching one of the two leads.

$(dI/dV)/G_0$ is depicted for $R=1,10,100$ as function of $eV/\Gamma$ for the $U/\Gamma=12, \e_d=-U/2$
in Fig.\ \ref{fig-8}(a) for $eV>0$. 
We added the  equilibrium spectral function as solid black line for comparison. 

We note that the differential conductance for $R=1$ traces the spectral function surprisingly good for small voltages $eV<T_K$. Taking into account  only
the derivatives of the Fermi functions in Eq.\ \eqref{eq:meir-wingreen-Qdot} 
$dI/dV(eV) \propto A_\sigma(\omega=eV/2)$ since $\mu_\alpha = \pm eV/2$.
For small bias, however, the spectral function entering Eq.\ \eqref{eq:meir-wingreen-Qdot}
changes as well: the Kondo resonance at $\omega=0$ is reduced with increasing bias, 
and spectral weight is redistributed to larger frequencies. Only at very large bias $eV\gg T_K$, the spectral function becomes nearly
bias independent and the $dI/dV \propto 2\pi\Gamma A_\sigma(2\w)$ as seen in the red curve in
 Fig.\ \ref{fig-8}(a). 

Note that the peak of the $dI/dV$ for $R=1$ at approximatively $eV=U$ (i.e., $\mu_L=eV/2=U/2$, aligning one lead with the Hubbard peak) is slightly higher than the
Hubbard peak in the spectral function at $T=0$. This is a consequence of redistributing spectral weight from the Kondo resonance into the Hubbard (or
charge excitation) peaks.

From Eq.\ \eqref{eq:f-eff} we obtain the relation $f_\text{eff}(\omega,R,V)= f_\text{eff}(\omega, 1/R,-V)$ leading to symmetry
transformation
\begin{equation}
\label{eq:41}
\frac{dI(R,V)}{dV}=\frac{dI(1/R,-V)}{dV}.
\end{equation}
In general, it is not sufficient to invert the bias voltage, but also requires the flipping of leads simultaneously to obtain an identical differential conductance.

Already the $dI/dV$ curve for $R=10$ agrees very well with the spectral function. We also checked the calculation
for negative $V$ and for the symmetry between $R$ and $1/R$ which essentially flips the coupling of the left and the right lead.
The results are depicted in Fig.\ \ref{fig-8}(b).  
$R\to 0$ or $R\to \infty$ implies that one of the two leads are increasingly detached from the \ac{QD} since
$\Gamma_L=R\Gamma_R$ while $\Gamma=\Gamma_L+\Gamma_R=const$ in our calculation. In both limits, the equilibrium single-lead
Kondo problem is recovered. The trivial reduction of the conductance  is included in the definition of $G_0$ to make
the difference $dI/dV$ comparable. Essentially, the tunneling regime is reached very fast, where the transport properties are
governed by the equilibrium spectral functions as illustrated in Fig.\ \ref{fig-8}(a).

\begin{figure}[t]
\begin{center}

 \includegraphics[width=0.45\textwidth]{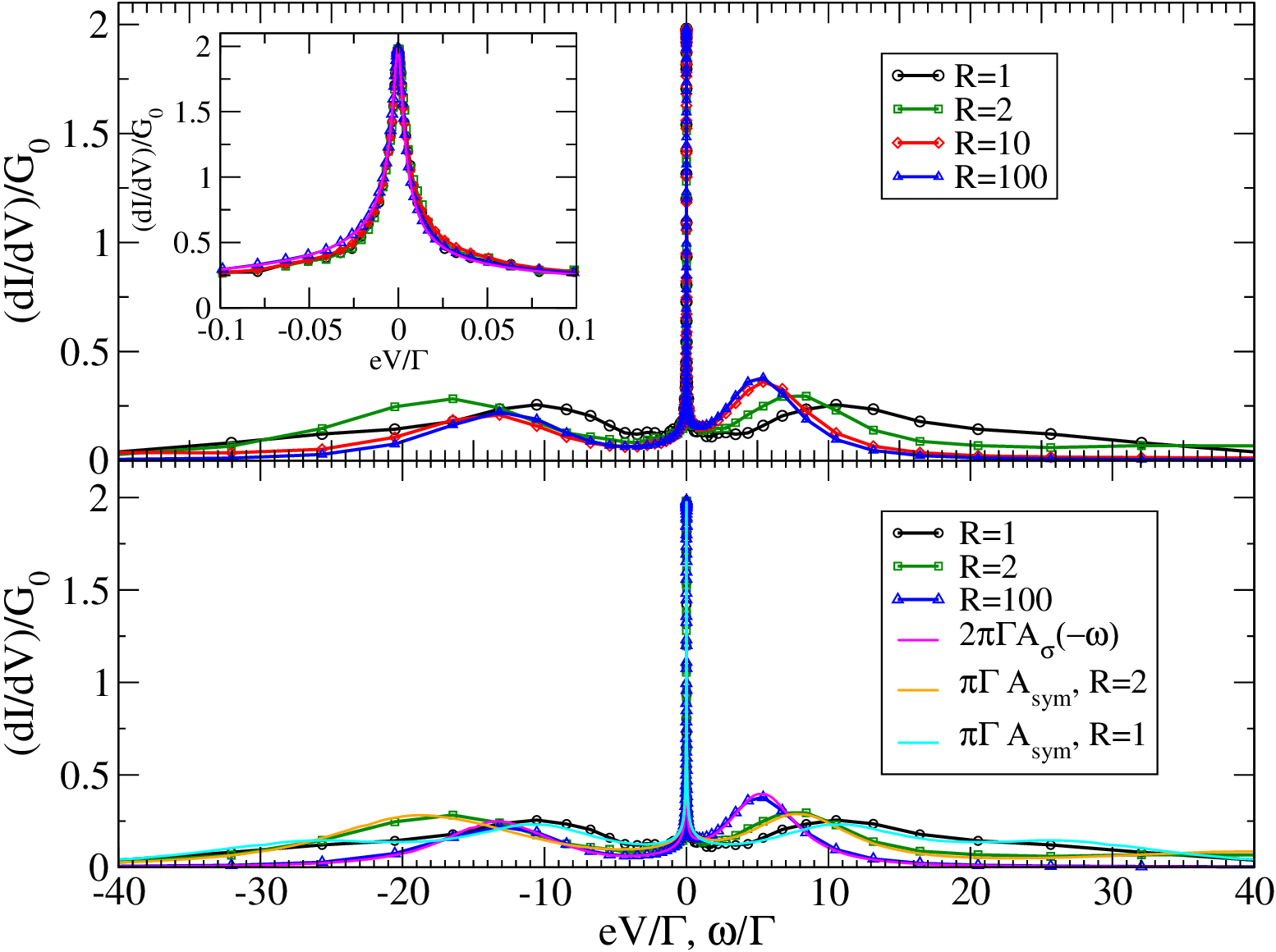}

\caption{ Top panel: Differential conductance $dI/dV$   
of  asymmetric junctions with $R=1,2,
10,100$ and local particle-hole asymmetry,  $U/\Gamma=20, \e_d/\Gamma=-6$ at $T=10^{-6}\Gamma$.  
Bottom panel compares the $dI/dV$ curves of the highly asymmetric case $R \gg 1$ with the equilibrium spectral function $A_\sigma(-\omega)$ , and the intermediate cases $R=1,2$ with a symmetrized spectral function $A_{\rm sym} (\omega)$. The inset focuses on the zero-bias conductance peak for different $R$, with the equilibrium spectral function $A_{\sigma}(-\omega)$ added for comparison. 
}

\label{fig-9}
\end{center}
\end{figure}

This asymmetry is enhanced in a particle-hole symmetry broken Hamiltonian. For $R=1$, the symmetry under bias inversion is maintained
in the $dI/dV$ curve as stated in Eq.\ \eqref{eq:41}. This is not the case for $R>1$. We plot $dI/dV$ data for $R=1,2,10,100$ in Fig.\ \ref{fig-9}.
For $R\to \infty$, the tunneling regime is approached in which the $dI/dV$ tracks the spectral function. 
We added $2\pi\Gamma A_\sigma(-\omega)$
to Fig.\ \ref{fig-9} for comparison and note an excellent agreement for $R=100$ as expected from the 
tunneling regime.
The $dI/dV$ for a general asymmetry can then be compared to a symmetrized spectral function,
\begin{equation}
A_{\rm sym} (\omega)= [A_\sigma (\omega \equiv \mu_L) + R\,A_\sigma (\omega \equiv \mu_R) ]/(R+1),
\end{equation}
 with asymmetry influencing the chemical potentials as $\mu_L=eV/(R+1)$ and $\mu_R=-eV\,R/(R+1)$. 
The $dI/dV$ calculations for various asymmetries $R$ from Fig. \ref{fig-9} shows good agreement with $A_{\rm sym}$ beyond the Kondo regime $eV\gg T_K$. 
We note that the Kondo peak in the zero-bias conductance remains
unchanged for different asymmetries, since $\Gamma=\Gamma_L+\Gamma_R$ is held constant. 
Consequently, $(dI/dV)/G_0$ in the Kondo regime is essentially unaffected by 
the asymmetry, except for the asymmetry of the individual lead chemical 
potentials [cf. inset of Fig. \ref{fig-9}].
 
\subsection{Magneto transport}
\label{sec:magneto-transport}

\subsubsection{Symmetric junction}
\begin{figure}[tb]
\begin{center}

 \includegraphics[width=0.45\textwidth,trim= 0 0 0 0,clip]{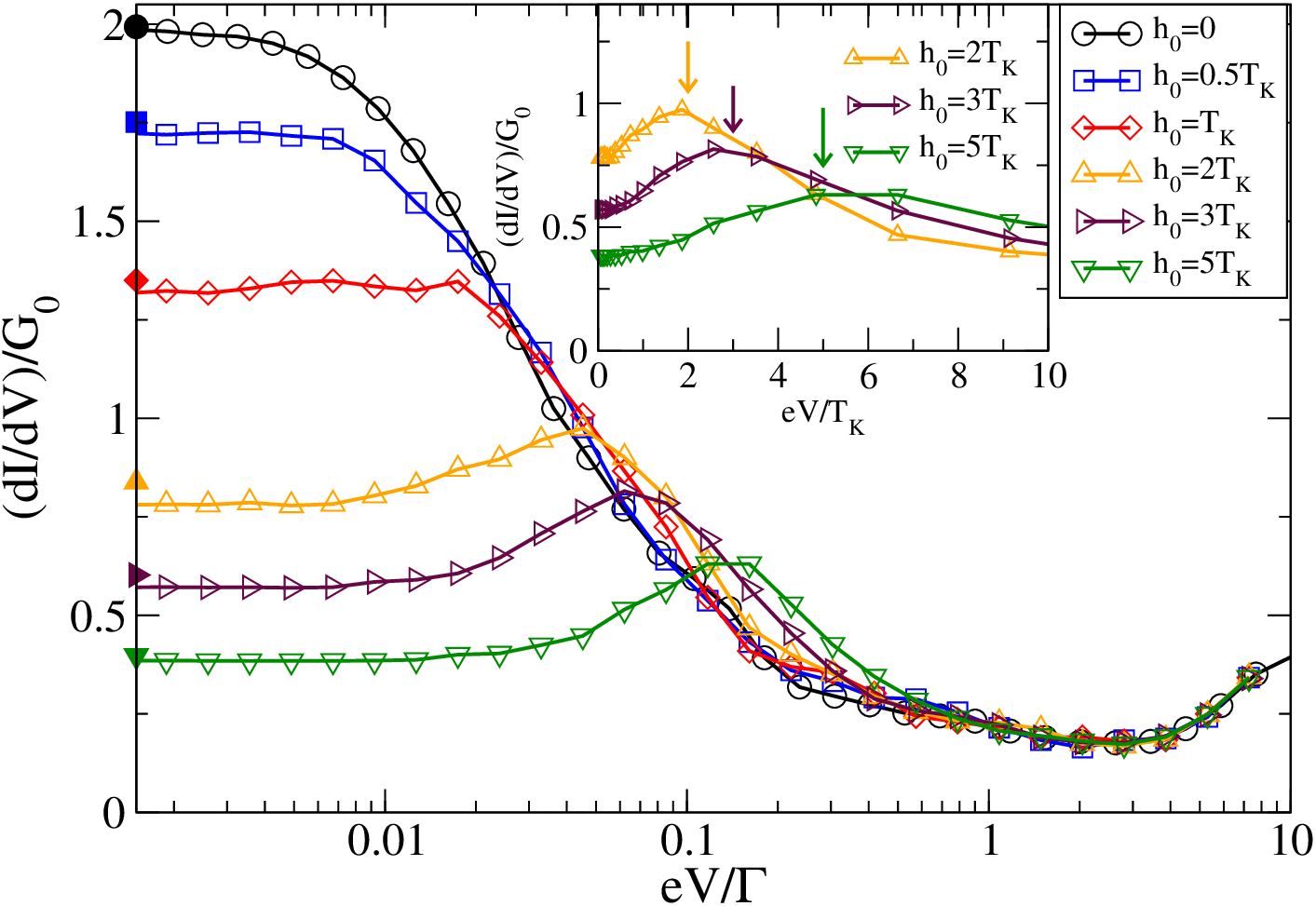}

\caption{Differential conductance $dI/dV$   
of a symmetric junction  $R=1$ and different magnetic fields $h_0= 0,0.5,1,2,3,5\,T_K$, $U/\Gamma=12, \e_d=-U/2$
at $T=10^{-6}\,\Gamma$.  
The solid symbols at $eV \to 0$ are the zero-bias conductance calculated with NRG for the same system parameters.
The inset shows the splitting of the Kondo peak along with the applied magnetic field $h_0$ shown by color-coded arrows.
}

\label{fig-10}
\end{center}
\end{figure}

\begin{figure}[tb]
\begin{center}

 \includegraphics[width=0.47\textwidth]{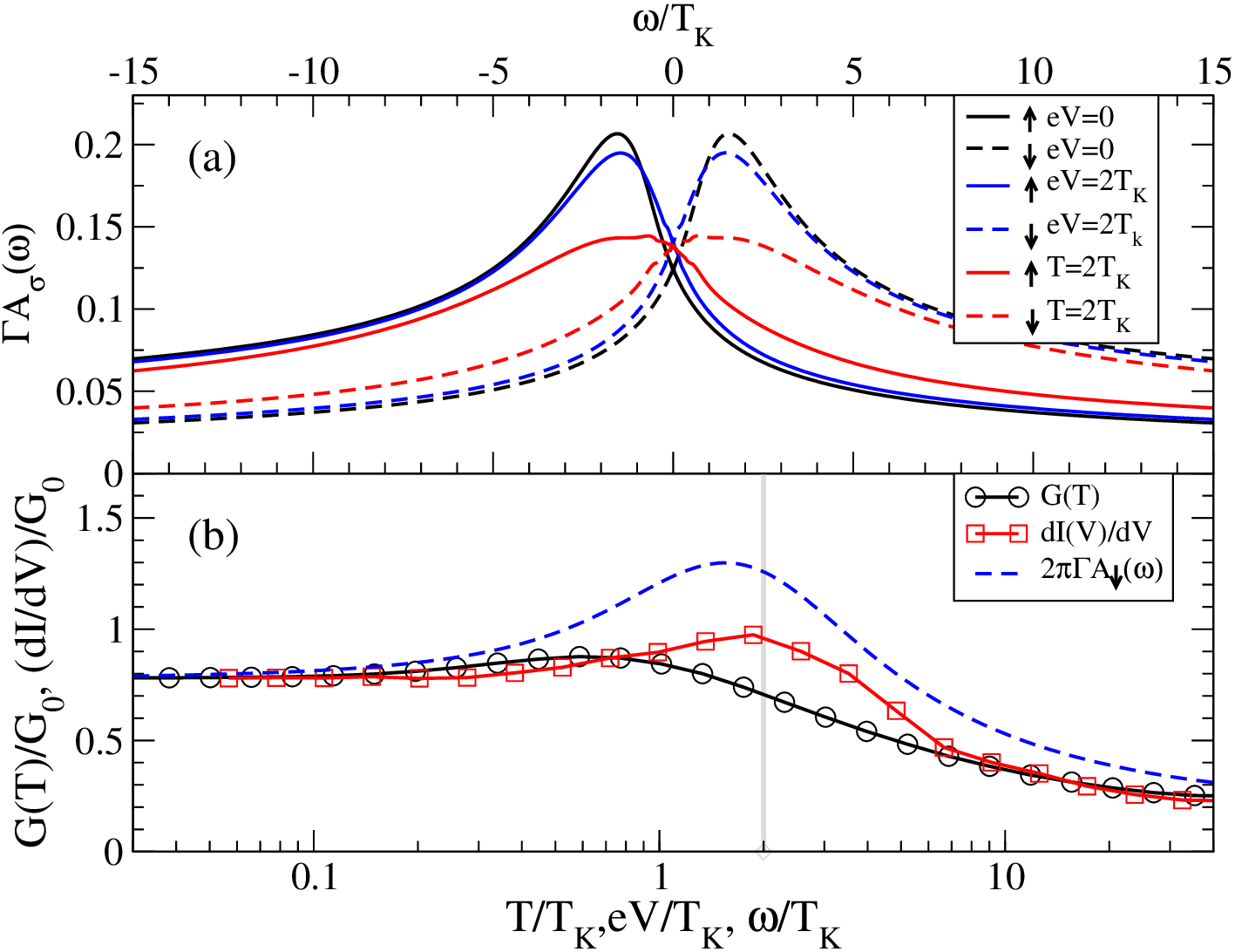}

\caption{(a): $A_\uparrow(\w)$ (solid lines) and $A_\downarrow(\w)$ (dashed lines)
at a finite magnetic field $h_0=2\,T_K$ and $U/\Gamma=12$, $\e_d/\Gamma=-6$.
The black lines show the spectrum at $T=10^{-6}\,\Gamma$ and zero bias (equilibrium);
the blue lines those at $T=10^{-6}\,\Gamma$ and finite bias $eV=2\,T_K$; and the red lines those
at zero bias but a finite temperature  $T=2\,T_K$.
(b): $G(T,V=0)$ and $dI(V,T=0)/dV$ at finite magnetic field $h_0=2\,T_K$ (indicated by the gray vertical line). 
For comparison, we overlay the normalized equilibrium spectral function $2\pi\,\Gamma A_\downarrow(\w)$ from panel (a). 
}

\label{fig-11}
\end{center}
\end{figure}

The effect of the magnetic field onto the zero-bias conductance as function of temperature was investigated by Costi more than 25 years ago \cite{costiMagFldSIAM00} using the equilibrium \ac{NRG} for the Kondo model. A finite magnetic field lead to a spin-dependent shift of the Kondo resonance. These Kondo resonances peak around $\omega \approx  \pm h_0$, where $h_0=g\mu_b B_0$ is the Zeeman energy.
Consequently, the total spectral function $A(\omega)=A_\uparrow(\omega) +A_\downarrow(\omega)$ exhibits a splitting of the zero-frequency resonance when $h_0>T_K$.

The Kondo temperature in Ref.\ \cite{costiMagFldSIAM00}  was defined by the \ac{HWHM}  of the Kondo resonance. In the case of the \ac{SIAM}, this is only a reasonable definition of a particle-hole symmetric problem.  We found $\omega_\text{HWHM}/\Gamma= 0.0392$ of the spectral function for $U/\Gamma=12, \e_d=-U/2$ and $T=10^{-6}\Gamma\to 0$ which correspond to $\omega_\text{HWHM}=1.6T_K$. Therefore, we expect an onset of splitting in the $dI/dV$ curve when the Zeeman energy exceeds approximately  $1.5T_K$.

The \ac{OC}-\ac{FDM} results for the voltage dependent $dI/dV$ in a finite magnetic field are depicted in Fig.\ \ref{fig-10}
for different values of $h_0$. The splitting of the $dI/dV$ maximum occurs when $h_0$ exceed $T_K$. The inset of Fig.\ \ref{fig-10}
shows that the peak in the $dI/dV$ develops around $eV\approx h_0$, so that the peak separation is $2h_0$. This is consistent with the differential conductance behavior obtained with the scattering states \ac{NRG} \cite{AndersSSnrg2008} and \ac{NRG}-\ac{TD-DMRG} \cite{SchwarzPRL-TransportQD2018}.

To illustrate  the different effect of the bias voltage and temperature at finite magnetic field  we present the spectral function
for $(T=10^{-6}\,\Gamma$, $eV=2\,T_K)$ (blue lines) and  $(T=2\,T_K$, $eV=0)$ (red lines) in Fig.\ \ref{fig-11}(a). We have also included the equilibrium spectral
functions for  $(T=10^{-6}\,\Gamma$, $eV=0)$ (black lines) for comparison. At low temperature, the effect of the voltage on the spectral function is small
but the resonance decreases significantly at zero bias and $T=2\,T_K$. 

Fig.\ \ref{fig-11}(b) shows the comparison between $G(T,V=0)$ and $dI(V,T=0)/dV$ calculated with the \ac{OC}-\ac{FDM} nonequilibrium approach.
We note that there is an excellent agreement between the spectral functions, $G(T,V=0)$ and $dI(V,T=0)/dV$ in the linear response regime, $\omega/T_K,\,eV/T_K,\,T/T_K\ll 1$.  The zero bias conductance $G(T)$ peaks roughly at $T=h_0/2$ which agrees with the data presented in Fig.\ 4 of Ref.\ \cite{costiMagFldSIAM00}, while the $T=0$ spectral function peaks around $\omega \approx h_0$, as can also be seen in Fig. 3\ of Ref.\ \cite{costiMagFldSIAM00} for the Kondo model. The reason for this shift in the peak is connected to the strong temperature dependency of the spectral function which enters the transport integral. Whereas, the $dI/dV$ peaks around $eV \approx h_0$, similar to the splitting of the Kondo peak in the equilibrium spectral function.

\subsubsection{Spin current in an asymmetric junction}

\begin{figure}
\begin{center}
\includegraphics[width=0.45\textwidth,clip]{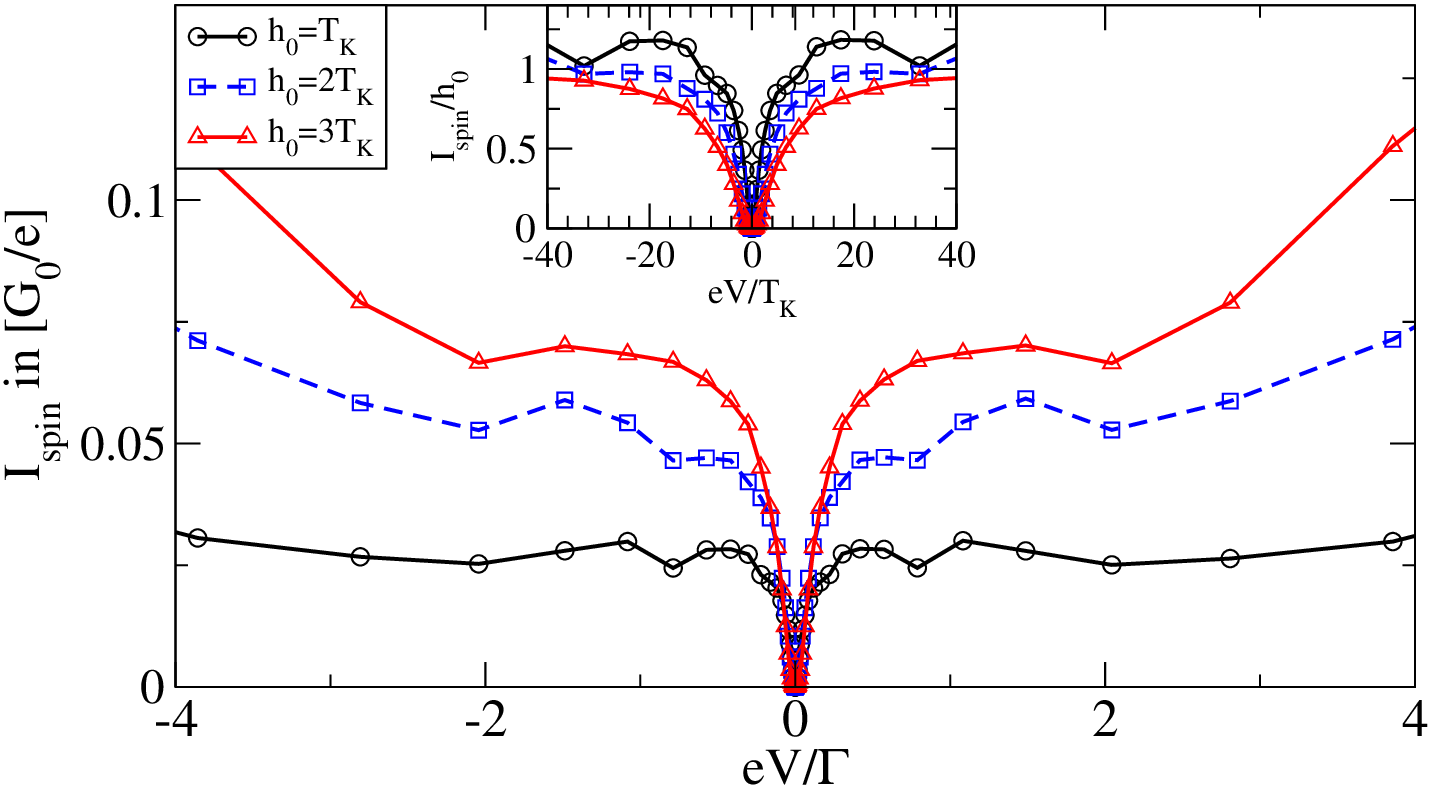}
\caption{Spin current $I_\text{spin}$ vs $eV$ for different magnetic fields $h_0=T_K,\,2\,T_K,\,3\,T_K$
at local particle-hole symmetry $U/\Gamma=12$, $\e_d/\Gamma=-6$
and $R=10$.  The inset shows data as in main panel, but focuses on the low-bias regime by scaling the bias with $T_K$ and scaling the spin current $I_\text{spin}$ with $h_0$.
}
\label{fig-12}
\end{center}
\end{figure}

So far, we investigated only the charge current $I=I_\text{charge} = I_\uparrow + I_\downarrow$, where the $I_\sigma$ are the individual contributions to
the trace in Eq.\ \eqref{eq:meir-wingreen-Qdot}. In a symmetry breaking external magnetic field, we can also introduce the spin current
\begin{equation}
\label{eq:42}
I_\text{spin} =I_\uparrow - I_\downarrow.
\end{equation}
For $R=1$, the spin current vanishes independent of the magnetic field for a particle-hole symmetric  quantum dot,
 since the bias window $f_L(\omega)-f_R(\omega)$ is symmetric. For a finite $R$, the bias window is asymmetric so that the Kondo splitting leads to different contributions to $I_\uparrow$ and $I_\downarrow$. 

The results for the spin current of a particle hole symmetric quantum dot is depicted for an asymmetric junction ($R=10$)
 in Fig.\ \ref{fig-12} for different magnetic field strength $h_0$. While the charge current changes sign with a sign change of the bias $V$, the spin current is always positive for positive $h_0$.
The origin of this fixed spin current direction is easily understood by inspecting Eq.\ \eqref{eq:42}. Changing the sign of $V$ reverses the currents $I_\sigma$,
but simultaneously, the bias window is flipped as well so the magnitude of $I_\uparrow$ and $I_\downarrow$ is flipped as well. Therefore, we obtain the same sign
for positive and negative bias voltage. 
The spin current $I_\text{spin}$ remains symmetric under the flipping of bias and increases in magnitude with applied magnetic field $h_0$.
Thus we note that $I_\text{spin}$ is an even function in $eV$, but odd in $h_0$ and changes sign at $R \to 1/R$.
The inset of Fig.~\ref{fig-12} shows that the
induced spin current per unit applied magnetic field $I_\text{spin}/h_0$ is larger for $h_0=T_K$ and goes down with increasing magnetic field.

\subsection{Thermoelectric transport}

\label{sec:thermo-current}

In this section, we discuss the charge transport through a \ac{QD} connected to two leads kept at different temperatures $T_L,T_R$. 
These boundary conditions also enter through $f_{\text{eff}}(\omega)$ defined in Eq.\ \eqref{eq:f-eff}, the rate equations and, therefore, the \ac{NEQ} spectral function. 
To focus on the essential effect, we first consider a symmetric junction ($R=1$) and zero-bias to keep the number of parameter limited.
We then primarily focus on the thermoelectric current, driven by a temperature gradient $\beta_L\neq \beta_R$ at zero bias ($eV=0$), in contrast to the previous sections, which used $\beta_L=\beta_R$ and a finite bias $eV \neq 0$.

For a finite temperature gradient $T_L\not = T_R$, the difference in the Fermi functions, defining the bias window in  Eq.~\eqref{eq:meir-wingreen-Qdot},
is given by
\begin{widetext}
\begin{eqnarray}
K(\omega,\mu_\alpha,\beta_\alpha) &=& 
f_L(\omega) - 
f_R(\omega) 
=
\frac{1}{2}\left(
\tanh\left(\frac{\beta_R(\omega-\mu_R)}{2}\right)
-
\tanh\left(\frac{\beta_L(\omega-\mu_L)}{2}\right)
\right).
\end{eqnarray}
\end{widetext}

At zero bias, $K(\omega,0,\beta_\alpha)=-K(-\omega,0,\beta_\alpha)$ is an odd function of $\omega$ and vanishes for $\beta_L=\beta_R$.
Therefore, a charge current gets induced by a temperature gradient between the two leads when 
the particle-hole symmetry is broken [cf. Eq.\ \eqref{eq:meir-wingreen-Qdot}] . $K(\omega,\mu_\alpha,\beta_\alpha)$
is exponentially suppressed for $|\omega|\to \infty$ and the width of the relevant spectral region is given by the larger temperature  $T_H=\text{max}\{T_L,T_R\} $ of the two lead temperatures.

We note that 
(i) $K(\omega,0,\beta_\alpha)$ is odd under the exchange $T_L\leftrightarrow T_R$ and
(ii) $f_{\text{eff}}(\omega)$ defined in Eq.\ \eqref{eq:f-eff} remains invariant for a symmetric junction $R=1$ and $V=0$. Since $f_{\text{eff}}(\omega)$
is unaltered, the nonequilibrium spectral function is invariant under the exchange $T_L\leftrightarrow T_R$ and the current only changes its sign. 
We parameterized the temperature gradient by the  ratio  $x_T = T_L/T_R$. Since $I(T_L,T_R) = -I(T_R,T_L)$, the relation  $I(T_R,x_T)= - I(x_T T_R, 1/x_T)$
must hold.

Using \ac{NRG}, Costi and Zlati\'{c} \cite{Costi2010} studied the linear response transport
coefficients including the linear response conductance $G_{\rm lin}$ and Seebeck
coefficient $S_{\rm lin}$ for an Anderson impurity in the strong coupling (Kondo)  regime.
For the sake of benchmarking, in our case when $V\to 0$, the bias-dependent change of the spectral function
vanishes, thus the contribution to the zero bias conductance only stems from the Fermi functions as $\partial K(\omega,\mu_\alpha,\beta_\alpha)/\partial V$,
\begin{eqnarray}
&&\left. \frac{\partial K(\omega,\mu_\alpha,\beta_\alpha)}{
\partial V} \right|_{V\to 0}\\
&&= 
\frac{e}{4}\left(
\frac{R}{1+R}
\frac{\beta_R}{\cosh^2 \left(\frac{\beta_R\omega}{2}\right)}
+
\frac{1}{1+R}
\frac{\beta_L}{\cosh^2 \left(\frac{\beta_L\omega}{2}\right)}
\right),
\nonumber
\end{eqnarray}
for an arbitrary asymmetry $R$. 
Clearly the zero-bias conductance is symmetric under the exchange of the temperature $T_L\leftrightarrow T_R$ for $R=1$, so that $G(T_R,x_T)= G(T_R\,x_T, 1/x_T)$.
Furthermore, $G(T_R,x_T)\ge 0$ for all parameters since $\partial K/\partial V>0$ and the \ac{NEQ} spectral functions remain positive.

The charge current $I_{\rm charge}$ depends on the  bias voltage as well as a finite temperature
gradient and vanishes in an open junction. In this 
configuration, a finite temperature gradient $\Delta T$ 
generates a finite bias voltage $\Delta V$ across
the junction in the absence of a  charge current.
The Seebeck coefficient $S$, also known as thermopower, connects
the finite  voltage $\Delta V$ and the $\Delta T$ and 
 is defined as the ratio
\begin{equation}
S= \left(\frac{-\Delta V}{\Delta T}\right)_{I_{\rm charge}=0}.
\end{equation}

In the linear response regime, where $\delta  T, \delta V\to 0$, the Seebeck coefficient $S_{\rm lin}$,
\begin{equation}
S_{\rm lin} =-\frac{1}{eT}\, \frac{L_1}{L_0},
\label{eq:Slin}
\end{equation}
is expressed by the ratio of two transport integrals, where
$L_n=-\tfrac{1}{h}\int d\w (\w -\mu)^n \tfrac{\partial f}{\partial \w} \mathcal{T}(\w)$ are the Onsager integrals,
with $\mathcal{T}(\w)$ being the transmission coefficient of the junction\cite{Costi2010}.
To make connection to the result in Ref.~ \cite{Costi2010}, we expand
$K(\omega,\mu_\alpha,\beta_\alpha)$ for $\delta T\to 0,\delta V=0$  and $\delta V\to 0,\delta T=0$ respectively,
\begin{eqnarray}
K(\omega,0,\beta_\alpha) &=& \frac{\beta }{4}\left(
\frac{1}{\cosh^2 \left(\frac{\beta\omega}{2}\right)}
\right)
\beta \omega \delta T + O( \delta T^2),
\non
K(\omega,\delta V,\beta)&=& 
\frac{\beta }{4}\left(
\frac{1}{\cosh^2 \left(\frac{\beta\omega}{2}\right)}
\right)e\delta V+ O( \delta V^2).
\end{eqnarray}
The current from these contributions vanishes according to Eq.~\eqref{eq:meir-wingreen-Qdot} yielding the form of Seebeck coefficient as in Eq.~\eqref{eq:Slin}.

\subsubsection{Broken particle-hole symmetry}

\begin{figure}
\begin{center}
 \includegraphics[width=0.45\textwidth,clip]{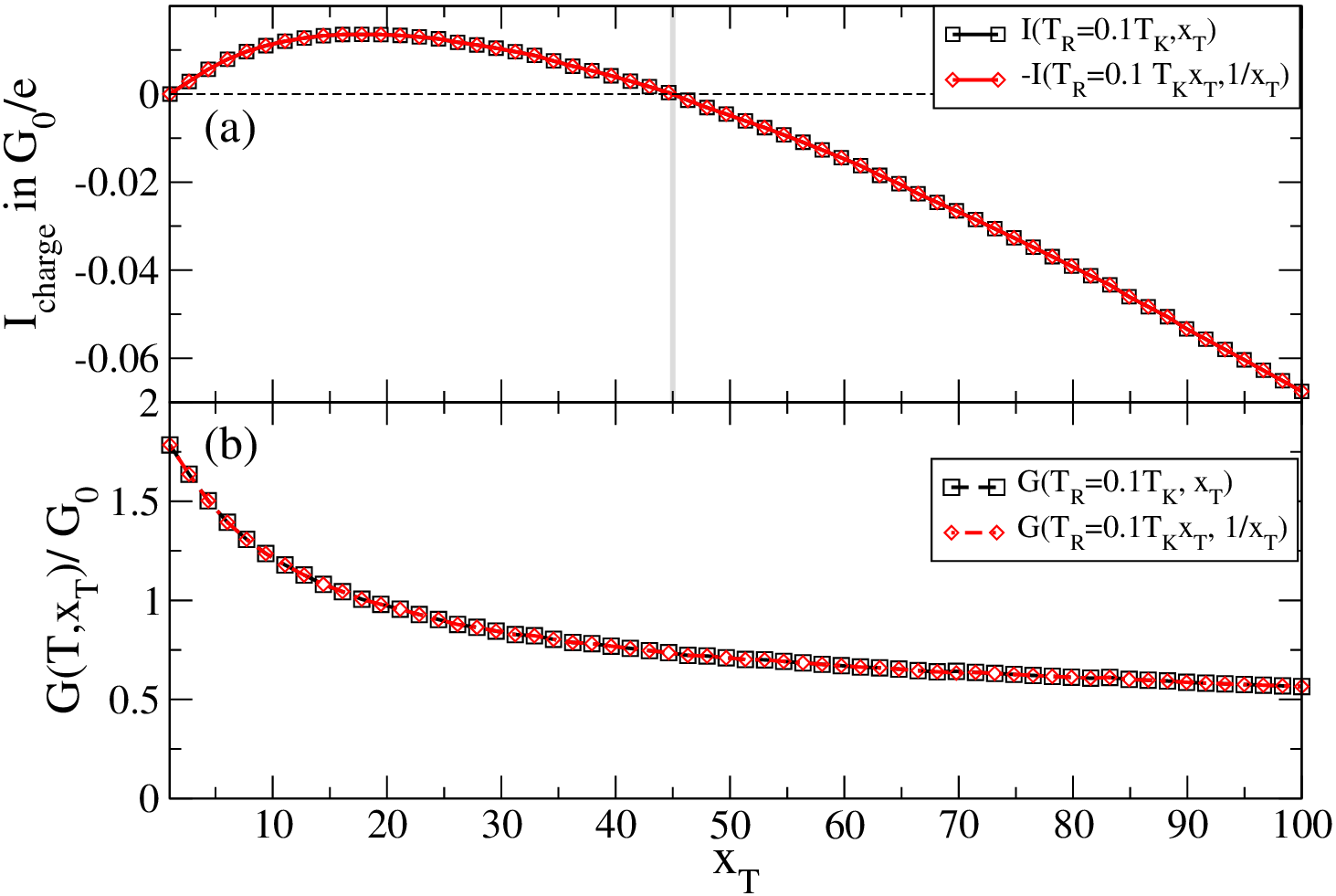}

\caption{(a) 
Thermoelectric charge currents $I(T_R=0.1\,T_K,x_T)$ and $I(T_R=0.1\,T_K\, x_T,1/x_T)$ as a function of the temperature asymmetry $x_T=T_L/T_R$ for $U/\Gamma=20,\e_d/\Gamma=-3$
at zero bias voltage and a symmetric junction $R=1$. The $x_T=45$ at which the thermoelectric current vanishes is indicated by the vertical gray line.
(b) Zero-bias conductance $G(T_R= 0.1\,T_K,x_T)$ and $G(0.1\, T_K x_T,1/x_T)$ for the same junction as in (a).
NRG parameter: $\Lambda=1.8,N_s=1000,D=10^3\Gamma, b=0.6,N_z=2,\rho_c=0.02$.
}

\label{fig-13}
\end{center}
\end{figure}

\begin{figure}[b]
\begin{center}
 \includegraphics[width=0.45\textwidth,clip]{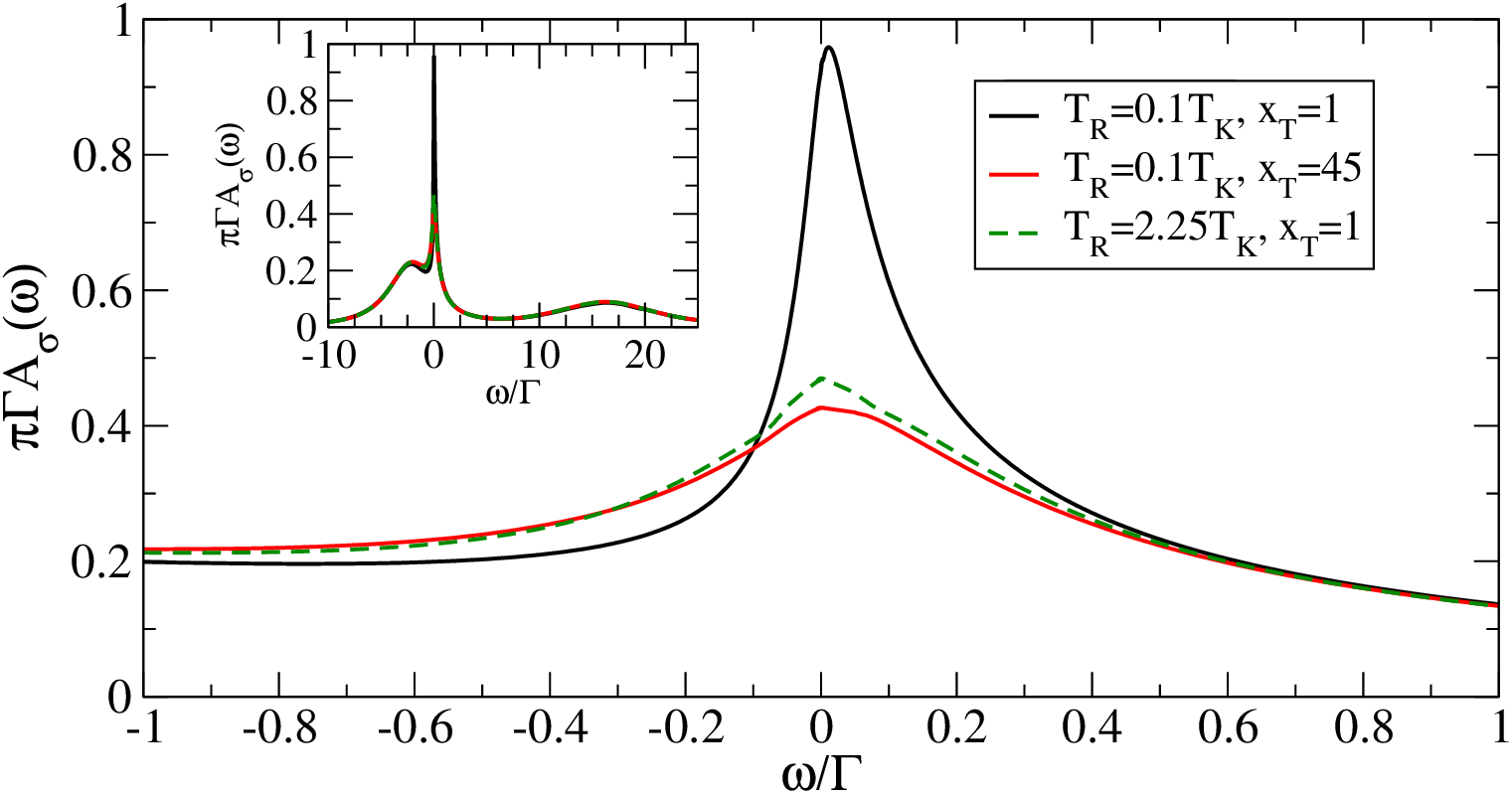}

\caption{Comparison of the equilibrium spectral function and the 
$A_\sigma(\omega)$ at $x_T=45$ at which the thermo-current vanishes. We added the equilibrium spectral function
at $T_R=T_L=2.25\,T_K$ as dashed green line for comparison.
NRG parameter: as in Fig \ref{fig-13}.
}

\label{fig-14}
\end{center}
\end{figure}

\begin{figure}
\begin{center}
 \includegraphics[width=0.45\textwidth,clip]{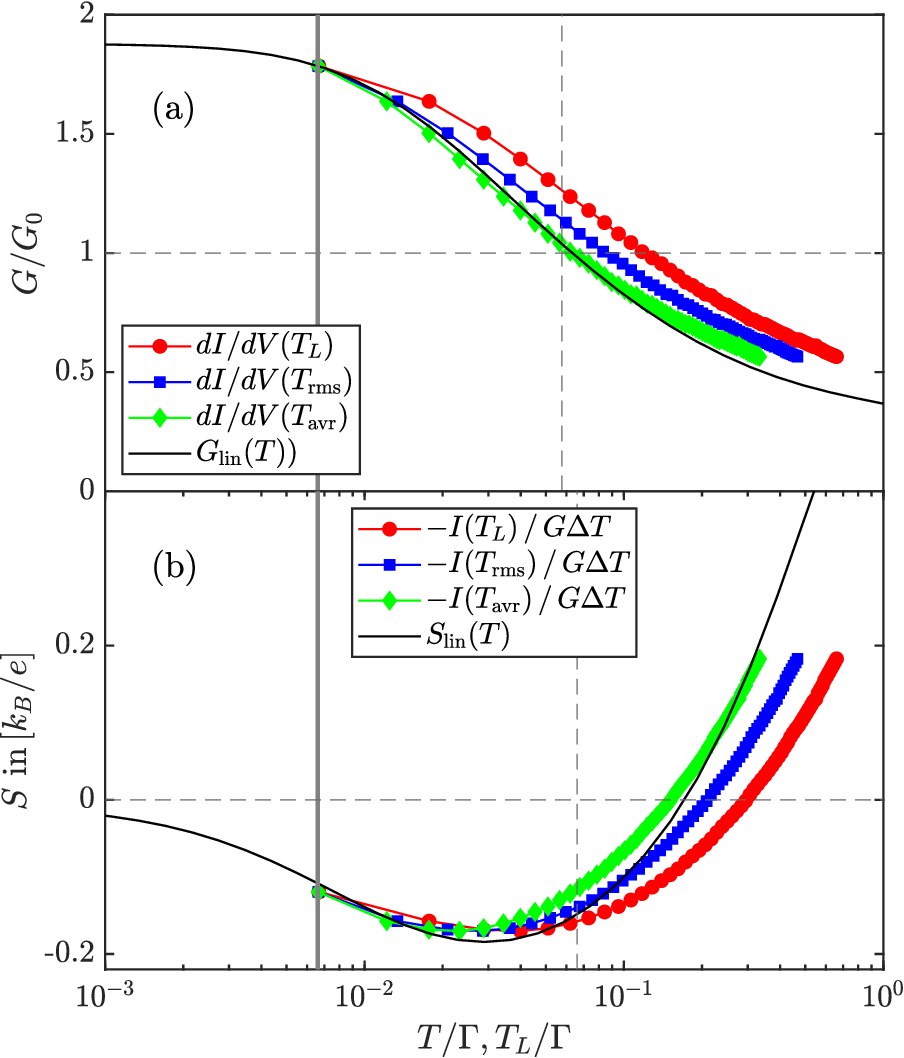}

\caption{The differential conductance (a) and the Seebeck effect (b) extracted from Fig.\ref{fig-13} as a function of $T_L$, and different effective temperatures $T_{\rm rms}, T_{\rm avr}$ compared with the equilibrium $G_{\rm lin}(T)$ and $S_{\rm lin}(T)$ calculated using \ac{NRG} (black curves). The vertical gray line indicates where both lead temperatures are in equilibrium $T_L=T_R=T$, the dashed vertical line shows the Kondo temperature $T=T_K$.
}

\label{fig-15}
\end{center}
\end{figure}

A  temperature gradient between the two leads drives a charge current through the junction for asymmetric \ac{QD} spectral functions.
This is exemplified for $\e_d/\Gamma=-3$ and $U/\Gamma=20$. The equilibrium transport Kondo temperature was
determined to $T_K=T_K(\text{trans})=0.066\Gamma$.

The charge transport current is depicted in Fig.\ \ref{fig-13}(a) as function of the temperature ratio $x_T$ for a base temperature $0.1\,T_K=\text{min}(T_L,T_R)$. 
The current shows a non-monotonic behavior. It increases starting  from   $x_T=1$, passes  through a maximum and changes sign around  $x_T\approx 45$. 
This sign change in the thermoelectric current is often referred to as 
a signature of the Kondo resonance \cite{Costi2010}.
The observed behavior is connected to a competition between the increase of the asymmetry in the function $K(\omega,0,\beta_\alpha)$ with increasing temperature asymmetry $x_T$ and the simultaneous reduction of the asymmetric Kondo resonance due to an effective temperature increase.
We also added the corresponding negative current generated by exchanging $T_L$ and $T_R$. 
The current $-I(T_R=x_T T_K/10,1/x_T)$
perfectly agrees with $I(T_R=T_K/10,x_T)$ as expected from the analytic discussion of $K(\omega,0,\beta_\alpha)$ above: by inverting the temperature gradient
the current is inverted in a symmetric junction $R=1$.
It has been conjectured  \cite{PhysRevB.111.035445}  that the effective temperature $T_\text{eff}\approx \sqrt{\tfrac{1}{2}(T_L^2 +T_R^2)} = \frac{T_R}{\sqrt{2}} \sqrt{1 + x_T^2}$
can be introduced which accounts for the change in $f_\text{eff}(\omega)$ when changing the asymmetry.  

In Fig.\ \ref{fig-14} we plotted the equilibrium spectral function for $T_R=T_L=T_K/10$ (black line) and the nonequilibrium spectral function (red line)  for $T_R=T_K/10$ at $x_T=45$  at which the  thermocurrent vanishes. Clearly, the nonequilibrium spectral function is much more symmetric than 
the equilibrium spectral function at $x_T=1$. With increasing $x_T$ spectral weight is transferred from the Kondo resonance above the chemical
potential to energies below zero. At $x_c=x_T=45$, the integral over the transport window $K(\omega,0,\beta_\alpha)$ vanished. Above $x_c$ there is more
spectral weight below the chemical potential so that there is a sign change in the thermo-current.
For comparison, the equilibrium spectral function for $T_R=T_L=2.25\,T_K$  has been added as dashed green line, which is smaller than $T_\text{eff}=3.18\,T_K$ expectation from Ref.~\cite{PhysRevB.111.035445}. This equilibrium spectral function appears to be reasonably close to the nonequilibrium spectrum.
The  zero bias conductance corresponding to the parameters used in Fig.\ \ref{fig-13}(a) are shown in Fig.\ \ref{fig-13}(b). 
$G(T_R,x_T)$ is positive as expected as well as identical  with $G(x_T T_R,1/x_T)$. It is also monotonically decreasing with  $x_T$.

To extend on the discussion of an effective equilibrium temperature recovering
the characteristics of a temperature gradient on the leads,
Fig.~\ref{fig-15} provides a comparison of the zero-bias conductance and Seebeck coefficient under a finite temperature gradient on the leads,
with linear response $G_{\rm lin}$ and $S_{\rm lin}$ calculations using NRG with equilibrium temperature $T$ (black curves).
The differential conductance from \ac{OC}-\ac{FDM} with finite temperature gradient
presented in panel  Fig.~\ref{fig-13}(a) shows good convergence with the linear response $G_{\rm lin}(T)$ from \ac{NRG} when $T_L \to T_R$.
With increase in just the left lead temperature $T_L$, the zero-bias conductance remains
larger than the $G_{\rm lin}$ with equilibrium temperature $T$.
This behavior is expected as the thermal fluctuations from heating up both 
leads will destroy the Kondo resonance
much more than the case of heating up just one lead.
The zero-bias conductance was rescaled using an effective temperature $T_{\rm eff}=T_{\rm rms}$ as observed in the \ac{NRG}-\ac{TD-DMRG} studies in Ref.~\cite{PhysRevB.111.035445} 
and an average of the lead temperatures $T_{\rm avr}$ to investigate
possible universal scaling as exhibited by $G_{\rm lin}(T)$.
Although not perfectly, results in Fig.~\ref{fig-15}(a) shows the scaling with $T_{\rm avr}$
tracks $G(T)$ much better than $T_{\rm rms}$. 

Since the current generated by a temperature gradient $\Delta T$ is small [cf. Fig.~\ref{fig-13} (a)],
we can assume a small enough voltage $\Delta V$ can compensate for the thermally induced charge current by asserting $|I_{\rm charge}| = G(\Delta T) \cdot |\Delta V|$.
Here, $G(\Delta T) \equiv G(T_R,x_T)$ is the zero-bias conductance in the case of different lead temperatures as shown in
Fig.~\ref{fig-13} (b).
Thus, Seebeck coefficient under a temperature gradient can be represented as,
\begin{equation}
S= \frac{-\Delta V}{\Delta T}=\frac{-I_{\rm charge}}{G\Delta T}.
\end{equation}
Figure~\ref{fig-15}(b) presents the Seebeck coefficient extracted from Fig.~\ref{fig-13} as described above. The linear response $S_{\rm lin} (T)$ calculated using \ac{NRG} is presented for comparison as solid black line.
The Seebeck coefficient also shows good convergence with equilibrium linear response results when $T_L \to T_R$. 
Since the thermoelectric charge current is a result of particle-hole asymmetric spectral function, 
influence of different lead temperatures in $f_{\rm eff}$ and in the Meir-Wingreen expression for current in Eq.\ \eqref{eq:meir-wingreen-Qdot}, one does not expect a universal scaling behavior. Nevertheless, for the sake of completeness, we have provided the Seebeck coefficient rescaled with respect to $T_{\rm rms}$ and $T_{\rm avr}$. Here as well, scaling with $T_{\rm avr}$ tracks the linear response slightly better in the two limits of $\Delta T$.

\subsubsection{Finite magnetic field}

\begin{figure}
\begin{center}
\includegraphics[width=0.45\textwidth,clip]{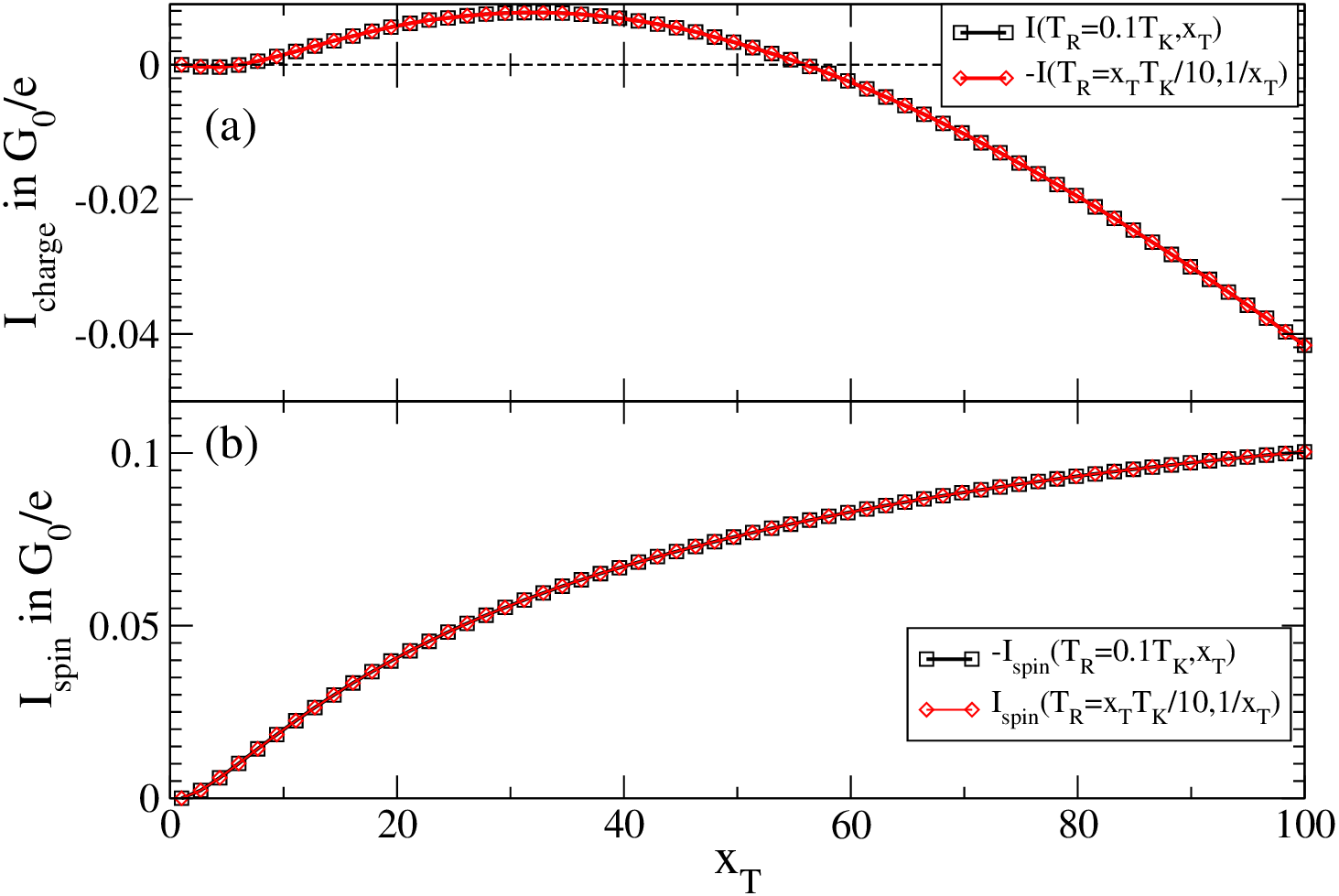}

\caption{Thermal gradient induced (a)  charge and (b) spin current through a junction with $U/\Gamma=20,\e_d/\Gamma=-3$ in a finite
magnetic field $h_0=2T_K$. 
NRG parameter: as in Fig \ref{fig-13}.
}

\label{fig-16}
\end{center}
\end{figure}

We investigated the influence of a magnetic field $h_0=2T_K$ 
onto the current through a junction defined by the parameters of   Fig \ref{fig-13}.
In addition to the modified thermoelectric charge current depicted in Fig.\ \ref{fig-16}(a)
we also find a finite spin current $I_\text{spin}$ displayed  in Fig.\ \ref{fig-16}(b). 
It also reveals current reversal upon swapping $T_L$ and $T_R$ shown as red line in Fig.\ \ref{fig-16}(a).

\begin{figure}
\begin{center}
\includegraphics[width=0.45\textwidth,clip]{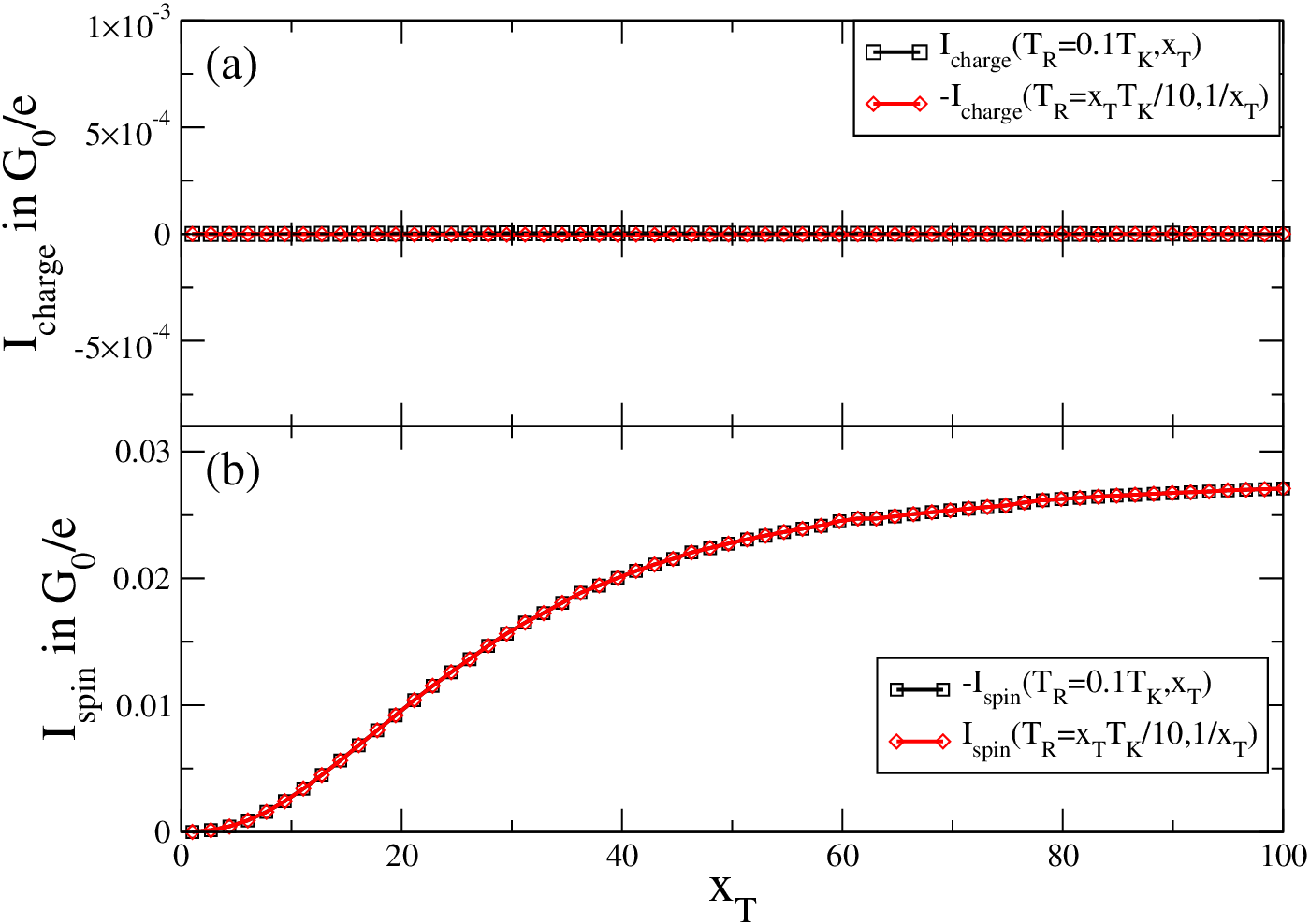}

\caption{
Charge (a) and spin current (b) as a function of $x_T$ for a particle-hole symmetric junction in a finite magnetic field
with $\e_d=-U/2,U/\Gamma=12, h_0=2T_K$.
 NRG parameter: as in Fig \ref{fig-13}.
}

\label{fig-17}
\end{center}
\end{figure}

Due to the finite splitting of the Kondo resonance above and below the chemical potential, $I_\uparrow$ and $I_\downarrow$ flow in opposite directions for a given temperature gradient. Therefore, the charge current $I_\uparrow+ I_\downarrow$ almost vanished
for small values of $|x_T-1|$ while the spin current increases linearly with $|x_T-1|$ as shown in Fig.\ \ref{fig-16}(b). Once  $x_T>10$, the Kondo resonance
is slowly reduced and a small charge current is observed. It is also interesting to note that
the Zeeman-splitting of the Kondo resonance shows up an additional sign change in the 
charge current for $x_T \approx 4$.

The spin-current, however, is significantly larger than the charge current due to finite magnetic field
which cause a significant asymmetry between the spectral functions of the two spin channels.

Since the charge current is driven by the particle-hole asymmetry, we can suppress the charge current completely in a magnetic field
for a particle-hole symmetric \ac{QD}, i.e., for $\e_d=-U/2$. We find that $I_\text{charge}=0$ for any $x_T$ as demonstrated in
Fig.\ \ref{fig-17}(a) while we detect a significant spin current shown in Fig.\ \ref{fig-17}(b)
Its direction is again dependent on the direction of the temperature gradient.

\section{Conclusion and outlook}

We presented an efficient 
hybrid  approach combining \ac{NRG} to accurately diagonalize the Hamiltonian
and a \ac{BR} approach by coupling a reservoir to each Wilson site.  It combines
the virtue of an accurate description of the Kondo effect and the strong coupling fixed point
with the strength of the \ac{BR} approach
in imposing the correct nonequilibrium boundary conditions imposed by the reservoir
to access the nonequilibrium steady state. The reservoir coupling functions 
are exactly constructed by the continued fraction expansion from the original coupling functions to the leads \cite{OpenChains2017,BoekerAnders2020}.

In this paper, we presented as a first application, an effective single lead implementation based on the total coupling
and the nonequilibrium effective occupation function. The combination of the lead coupling functions
and the effective lead occupation function $f_{\text{eff}}(\omega)$
completely determines the influence of the boundary condition
onto the nonequilibrium dynamics of the interacting region. The resulting nonequilibrium
Green's function enters the transport integral. Assuming 
$\Gamma^L(\omega)= R\Gamma^R(\omega)$, it is sufficient to calculate the retarded Green's function \cite{MeirWingreen1992}.
The  implementation of the effective lead \ac{OC}-\ac{FDM} nonequilibrium steady state
approach is numerically very fast. 
With the effective Wilson chain length, and hence the runtime, set by the bias and temperature,
a current data point is obtained in 1-2 mins on an off-the-shelf laptop with M1 ARM CPU  in contrast to the recent inchworm
 \ac{QMC} approach \cite{CohenGullQMC2023} which requires about $10^4$ core-hours CPU time on a HPC supercomputer.
The equilibrium spectral functions obtained by the \ac{OC}-\ac{FDM} approach are nearly identical to
the \ac{FDM} approach \cite{WeichselbaumDelft2007} while the
destruction of the Kondo effect with increasing voltage can be seen in the finite bias spectrum.
The voltage dependency of the \ac{QD}
double occupancy $D_o$  follows very closely the RT-\ac{QMC} data of Ref.\ \cite{DirksSchmitt2013}.

Moreover, we have investigated the differential conductance through a \ac{SIAM} in the presence of finite bias, 
recovering the zero-bias conductance peak for Kondo energy scales down to
$T_K/D = 3.3 \times 10^{-8}$ and showing good agreement with the Fermi-liquid theory predictions from Oguri~\cite{OguriFermiLiquid2005} and nonequilibrium Kondo energy scale ratio from Schwarz et al.~\cite{SchwarzPRL-TransportQD2018}.
Further, we discuss the influence of particle-hole asymmetry, asymmetric coupling to the leads and an external magnetic field on the nonequilibrium conductance through the \ac{SIAM}. 
We also show that a finite bias-driven spin current can flow 
in the case of a Zeeman-split Kondo resonance and asymmetric coupling to the leads.
In the presence of temperature gradient, we investigated the zero-bias conductance,
thermoelectric current and consequently the Seebeck coefficient, which showed convergence to equilibrium \ac{NRG} calculations when the temperature gradient approached the linear response regime. Additionally, we observed finite spin currents in the presence of a magnetic field and finite temperature gradient.

In the future, we plan to apply the \ac{OC}-\ac{FDM} approach to
a two-lead setup  to access the low energy excitations in each lead individually which are separated by an energy $E=eV$. 
Although numerically much more expensive, it is straightforward to 
implement multiple leads as outlined in Sec.\ \ref{sec:NRG+BRT}.
This implementation would also prove useful in resolving any fine structure in the 
spectral function near the bias window.

Recently Child et al \cite{Child-QPC-Entropy2022}
applied a robust protocol for entropy measurement in mesoscopic circuits
\cite{Child-A-2022} to \ac{QD} and compared the experimental results to
the prediction of the equilibrium NRG. The discrepancy between strong coupling \ac{NRG} results
and the experiment were later attributed to backaction \cite{MaMeirQPC2023,Sankar2025}
of the \ac{QPC} used to detect the charge occupation
onto the  \ac{QD}. We plan to employ our \ac{OC}-\ac{FDM}  
approach to this geometry to calculate the influence of
the charge fluctuation in the  \ac{QPC} onto \ac{QD} entropy 
in a two-lead, three-chemical-potential setup. The \ac{OC}-\ac{FDM}  can
naturally include additional 
local heating effects due to the finite current and resulting temperature gradients.
Specifically, we will address whether the Maxwell relation \cite{Child-QPC-Entropy2022} 
derived from equilibrium thermodynamics requires modification in the presence of a \ac{QPC} 
whose back-action drives the \ac{QD} out of equilibrium.

\section{Acknowledgments}
We acknowledge fruitful discussions with Jong Han and Yigal Meir as well 
as financial support by Deutsche Forschungsgemeinschaft via
 Grant AN 275/10-1.

\appendix

\section{Numerical implementation of the OC-FDM approach}

\tikzstyle{block} = [draw, fill=white, rectangle, 
    minimum height=3em, minimum width=6em]
\tikzstyle{sum} = [draw, fill=white, circle, node distance=1cm]
\tikzstyle{output} = [coordinate]
\tikzstyle{pinstyle} = [pin edge={to-,thin,black}]

\begin{figure}[htbp]
\begin{center}

\begin{tikzpicture}[auto, node distance=2cm,>=latex']

    \node [block] (init) {INIT parameters; $\beta_\alpha,\mu_\alpha,\cdots$};
    \node [block, below of =init, node distance=1.5cm] (himp) {setup $H_{imp},m=-1$};
     \node [sum, below of=himp, node distance=1.5cm](sum2){};
      \node [block, below of =sum2, node distance=1.5cm] (NRG) {NRG iteration $m$};
       \node [block, below of =NRG, node distance=1.5cm] (BRT) {BRT: $\rho_\text{steady state}\to \bar \rho^K_m$ };
         \node [block, below of =BRT, node distance=1.5cm] (rho) {store $\rho_{ll}^{red}=\bar \rho^K_{m-1} \rho_{ll}^{m}$ };
       \node [shape aspect=3,diamond,draw, below of =rho, node distance=1.5cm] (cutoff) { $ \rho^K_m>\rho_c$ and $m < N_\text{max}$? };
       \node [block, below of =cutoff, node distance=2cm] (backward) {GF backward iteration $m\to m-1$ };
     \node [block, below of =backward, node distance=1.5cm] (stop) {stop at $m=m_\text{min}$ };

 \coordinate [left of=cutoff,node distance=3cm] (left) {};
 \coordinate [left of=sum2,node distance=3cm] (leftsum) {};

      \draw [->] 	(init) --(himp);
     \draw [->] 	(himp)--(sum2);
     \draw [->] 	(sum2)--node{$m\to m+1$}(NRG);
      \draw [->] 	(NRG)--(BRT);
     \draw [->] 	(BRT)--(rho);
     \draw [->] 	(rho)--(cutoff);
     \draw[-](cutoff)--(left);
     \draw[-](cutoff)--(left);
     \draw[->] (left) -I (leftsum) -l node [near end] {yes} (sum2);
          \draw [->] 	(cutoff)--(backward);
               \draw [->] 	(backward)--(stop);
 \end{tikzpicture}

\caption{Schematic box flow diagram of the \ac{OC-FDM} algorithm}
\label{fig:boxflow}
\end{center}
\end{figure}

\begin{figure}[t]
\begin{center}
\includegraphics[width=0.95\columnwidth, trim=0 0 0 0, clip]{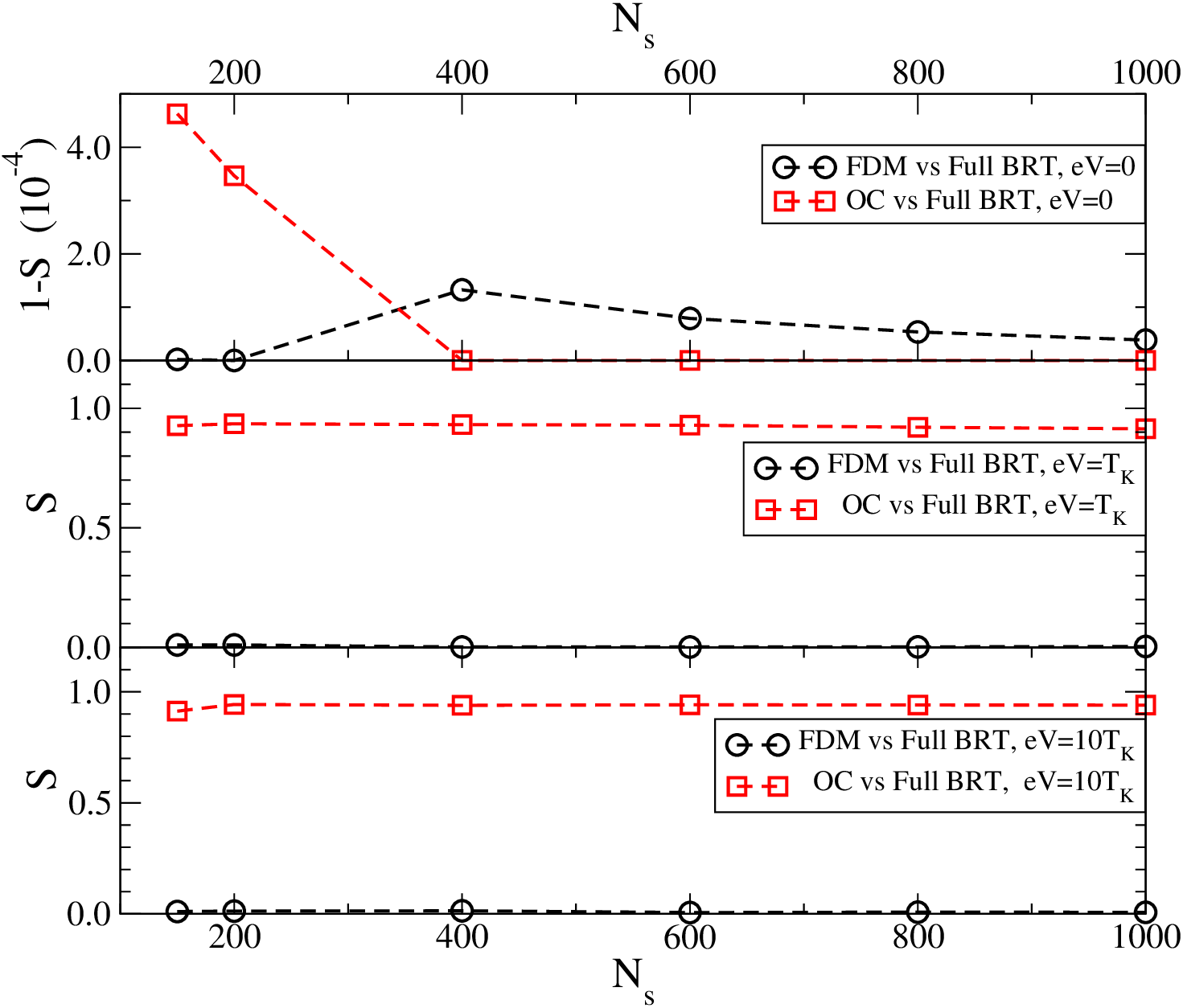}
\caption{
Comparison of the overlap $S$ between density matrices from different approaches (FDM, OC-FDM, full-BRT) as a function of the number of kept states $N_s$ at each NRG iteration: In the top panel we compare
$1-S$ for equilibrium parameter ($eV=0$ ), in the middle panel $S$ for $eV=T_K$ and the bottom panel  shows $S$ for $eV=10T_K$.
NRG parameters: $\Lambda=2.5, D/\Gamma=100, U/\Gamma=12, \e_d/\Gamma=-6, N_z=1$.
}
\label{fig:A2}
\end{center}
\end{figure}

The  \ac{OC-FDM} algorithm is embedded in the standard approach for calculating spectral functions within the NRG
\cite{PetersPruschkeAnders2006,WeichselbaumDelft2007}. 
The numerical implementation of the method described in Sec.\ \ref{sec:III} is schematically depicted in Fig.\ \ref{fig:boxflow}.
Once the couplings along the Wilson chain and to the high-energy reservoirs are calculated according
to the continued fraction expansion, we start the NRG iterations with the impurity Hamiltonian as the $m=-1$ iteration.
After each NRG iteration step $m$, the discarded states are determined by the Wilson  energy criteria \cite{BullaCostiPruschke2008}.
Then the \ac{BRT} is calculated and the corresponding normalized steady state density matrix via a Lanczos algorithm.
The eigenvector is used to determine $\rho^m_{ll}$ and  $\bar \rho_{m}^K$. We save all relevant data, and proceed to the
next NRG iteration provided that  $\bar \rho_{m}^K>\rho_c$ and $m < N_\text{max}$.

In the equilibrium NRG, typically  $N=N_\text{max}$ is provided defining the smallest Wilson shell energy.
Typically, $\beta_N$ is selected as the temperature. In the FDM approach \cite{WeichselbaumDelft2007}, however,
$\omega_N \ll T$, so that $N=N_\text{max}$  must be chosen such that the spectral weights $w_m$ are still finite.
In our \ac{NEQ} approach we have two different chemical potentials and potentially two different temperatures $T_\alpha$
for both leads. Therefore, we use $\bar \rho_{m}^K$ as an indicator how much spectral weight the remaining iteration
can contribute to the reduced density matrix. If $\bar \rho_{m}^K$ is below some threshold, essentially the rest of
the chain decouples from the impurity and one must stop the iteration.

Once the NRG iterations stop, we proceed with a standard backward NRG run, which
(i) recursively calculates the reduced density matrix $\rho^\text{red}_{k,k'}(m)$ in the kept-kept sector - for more details
see \cite{AndersSchiller2005,AndersSchiller2006,PetersPruschkeAnders2006} -- (ii) include the 
$\bar \rho_{m}^K\rho^m_{ll}$  as diagonal matrix elements in the discarded sector, and (iii) calculate the
contribution of the Wilson shell $m$ to the spectral functions. We use about $250-500$ bins for the Lehmann representation of the
spectral weights and a
log-Gaussian to Gaussian broadening as described in the supplement of Ref.\ \cite{WeichselbaumDelft2007}. 

We use the equation of motion approach \cite{BullaHewsonPruschke98}
to obtain an accurate representation of the retarded \ac{GF} which is only slightly dependent on the broadening.
The transport currents are then calculated using Meir-Wingreen formula and the differential conductance $dI/dV$ from numerical differentiation.

\section{Comparison between the OC-FDM and the Full BRT approach}

\begin{figure}[ht]
\begin{center}
\includegraphics[width=0.95\columnwidth, trim=0 0 0 0, clip]{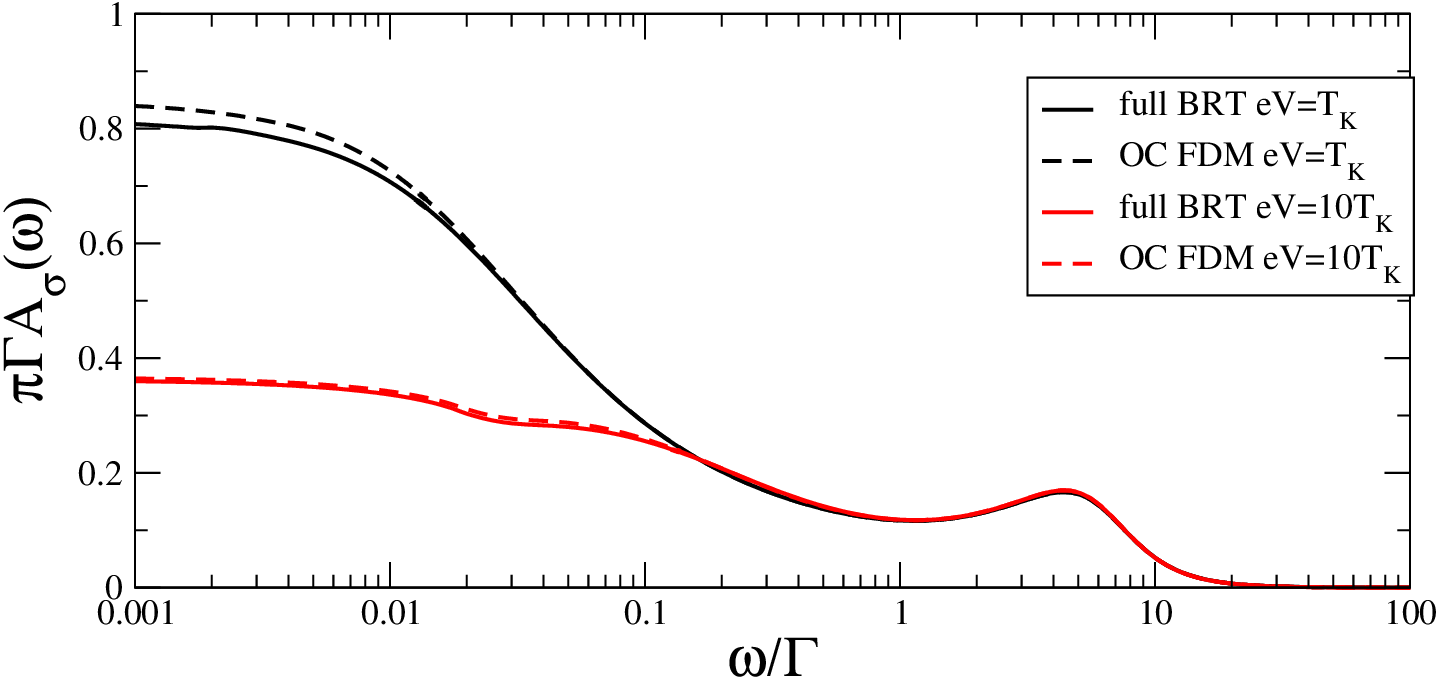}
\caption{
Comparison of the spectral functions calculated by the \ac{OC-FDM} approach and the full \ac{BRT} approach
for $eV=T_K$ (black lines) and for $eV=10\,T_K$ (red lines) using $N_s=1000$ kept states, no z-averaging and 
the same NRG parameters as in
Fig.\ \ref{fig:A2}.
}

\label{fig:A3}
\end{center}
\end{figure}

In this Appendix, we provide a quantitative comparison between
the full BRT approach of Ref.~\cite{BoekerAnders2020} and the simplified but numerically much
more efficient \ac{OC-FDM} algorithm proposed in this paper for NEQ quantum transport.

The numerical implementation of the  full BRT approach  follows the same box flow diagram as illustrated  in Fig.~\ref{fig:boxflow}. In addition to the accumulation of the on-shell  \ac{BRT} matrix elements $\Gamma_{l1,l2}(m_1,m_1)$ which are given by the subset of the \ac{OC-FDM} \ac{BRT}, also  the additional inter-shell contributions $\Gamma_{l1,l2}(m_1,m_1-1),\Gamma_{l2,l1}(m_1-1,m_1)$ are calculated.  For $m_1<m_2$, the discarded states $|l_2,e_2,m_2\rangle$ in $X^{\tilde m\alpha\nu}_{l_1,l_2}(m_1,m_2) $ defined in Eq.~\eqref{eq:14d}  must be expanded in the kept states of $m_1$ which requires an expensive summation over all different overlaps. The technical detail can be found in Sec III.C.2 of Ref.~\cite{BoekerAnders2020}.   

After the NRG iterations stop, we take the full \ac{BRT}, which is a sparse matrix, and calculate its eigenvalues and the eigenstates
in a small Krylov space with the dimension $N_k\le 100$. We always find a zero eigenvalue whose corresponding eigenvector is the steady
state reduced density matrix. Using this as the steady state density matrix, the standard NRG algorithm \cite{PetersPruschkeAnders2006,WeichselbaumDelft2007}  yields the spectral functions.

To keep the CPU times reasonable, we restricted our calculations to $\Lambda=2.5$
to reduce the number of NRG iterations. We also did not add intershell matrix elements for $|m_1-m_2|>1$,
and chose $T=10^{-4}\Gamma\ll T_K$. For the equilibrium benchmark settings $eV=0$, we stop NRG after 40 iterations
corresponding to a lowest energy scale $\omega_N$ reached after $N$ NRG iterations, $\omega_N/T\approx  0.012$.
The dimension of the full BRT was $L^2=103252^2$, with $L$ being the total number of discarded states (without degeneracy) 
for $N_s=1000$ kept states, which cannot be exactly diagonalized. 
Hence we use a bidirectional Lanczos algorithm \cite{SaadSparseLinearSystemsBook2003}  to calculate the steady state solution of this  full \ac{BRT}.

We calculate the three different density matrices $\vec{\rho}_\text{FDM}$, $\vec{\rho}_\text{OC-FDM}$ and $\vec{\rho}_\text{Full-BRT}$,
where  $\vec{\rho}_\text{FDM}$ \cite{WeichselbaumDelft2007}  is only a useful quantity in equilibrium at $eV=0$, but can serve as reference and as a measure of the deviation from equilibrium.  $\vec{\rho}_\text{OC-FDM}$ is the steady-state density matrix obtained from the OC-FDM approach described in this work and 
$\vec{\rho}_\text{Full-BRT}$ is the steady-state solution obtained from the implementation of the full BRT. These vectors are elements of the space spanned by all discarded states, and their coordinates define diagonal matrix elements of the reduced  density matrix $\rho^\text{red}_{ll}(m)$ \cite{BoekerAnders2020}. 

We define the overlap S
\begin{eqnarray}
S(\vec{\rho}_1,\vec{\rho}_2)&=& \frac{\vec{\rho_1} \vec{\rho_2}}{||\vec{\rho_1}|| \cdot ||\vec{\rho_2}||}
\label{Eq:overlap}
\end{eqnarray}
as a measure of agreement between the different density operators. 
The results for $1-S$ at $eV=0$ are
shown in the upper panel of Fig.\ \ref{fig:A2}. 
As pointed out above, the $\vec{\rho}_\text{FDM}$  is the 
analytical steady state  solution of the  full BRT in equilibrium.
The reason of the deviation  $1-S(\vec{\rho}_\text{FDM},\vec{\rho}_\text{Full BRT})$
at $eV=0$ is related to the numerical representation of the full BRT and the accuracy of the Lanczos approach. When using   $\vec{\rho}_\text{FDM}$
as a starting vector of the Lanczos algorithm always yields $\vec{\rho}_\text{Full BRT} = \vec{\rho}_\text{FDM}$, hence $S(\vec{\rho}_\text{FDM},\vec{\rho}_\text{Full BRT})=1$. 
Using $\vec{\rho}_\text{OC-FDM}$ as a starting vector yields $\vec{\rho}_\text{Full BRT}= \vec{\rho}_\text{OC-FDM}$ 
at $eV=0$, since the numerical differences are too small to generate a larger Krylov space \cite{SaadSparseLinearSystemsBook2003} of independent base vectors. $\vec{\rho}_\text{OC-FDM}$ agrees within the accuracy of the Lanczos approach with  $\vec{\rho}_\text{FDM}$ and $\vec{\rho}_\text{Full BRT}$ at equilibrium. All these density matrices yield the same spectral function
for $N_s\ge 400$, not shown here.

\begin{figure}[t]
\begin{center}
\includegraphics[width=0.85\columnwidth, trim=0 0 0 0, clip,angle=270]{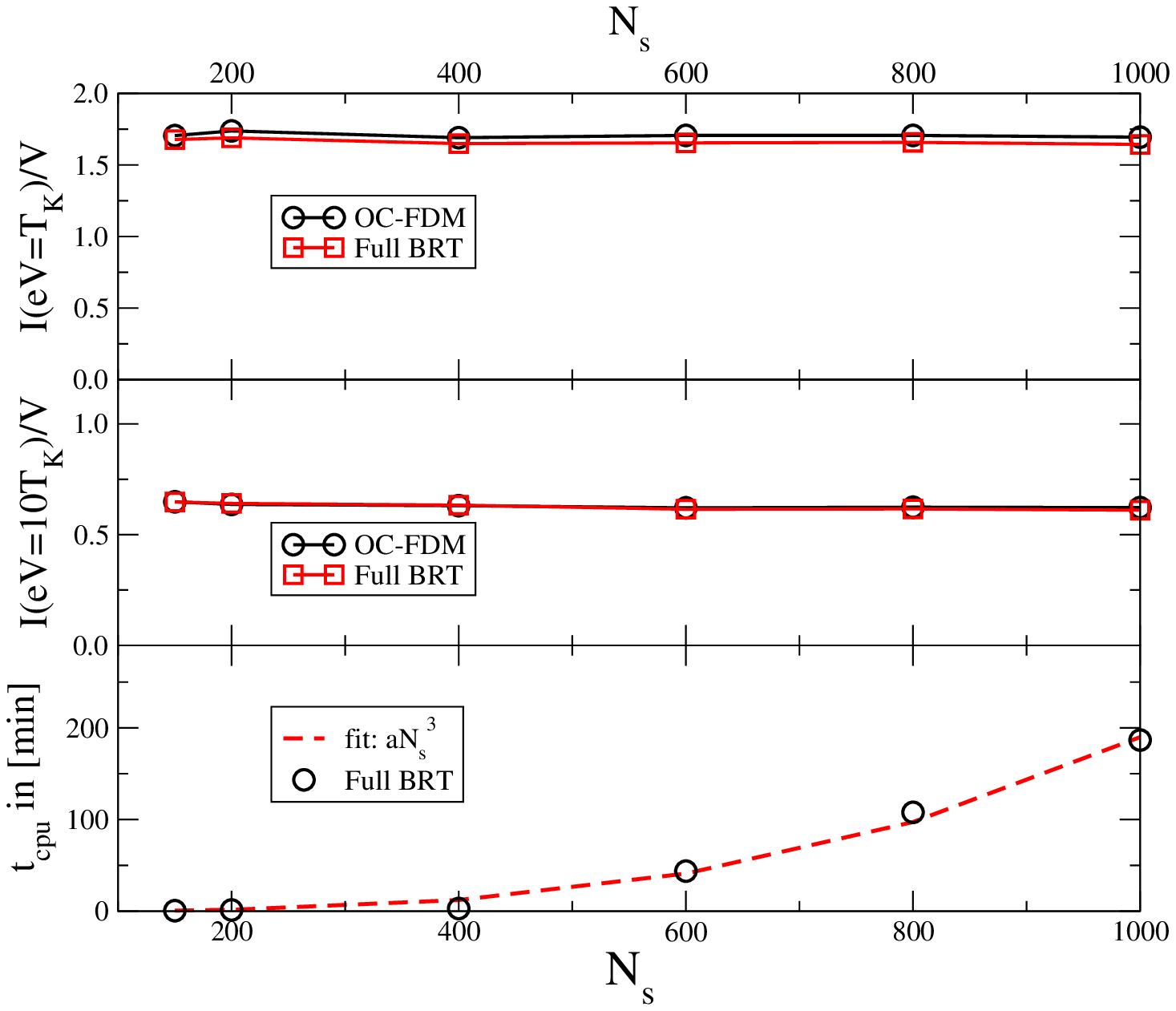}
\caption{
Comparison of currents calculated using different approaches as a function of $N_s$ for $eV=T_K$ (upper panel) and for $eV=10T_K$ (middle panel)
 in  units $[G_0]/e$.  NRG parameters same as in
Fig.\ \ref{fig:A2}. 
The computational time $t_\text{cpu}$ for full BRT approach as a function of $N_s$ is presented in the bottom panel.
}

\label{fig:A4}
\end{center}
\end{figure}

At finite $eV$ the equilibrium density matrix  $\vec{\rho}_\text{FDM}$ must differ from the $\vec{\rho}_\text{OC-FDM}$ and $\vec{\rho}_\text{Full BRT}$ solution. $S(\vec{\rho}_\text{FDM},\vec{\rho}_\text{Full BRT})\to 0$ while the overlap between $\vec{\rho}_\text{OC-FDM}$ and $\vec{\rho}_\text{Full BRT}$ remains finite and about $0.93$ as shown in the middle and the lower panel of  Fig.\ \ref{fig:A2}.  The comparison between the different spectral functions is shown in Fig.~\ref{fig:A3}. The \ac{OC-FDM} spectrum reproduces  the spectrum based on $\vec{\rho}_\text{Full BRT}$ very well but at a fraction of the CPU time costs.

Ultimately, we are interested in the observables. The comparison of the total currents obtained from the  two different \ac{BRT} approaches are shown
as function of the kept NRG states $N_s$ in the upper ($eV=T_K$) and middle panel ($eV=10T_K$)  of Fig.\ \ref{fig:A4}. The small difference in the overlap
$S(\vec{\rho}_\text{OC-FDM},\vec{\rho}_\text{Full BRT})$ translates into a difference in the currents of approximately $3\%$. Since we employed the equation of motion method \cite{BullaCostiPruschke2008}, the dependence on $N_s$ is very weak. Meanwhile, the CPU time $t_\text{cpu}$ increases for the full \ac{BRT} approach with a power law  $\propto N_s^3$  as illustrated in the lower panel of Fig.\ \ref{fig:A4}. While the \ac{OC-FDM} required about 15sec for a run, we need about 3h for a full \ac{BRT} run with typically $(10^5)^2$ BRT matrix dimension. Note that we did not perform  z-averaging in this section, and  the full \ac{BRT} code was not optimized for  massive parallelization: An additional speedup is possible but the scaling of the CPU time with $\propto N_s^3$ 
remains a massive bottleneck.

\bibliography{references-paper}

@string{prb = {Phys.~Rev.~B}}

@string{prl = {Phys.~Rev.~Lett.}}

@article{han2026,
	author = {Han, Jong E.},
	doi = {10.1103/zl53-6bp4},
	issue = {4},
	journal = {Phys. Rev. B},
	month = {Jan},
	numpages = {7},
	pages = {045141},
	publisher = {American Physical Society},
	title = {{Nonequilibrium statistics of a biased Kondo resonance}},
	url = {https://link.aps.org/doi/10.1103/zl53-6bp4},
	volume = {113},
	year = {2026}}

@article{PhysRevB.89.165105,
  title = {Auxiliary master equation approach to nonequilibrium correlated impurities},
  author = {Dorda, Antonius and Nuss, Martin and von der Linden, Wolfgang and Arrigoni, Enrico},
  journal = {Phys. Rev. B},
  volume = {89},
  issue = {16},
  pages = {165105},
  numpages = {17},
  year = {2014},
  month = {Apr},
  publisher = {American Physical Society},
  doi = {10.1103/PhysRevB.89.165105},
  url = {https://link.aps.org/doi/10.1103/PhysRevB.89.165105}
}

@article{Sankar2025,
  title = {Back-action effects in charge detection},
  author = {Sankar, Sarath and Lotem, Matan and Folk, Joshua and Sela, Eran and Meir, Yigal},
  journal = {Phys. Rev. B},
  volume = {112},
  issue = {7},
  pages = {075151},
  numpages = {13},
  year = {2025},
  month = {Aug},
  publisher = {American Physical Society},
  doi = {10.1103/mdsg-n9nk},
  url = {https://link.aps.org/doi/10.1103/mdsg-n9nk}
}

@article{Eckern_2020,
	author = {Eckern, Ulrich and Wysoki{\'n}ski, Karol I},
	doi = {10.1088/1367-2630/ab6874},
	journal = {New Journal of Physics},
	month = {jan},
	number = {1},
	pages = {013045},
	publisher = {IOP Publishing},
	title = {{Two- and three-terminal far-from-equilibrium thermoelectric nano-devices in the Kondo regime}},
	url = {https://doi.org/10.1088/1367-2630/ab6874},
	volume = {22},
	year = {2020}}

@article{ChargeHeatTransportCorrelatedHopping2021,
	author = {Eckern, Ulrich and Wysoki\ifmmode \acute{n}\else \'{n}\fi{}ski, Karol I.},
	doi = {10.1103/PhysRevResearch.3.043003},
	issue = {4},
	journal = {Phys. Rev. Res.},
	month = {Oct},
	numpages = {18},
	pages = {043003},
	publisher = {American Physical Society},
	title = {Charge and heat transport through quantum dots with local and correlated-hopping interactions},
	url = {https://link.aps.org/doi/10.1103/PhysRevResearch.3.043003},
	volume = {3},
	year = {2021}}

@article{SlaveBosonQuantumTransport2017,
	author = {Sierra, Miguel A. and L\'opez, Rosa and S\'anchez, David},
	doi = {10.1103/PhysRevB.96.085416},
	issue = {8},
	journal = {Phys. Rev. B},
	month = {Aug},
	numpages = {14},
	pages = {085416},
	publisher = {American Physical Society},
	title = {{Fate of the spin-$\frac{1}{2}$ Kondo effect in the presence of temperature gradients}},
	url = {https://link.aps.org/doi/10.1103/PhysRevB.96.085416},
	volume = {96},
	year = {2017}}

@article{PhysRevB.100.201104,
	author = {Krivenko, Igor and Kleinhenz, Joseph and Cohen, Guy and Gull, Emanuel},
	doi = {10.1103/PhysRevB.100.201104},
	issue = {20},
	journal = {Phys. Rev. B},
	month = {Nov},
	numpages = {6},
	pages = {201104},
	publisher = {American Physical Society},
	title = {{Dynamics of Kondo voltage splitting after a quantum quench}},
	url = {https://link.aps.org/doi/10.1103/PhysRevB.100.201104},
	volume = {100},
	year = {2019}}

@article{Anders1995,
	author = {F B Anders},
	doi = {10.1088/0953-8984/7/14/018},
	journal = {Journal of Physics: Condensed Matter},
	month = {apr},
	number = {14},
	pages = {2801},
	title = {An enhanced perturbational study on spectral properties of the Anderson model},
	url = {https://doi.org/10.1088/0953-8984/7/14/018},
	volume = {7},
	year = {1995}}

@article{Kuramoto83,
	author = {Kuramoto, Y.},
	date = {1983/03/01},
	doi = {10.1007/BF01578246},
	id = {Kuramoto1983},
	isbn = {1431-584X},
	journal = {Zeitschrift f{\"u}r Physik B Condensed Matter},
	number = {1},
	pages = {37--52},
	title = {Self-consistent perturbation theory for dynamics of valence fluctuations},
	url = {https://doi.org/10.1007/BF01578246},
	volume = {53},
	year = {1983}}

@article{Hartman:2018hw,
	author = {Hartman, Nikolaus and Olsen, Christian and L{\"u}scher, Silvia and Samani, Mohammad and Fallahi, Saeed and Gardner, Geoffrey C. and Manfra, Michael and Folk, Joshua},
	date = {2018/11/01},
	doi = {10.1038/s41567-018-0250-5},
	id = {Hartman2018},
	isbn = {1745-2481},
	journal = {Nature Physics},
	number = {11},
	pages = {1083--1086},
	title = {Direct entropy measurement in a mesoscopic quantum system},
	url = {https://doi.org/10.1038/s41567-018-0250-5},
	volume = {14},
	year = {2018}}

@article{PhysRevLett.123.147702,
	author = {Sela, Eran and Oreg, Yuval and Plugge, Stephan and Hartman, Nikolaus and L\"uscher, Silvia and Folk, Joshua},
	doi = {10.1103/PhysRevLett.123.147702},
	issue = {14},
	journal = {Phys. Rev. Lett.},
	month = {Oct},
	numpages = {5},
	pages = {147702},
	publisher = {American Physical Society},
	title = {Detecting the Universal Fractional Entropy of Majorana Zero Modes},
	url = {https://link.aps.org/doi/10.1103/PhysRevLett.123.147702},
	volume = {123},
	year = {2019}}

@article{PhysRevB.111.035445,
	author = {Manaparambil, Anand and Weichselbaum, Andreas and von Delft, Jan and Weymann, Ireneusz},
	doi = {10.1103/PhysRevB.111.035445},
	issue = {3},
	journal = {Phys. Rev. B},
	month = {Jan},
	numpages = {15},
	pages = {035445},
	publisher = {American Physical Society},
	title = {{Nonequilibrium steady-state thermoelectrics of Kondo-correlated quantum dots}},
	url = {https://link.aps.org/doi/10.1103/PhysRevB.111.035445},
	volume = {111},
	year = {2025}}

@article{DMRGwithDMFT2004,
	author = {Garc\'{\i}a, Daniel J. and Hallberg, Karen and Rozenberg, Marcelo J.},
	doi = {10.1103/PhysRevLett.93.246403},
	issue = {24},
	journal = {Phys. Rev. Lett.},
	month = {Dec},
	numpages = {4},
	pages = {246403},
	publisher = {American Physical Society},
	title = {Dynamical Mean Field Theory with the Density Matrix Renormalization Group},
	url = {https://link.aps.org/doi/10.1103/PhysRevLett.93.246403},
	volume = {93},
	year = {2004}}

@article{DiazHan2023,
	author = {D\'{\i}az, Manuel I. and Han, Jong E. and Aron, Camille},
	doi = {10.1103/PhysRevB.107.195148},
	issue = {19},
	journal = {Phys. Rev. B},
	month = {May},
	numpages = {14},
	pages = {195148},
	publisher = {American Physical Society},
	title = {Electrically driven insulator-to-metal transition in a correlated insulator: Electronic mechanism and thermal description},
	url = {https://link.aps.org/doi/10.1103/PhysRevB.107.195148},
	volume = {107},
	year = {2023}}

@article{LeeSung2025,
	author = {Lee, Jae Sung and Lee, Shinbuhm and Noh, Tae Won},
	doi = {10.1063/1.4929512},
	issn = {1931-9401},
	journal = {Applied Physics Reviews},
	month = {08},
	number = {3},
	pages = {031303},
	rrrreprint = {https://pubs.aip.org/aip/apr/article-pdf/doi/10.1063/1.4929512/13398692/031303_1_online.pdf},
	title = {Resistive switching phenomena: A review of statistical physics approaches},
	url = {https://doi.org/10.1063/1.4929512},
	volume = {2},
	year = {2015}}

@article{SAWA2008,
	author = {Akihito Sawa},
	doi = {https://doi.org/10.1016/S1369-7021(08)70119-6},
	issn = {1369-7021},
	journal = {Materials Today},
	number = {6},
	pages = {28-36},
	title = {Resistive switching in transition metal oxides},
	url = {https://www.sciencedirect.com/science/article/pii/S1369702108701196},
	volume = {11},
	year = {2008}}

@article{AdatomKondo1998,
	author = {V. Madhavan and W. Chen and T. Jamneala and M. F. Crommie and N. S. Wingreen},
	doi = {10.1126/science.280.5363.567},
	journal = {Science},
	number = {5363},
	pages = {567-569},
	rrrrbdsk-url-1 = {https://www.science.org/doi/abs/10.1126/science.280.5363.567},
	rrrrbdsk-url-2 = {https://doi.org/10.1126/science.280.5363.567},
	rrrreprint = {https://www.science.org/doi/pdf/10.1126/science.280.5363.567},
	title = {{Tunneling into a Single Magnetic Atom: Spectroscopic Evidence of the Kondo Resonance}},
	url = {https://www.science.org/doi/abs/10.1126/science.280.5363.567},
	volume = {280},
	year = {1998}}

@article{OguriFermiLiquid2005,
	author = {Oguri ,Akira},
	doi = {10.1143/JPSJ.74.110},
	eprint = {https://doi.org/10.1143/JPSJ.74.110},
	journal = {Journal of the Physical Society of Japan},
	number = {1},
	pages = {110-117},
	title = {{Fermi Liquid Theory for the Nonequilibrium Kondo Effect at Low Bias Voltages}},
	url = {https://doi.org/10.1143/JPSJ.74.110},
	volume = {74},
	year = {2005}}

@article{SchwarzPRL-TransportQD2018,
	author = {Schwarz, F. and Weymann, I. and von Delft, J. and Weichselbaum, A.},
	doi = {10.1103/PhysRevLett.121.137702},
	issue = {13},
	journal = {Phys. Rev. Lett.},
	month = {Sep},
	numpages = {7},
	pages = {137702},
	publisher = {American Physical Society},
	title = {Nonequilibrium Steady-State Transport in Quantum Impurity Models: A Thermofield and Quantum Quench Approach Using Matrix Product States},
	url = {https://link.aps.org/doi/10.1103/PhysRevLett.121.137702},
	volume = {121},
	year = {2018}}

@article{GlazmanRaikh1988,
	author = {L. I. Glazman and M. E. Raikh},
	XXXjournal = {JETP Lett.},
	journal = {Pis’ma Zh. Eksp. Teor. Fiz. },
	pages = {378},
	title = {{Resonant Kondo transparency of barrier with quasilocal impurity states}},
	volume = {47},
	year = {1988}}

@article{BoekerGf2022,
	author = {B\"oker, Jan and Anders, Frithjof B.},
	doi = {10.1103/PhysRevB.105.235127},
	issue = {23},
	journal = {Phys. Rev. B},
	month = {Jun},
	numpages = {14},
	pages = {235127},
	publisher = {American Physical Society},
	title = {{Open Wilson chain numerical renormalization group approach to Green's functions}},
	url = {https://link.aps.org/doi/10.1103/PhysRevB.105.235127},
	volume = {105},
	year = {2022}}

@article{PhysRevE.76.031115,
	author = {Wichterich, Hannu and Henrich, Markus J. and Breuer, Heinz-Peter and Gemmer, Jochen and Michel, Mathias},
	doi = {10.1103/PhysRevE.76.031115},
	issue = {3},
	journal = {Phys. Rev. E},
	month = {Sep},
	numpages = {7},
	pages = {031115},
	publisher = {American Physical Society},
	title = {Modeling heat transport through completely positive maps},
	url = {https://link.aps.org/doi/10.1103/PhysRevE.76.031115},
	volume = {76},
	year = {2007}}

@article{CohenGullQMC2023,
	author = {Erpenbeck, A. and Gull, E. and Cohen, G.},
	doi = {10.1103/PhysRevLett.130.186301},
	issue = {18},
	journal = {Phys. Rev. Lett.},
	month = {May},
	numpages = {7},
	pages = {186301},
	publisher = {American Physical Society},
	title = {{Quantum Monte Carlo Method in the Steady State}},
	url = {https://link.aps.org/doi/10.1103/PhysRevLett.130.186301},
	volume = {130},
	year = {2023}}

@phdthesis{MayPhD2020,
	author = {Daniel May},
	month = {Januar},
	school = {TU Dortmund university},
	title = {Numerical Study of Magnetic Impurities in Graphene and Steady-State Transport in Quantum Impurity Systems},
	year = {2020}}

@article{RevModPhys.NEQ-QS-2022,
	author = {Landi, Gabriel T. and Poletti, Dario and Schaller, Gernot},
	doi = {10.1103/RevModPhys.94.045006},
	issue = {4},
	journal = {Rev. Mod. Phys.},
	month = {Dec},
	numpages = {58},
	pages = {045006},
	publisher = {American Physical Society},
	title = {Nonequilibrium boundary-driven quantum systems: Models, methods, and properties},
	url = {https://link.aps.org/doi/10.1103/RevModPhys.94.045006},
	volume = {94},
	year = {2022}}

@article{LotemWeichselbaum2020,
	author = {Lotem, Matan and Weichselbaum, Andreas and von Delft, Jan and Goldstein, Moshe},
	doi = {10.1103/PhysRevResearch.2.043052},
	issue = {4},
	journal = {Phys. Rev. Res.},
	month = {Oct},
	numpages = {17},
	pages = {043052},
	publisher = {American Physical Society},
	title = {{Renormalized Lindblad driving: A numerically exact nonequilibrium quantum impurity solver}},
	url = {https://link.aps.org/doi/10.1103/PhysRevResearch.2.043052},
	volume = {2},
	year = {2020}}

@misc{conductanceDMRG2010,
	author = {Alexander Bransch{\"a}del and Guenter Schneider and Peter Schmitteckert},
	howpublished = {arXiv:1004.4178},
	title = {{Conductance of correlated systems: real-time dynamics in finite systems}},
	year = {2010}}

@article{Child-A-2022,
	article-number = {417},
	author = {Child, Timothy and Sheekey, Owen and L{\"u}scher, Silvia and Fallahi, Saeed and Gardner, Geoffrey C. and Manfra, Michael and Folk, Joshua},
	doi = {10.3390/e24030417},
	issn = {1099-4300},
	journal = {Entropy},
	number = {3},
	pubmedid = {35327927},
	title = {A Robust Protocol for Entropy Measurement in Mesoscopic Circuits},
	url = {https://www.mdpi.com/1099-4300/24/3/417},
	volume = {24},
	year = {2022}}

@article{Child-QPC-Entropy2022,
	author = {Child, Timothy and Sheekey, Owen and L\"uscher, Silvia and Fallahi, Saeed and Gardner, Geoffrey C. and Manfra, Michael and Mitchell, Andrew and Sela, Eran and Kleeorin, Yaakov and Meir, Yigal and Folk, Joshua},
	doi = {10.1103/PhysRevLett.129.227702},
	issue = {22},
	journal = {Phys. Rev. Lett.},
	month = {Nov},
	numpages = {5},
	pages = {227702},
	publisher = {American Physical Society},
	title = {Entropy Measurement of a Strongly Coupled Quantum Dot},
	url = {https://link.aps.org/doi/10.1103/PhysRevLett.129.227702},
	volume = {129},
	year = {2022}}

@article{MaMeirQPC2023,
	author = {Ma, Zhanyu and Han, Cheolhee and Meir, Yigal and Sela, Eran},
	doi = {10.1103/PhysRevLett.131.126502},
	issue = {12},
	journal = {Phys. Rev. Lett.},
	month = {Sep},
	numpages = {6},
	pages = {126502},
	publisher = {American Physical Society},
	title = {Identifying an Environment-Induced Localization Transition from Entropy and Conductance},
	url = {https://link.aps.org/doi/10.1103/PhysRevLett.131.126502},
	volume = {131},
	year = {2023}}

@article{Oliveira1994,
	author = {Oliveira, Wanda C. and Oliveira, Luiz N.},
	doi = {10.1103/PhysRevB.49.11986},
	issue = {17},
	journal = {Phys. Rev. B},
	month = {May},
	numpages = {0},
	pages = {11986--11994},
	publisher = {American Physical Society},
	title = {Generalized numerical renormalization-group method to calculate the thermodynamical properties of impurities in metals},
	url = {https://link.aps.org/doi/10.1103/PhysRevB.49.11986},
	volume = {49},
	year = {1994}}

@article{KrishWilWilson80b,
	author = {Krishna-murthy, H. R. and Wilkins, J. W. and Wilson, K. G.},
	doi = {10.1103/PhysRevB.21.1044},
	issue = {3},
	journal = {Phys. Rev. B},
	month = {Feb},
	numpages = {0},
	pages = {1044--1083},
	publisher = {American Physical Society},
	title = {{Renormalization-group approach to the Anderson model of dilute magnetic alloys. II. Static properties for the asymmetric case}},
	url = {https://link.aps.org/doi/10.1103/PhysRevB.21.1044},
	volume = {21},
	year = {1980}}

@article{KrishWilWilson80a,
	author = {Krishna-murthy, H. R. and Wilkins, J. W. and Wilson, K. G.},
	doi = {10.1103/PhysRevB.21.1003},
	issue = {3},
	journal = {Phys. Rev. B},
	month = {Feb},
	numpages = {0},
	pages = {1003--1043},
	publisher = {American Physical Society},
	title = {{Renormalization-group approach to the Anderson model of dilute magnetic alloys. I. Static properties for the symmetric case}},
	url = {https://link.aps.org/doi/10.1103/PhysRevB.21.1003},
	volume = {21},
	year = {1980}}

@phdthesis{BoekerPhD2021,
	address = {44221 Dortmund, Germany},
	author = {Jan Oliver Boeker},
	school = {TU Dortmund university},
	title = {A Novel Hybrid Numerical Renormalization Group Approach to Non-Equilibrium Dynamics and Spectral Functions},
	year = {2021}}

@article{BoekerAnders2020,
	author = {B\"oker, Jan and Anders, Frithjof B.},
	doi = {10.1103/PhysRevB.102.075149},
	issue = {7},
	journal = {Phys. Rev. B},
	month = {Aug},
	numpages = {25},
	pages = {075149},
	publisher = {American Physical Society},
	title = {Restoring the continuum limit in the time-dependent numerical renormalization group approach},
	url = {https://link.aps.org/doi/10.1103/PhysRevB.102.075149},
	volume = {102},
	year = {2020}}

@article{KondoModelComparisonGebhard2020,
	author = {Barcza, Gergely and Bauerbach, Kevin and Eickhoff, Fabian and Anders, Frithjof B. and Gebhard, Florian and Legeza, \"Ors},
	doi = {10.1103/PhysRevB.101.075132},
	issue = {7},
	journal = {Phys. Rev. B},
	month = {Feb},
	numpages = {35},
	pages = {075132},
	publisher = {American Physical Society},
	title = {Symmetric single-impurity Kondo model on a tight-binding chain: Comparison of analytical and numerical ground-state approaches},
	url = {https://link.aps.org/doi/10.1103/PhysRevB.101.075132},
	volume = {101},
	year = {2020}}

@article{BullaCostiPruschke2008,
	author = {R. Bulla and T. A. Costi and T. Pruschke},
	journal = {Rev.~Mod.~Phys.},
	optnumber = {and cond-mat/0701105},
	optpages = {and cond-mat/0701105},
	pages = {395},
	title = {The numerical renormalization group method for quantum impurity systems},
	volume = {80},
	year = {2008}}

@article{AndersSSnrg2008,
	author = {Frithjof B. Anders},
	doi = {10.1103/PhysRevLett.101.066804},
	eid = {066804},
	journal = {Phys. Rev. Lett.},
	number = {6},
	numpages = {4},
	pages = {066804},
	publisher = {APS},
	title = {Steady-State Currents through Nanodevices: A Scattering-States Numerical Renormalization-Group Approach to Open Quantum Systems},
	url = {http://link.aps.org/abstract/PRL/v101/e066804},
	volume = {101},
	year = {2008}}

@article{AndersSchiller2006,
	author = {F. B. Anders and A. Schiller},
	journal = PRB,
	pages = {245113},
	title = {{Spin precession and real-time dynamics in the Kondo model: Time-dependent numerical renormalization-group study}},
	volume = {74},
	year = {2006}}

@article{AndersSchiller2005,
	author = {F. B. Anders and A. Schiller},
	journal = PRL,
	opthowpublished = {cond-mat/0505553},
	pages = {196801},
	title = {{Time-dependent Numerical Renormalization Group Approach to non-Equilibrium Dynamics of Quantum Impurity Systems}},
	volume = {95},
	year = {2005}}

@article{AronWeberKotliar2013,
	author = {Aron, Camille and Weber, Cedric and Kotliar, Gabriel},
	doi = {10.1103/PhysRevB.87.125113},
	issue = {12},
	journal = {Phys. Rev. B},
	month = {Mar},
	numpages = {5},
	pages = {125113},
	publisher = {American Physical Society},
	title = {Impurity model for non-equilibrium steady states},
	url = {http://link.aps.org/doi/10.1103/PhysRevB.87.125113},
	volume = {87},
	year = {2013}}

@article{OpenChains2017,
	author = {Bruognolo, B. and Linden, N.-O. and Schwarz, F. and Lee, S.-S. B. and Stadler, K. and Weichselbaum, A. and Vojta, M. and Anders, F. B. and von Delft, J.},
	doi = {10.1103/PhysRevB.95.121115},
	issue = {12},
	journal = {Phys. Rev. B},
	month = {Mar},
	numpages = {5},
	pages = {121115(R)},
	publisher = {American Physical Society},
	title = {Open Wilson chains for quantum impurity models: Keeping track of all bath modes},
	url = {https://link.aps.org/doi/10.1103/PhysRevB.95.121115},
	volume = {95},
	year = {2017}}

@article{BullaHewsonPruschke98,
	author = {R. Bulla and A. C. Hewson and Th. Pruschke},
	journal = {J. Phys.: Condens. Matter},
	pages = {8365},
	title = {{Numerical renormalization group calculations for the self-energy of the impurity Anderson model}},
	volume = {10},
	year = {1998}}

@article{GuyGullMillis2014,
	author = {Cohen, Guy and Gull, Emanuel and Reichman, David R. and Millis, Andrew J.},
	doi = {10.1103/PhysRevLett.112.146802},
	issue = {14},
	journal = {Phys. Rev. Lett.},
	month = {Apr},
	numpages = {5},
	pages = {146802},
	publisher = {American Physical Society},
	title = {Green's Functions from Real-Time Bold-Line Monte Carlo Calculations: Spectral Properties of the Nonequilibrium Anderson Impurity Model},
	url = {http://link.aps.org/doi/10.1103/PhysRevLett.112.146802},
	volume = {112},
	year = {2014}}

@article{costiMagFldSIAM00,
	author = {Costi, T. A.},
	doi = {10.1103/PhysRevLett.85.1504},
	journal = {Phys. Rev. Lett.},
	month = {Aug},
	number = {7},
	numpages = {3},
	pages = {1504--1507},
	publisher = {American Physical Society},
	title = {{Kondo Effect in a Magnetic Field and the Magnetoresistivity of Kondo Alloys}},
	volume = {85},
	year = {2000}}

@article{DirksSchmitt2013,
	author = {Dirks, A. and Schmitt, S. and Han, J. E. and Anders, F. and Werner, P. and Pruschke, T.},
	doi = {10.1209/0295-5075/102/37011},
	journal = {Europhysics Letters},
	month = {may},
	number = {3},
	pages = {37011},
	publisher = {EDP Sciences, IOP Publishing and Societ{\`a} Italiana di Fisica},
	title = {Double occupancy and magnetic susceptibility of the Anderson impurity model out of equilibrium},
	url = {https://doi.org/10.1209/0295-5075/102/37011},
	volume = {102},
	year = {2013}}

@article{DoraArrigioni2015,
	author = {Dorda, Antonius and Ganahl, Martin and Evertz, Hans Gerd and von der Linden, Wolfgang and Arrigoni, Enrico},
	doi = {10.1103/PhysRevB.92.125145},
	issue = {12},
	journal = {Phys. Rev. B},
	month = {Sep},
	numpages = {13},
	pages = {125145},
	publisher = {American Physical Society},
	title = {{Auxiliary master equation approach within matrix product states: Spectral properties of the nonequilibrium Anderson impurity model}},
	url = {http://link.aps.org/doi/10.1103/PhysRevB.92.125145},
	volume = {92},
	year = {2015}}

@book{Feynman72,
	address = {Reading, MA, USA},
	author = {R. P. Feynman},
	publisher = {Benjamin},
	title = {Statistical Mechanics, A Set of Lectures},
	year = {1972}}

@article{goldhaberSET98,
	author = {Goldhaber-Gordon, D. and G\"ores, J. and Kastner, M. A. and Shtrikman, Hadas and Mahalu, D. and Meirav, U.},
	doi = {10.1103/PhysRevLett.81.5225},
	journal = {Phys. Rev. Lett.},
	pages = {5225--5228},
	publisher = {American Physical Society},
	title = {{From the Kondo Regime to the Mixed-Valence Regime in a Single-Electron Transistor}},
	volume = {81},
	year = {1998}}

@article{NatureGoldhaberGordon1998,
	author = {D. Goldhaber-Gordon and H. Shtrikman and D. Mahalu and D. Abusch-Magder and U. Meirav and M.A. Kastner},
	journal = {Nature},
	pages = {156},
	title = {{Kondo effect in a single-electron transistor}},
	volume = {391},
	year = {1998}}

@article{Grewe83,
	author = {N. Grewe},
	journal = {Z. Phys. B},
	pages = {193},
	title = {{Perturbation expansions for systems with strong local correlation}},
	volume = {52},
	year = {1983}}

@article{GuettgeAndersSchiller2013,
	author = {Fabian Guettge and Frithjof B. Anders and Ulrich Schollwoeck and Eitan Eidelstein and Avraham Schiller},
	journal = {Phys. Rev. B},
	pages = {115115},
	title = {{Hybrid NRG-DMRG approach to real-time dynamics of quantum impurity systems}},
	volume = {87},
	year = {2013}}

@article{GulletAl2011,
	author = {Gull, Emanuel and Millis, Andrew J. and Lichtenstein, Alexander I. and Rubtsov, Alexey N. and Troyer, Matthias and Werner, Philipp},
	doi = {10.1103/RevModPhys.83.349},
	journal = {Rev. Mod. Phys.},
	month = {May},
	number = {2},
	numpages = {55},
	pages = {349--404},
	publisher = {American Physical Society},
	title = {Continuous-time Monte Carlo methods for quantum impurity models},
	volume = {83},
	year = {2011}}

@article{Heidrich-Meisner-SIAM-transport2009,
	author = {Heidrich-Meisner, F. and Feiguin, A. E. and Dagotto, E.},
	doi = {10.1103/PhysRevB.79.235336},
	issue = {23},
	journal = {Phys. Rev. B},
	month = {Jun},
	numpages = {6},
	pages = {235336},
	publisher = {American Physical Society},
	title = {{Real-time simulations of nonequilibrium transport in the single-impurity Anderson model}},
	url = {http://link.aps.org/doi/10.1103/PhysRevB.79.235336},
	volume = {79},
	year = {2009}}

@article{Hershfield1993,
	author = {S. Hershfield},
	doi = {10.1103/PhysRevLett.70.2134},
	journal = PRL,
	pages = {2134},
	title = {Reformulation of steady state nonequilibrium quantum statistical mechanics},
	url = {http://link.aps.org/doi/10.1103/PhysRevLett.70.2134},
	volume = {70},
	year = {1993}}

@article{HershfieldDaviesWilkins1991,
	author = {Selman Hershfield and John H. Davies and John W. Wilkins},
	journal = PRL,
	pages = {003720},
	title = {{Probing the Kondo resonance by resonant tunneling through an Anderson impurity}},
	volume = {67},
	year = {1991}}

@article{JauhoWingreenMeir1994,
	author = {Antti-Pekka Jauho and Ned S. Wingreen and Yigal Meir},
	journal = PRB,
	pages = {005528},
	title = {Time-dependent transport in interacting and noninteracting resonant-tunneling systems},
	volume = {50},
	year = {1994}}

@book{MayKuehn2000,
	address = {Berlin},
	author = {V. May and O. K{\"u}hn},
	publisher = {Wiley-VCH},
	title = {Charge and Energy Transfer Dynamics in Molecular Systems},
	year = {2000}}

@article{MeirWingreen1992,
	author = {Yigal Meir and Ned S. Wingreen},
	journal = PRL,
	pages = {2512},
	title = {Landauer formula for the current through an interacting electron region},
	volume = {68},
	year = {1992}}

@article{meirTransportSIAM93,
	author = {Meir, Yigal and Wingreen, Ned S. and Lee, Patrick A.},
	doi = {10.1103/PhysRevLett.70.2601},
	journal = {Phys. Rev. Lett.},
	month = {Apr},
	number = {17},
	numpages = {3},
	pages = {2601--2604},
	publisher = {American Physical Society},
	title = {{Low-temperature transport through a quantum dot: The Anderson model out of equilibrium}},
	volume = {70},
	year = {1993}}

@article{Oguri2007,
	author = {Akira Oguri},
	doi = {10.1103/PhysRevB.75.035302},
	journal = PRB,
	pages = {035302},
	title = {{Mixed-state aspects of an out-of-equilibrium Kondo problem in a quantum dot}},
	url = {http://link.aps.org/doi/10.1103/PhysRevB.75.035302},
	volume = {75},
	year = {2007}}

@article{PetersPruschkeAnders2006,
	author = {R. Peters and Th. Pruschke and F. B. Anders},
	journal = PRB,
	pages = {245114},
	title = {{A Numerical Renormalization Group approach to Green's Functions for Quantum Impurity Models}},
	volume = {74},
	year = {2006}}

@article{RoschKrohaWoelfe2001,
	author = {A. Rosch and J. Kroha and P. W\"olfle},
	journal = PRL,
	pages = {156802},
	title = {Kondo Effect in Quantum Dots at High Voltage: Universality and Scaling},
	volume = {87},
	year = {2001}}

@article{RoschPaaskeKrohaWoelfe2003,
	author = {A. Rosch and J. Paaske and J. Kroha and P. W\"olfle},
	journal = PRL,
	pages = {076804},
	title = {Nonequilibrium Transport through a Kondo Dot in a Magnetic Field: Perturbation Theory and Poor Man's Scaling},
	volume = {90},
	year = {2003}}

@book{SaadSparseLinearSystemsBook2003,
	author = {Yousef Saad},
	publisher = {Society for Industrial and Applied Mathematics, Philadelphia, USA},
	title = {Iterative Methods for Sparse Linear Systems},
	year = {2003}}

@article{SchillerHershfield96,
	author = {A. Schiller and S. Hershfield},
	journal = PRL,
	pages = {1821},
	title = {Solution of an ac Kondo Model},
	volume = {77},
	year = {1996}}

@article{Schmitteckert2004,
	author = {Peter Schmitteckert},
	journal = PRB,
	pages = {121302(R)},
	title = {Nonequilibrium electron transport using the density matrix renormalization group method},
	volume = {70},
	year = {2004}}

@article{Schmitteckert2013,
	author = {Schmitteckert, Peter and Carr, Sam T. and Saleur, Hubert},
	doi = {10.1103/PhysRevB.89.081401},
	issue = {8},
	journal = {Phys. Rev. B},
	month = {Feb},
	numpages = {5},
	pages = {081401},
	publisher = {American Physical Society},
	title = {Transport through nanostructures: Finite time versus finite size},
	url = {http://link.aps.org/doi/10.1103/PhysRevB.89.081401},
	volume = {89},
	year = {2014}}

@article{Schollwoeck2011,
	author = {Ulrich Schollw\"ock},
	doi = {10.1016/j.aop.2010.09.012},
	issn = {0003-4916},
	journal = {Ann. Phys.},
	number = {1},
	pages = {96 - 192},
	title = {The density-matrix renormalization group in the age of matrix product states},
	url = {http://www.sciencedirect.com/science/article/pii/S0003491610001752},
	volume = {326},
	year = {2011}}

@article{Schollwoeck-2005,
	author = {U. Schollw\"ock},
	journal = {Rev. Mod. Phys.},
	pages = {259},
	title = {The density-matrix renormalization group},
	volume = {77},
	year = {2005}}

@article{LindbladContinuumVonDelft2016,
	author = {Schwarz, F. and Goldstein, M. and Dorda, A. and Arrigoni, E. and Weichselbaum, A. and von Delft, J.},
	doi = {10.1103/PhysRevB.94.155142},
	issue = {15},
	journal = {Phys. Rev. B},
	month = {Oct},
	numpages = {21},
	pages = {155142},
	publisher = {American Physical Society},
	title = {Lindblad-driven discretized leads for nonequilibrium steady-state transport in quantum impurity models: Recovering the continuum limit},
	url = {http://link.aps.org/doi/10.1103/PhysRevB.94.155142},
	volume = {94},
	year = {2016}}

@article{silvaDagotto2008,
	author = {Luis G. G. V. Dias da Silva and F. Heidrich-Meisner and A. E. Feiguin and C. A. B\"{u}sser and G. B. Martins and E. V. Anda and E. Dagotto},
	doi = {10.1103/PhysRevB.78.195317},
	eid = {195317},
	journal = {Phys. Rev. B},
	number = {19},
	numpages = {9},
	pages = {195317},
	publisher = {APS},
	title = {{Transport properties and Kondo correlations in nanostructures: Time-dependent DMRG method applied to quantum dots coupled to Wilson chains}},
	url = {http://link.aps.org/abstract/PRB/v78/e195317},
	volume = {78},
	year = {2008}}

@article{spataruGwMillis2009,
	author = {Catalin D. Spataru and Mark S. Hybertsen and Steven G. Louie and Andrew J. Millis},
	doi = {10.1103/PhysRevB.79.155110},
	eid = {155110},
	journal = PRB,
	number = {15},
	pages = {155110},
	publisher = {APS},
	title = {{GW approach to Anderson model out of equilibrium: Coulomb blockade and false hysteresis in the I-V characteristics}},
	volume = {79},
	year = {2009}}

@article{thygesenNonEqGW07,
	author = {Kristian S. Thygesen and Angel Rubio},
	doi = {10.1063/1.2565690},
	journal = {J. Chem. Phys.},
	pages = {091101},
	title = {Nonequilibrium GW approach to quantum transport in nano-scale contacts},
	volume = {126},
	year = {2007}}

@article{WeichselbaumDelft2007,
	author = {A. Weichselbaum and J. von Delft},
	journal = PRL,
	pages = {076402},
	title = {{Sum-Rule Conserving Spectral Functions from the Numerical Renormalization Group}},
	volume = {99},
	year = {2007}}

@article{Werner09,
  title = {{Diagrammatic Monte Carlo simulation of nonequilibrium systems}},
  author = {Werner, Philipp and Oka, Takashi and Millis, Andrew J.},
  journal = {Phys. Rev. B},
  volume = {79},
  issue = {3},
  pages = {035320},
  numpages = {18},
  year = {2009},
  month = {Jan},
  publisher = {American Physical Society},
  doi = {10.1103/PhysRevB.79.035320},
  url = {https://link.aps.org/doi/10.1103/PhysRevB.79.035320}
}

@article{Kouwenhoven2000,
	author = {W. G. van der Wiel and S. De Franceschi and T. Fujisawa J.M. Elzerman and S. Tarucha and L. P. Kouvenhoven},
	journal = {Science},
	pages = {2105},
	title = {The Kondo effect in the unitary limit},
	volume = {289},
	year = {2000}}

@article{Wilson75,
	author = {K. G. Wilson},
	journal = {Rev. Mod. Phys.},
	pages = {773},
	title = {{The renormalization group: Critical phenomena and the Kondo problem}},
	volume = {47},
	year = {1975}}

@article{MeirWingreen1994,
	author = {N. S. Wingreen and Y. Meir},
	journal = PRB,
	number = {16},
	pages = {11040},
	title = {Anderson model out of equilibrium: Noncrossing-approximation approach to transport through a quantum dot},
	volume = {49},
	year = {1994}}

@article{YoshidaWithakerOliveira1990,
	author = {M. Yoshida and M. A. Whitaker and L. N. Oliveira},
	journal = PRB,
	pages = {9403},
	title = {Renormalization-group calulation of excitation properties for impurity models},
	volume = {41},
	year = {1990}}

@article{Dzhioev_2012,
	author = {Alan A Dzhioev and D S Kosov},
	doi = {10.1088/0953-8984/24/22/225304},
	journal = {Journal of Physics: Condensed Matter},
	month = {may},
	number = {22},
	pages = {225304},
	publisher = {{IOP} Publishing},
	title = {{Nonequilibrium perturbation theory in Liouville{\textendash}Fock space for inelastic electron transport}},
	url = {https://doi.org/10.1088%2F0953-8984%2F24%2F22%2F225304},
	volume = {24},
	year = {2012}}

@article{Pletyukhov2012,
	author = {Pletyukhov, Mikhail and Schoeller, Herbert},
	doi = {10.1103/PhysRevLett.108.260601},
	journal = {Phys. Rev. Lett.},
	pages = {260601},
	title = {Nonequilibrium {Kondo} Model: Crossover from Weak to Strong Coupling},
	volume = {108},
	year = {2012}}

@article{Manaparambil2022,
	author = {Manaparambil, Anand and Weichselbaum, Andreas and von Delft, Jan and Weymann, Ireneusz},
	doi = {10.1103/PhysRevB.106.125413},
	journal = {Phys. Rev. B},
	pages = {125413},
	title = {Nonequilibrium spintronic transport through {Kondo} impurities},
	volume = {106},
	year = {2022}}

@article{Costi2010,
	author = {Costi, T. A. and Zlati\'{c}, V.},
	doi = {10.1103/PhysRevB.81.235127},
	journal = {Phys. Rev. B},
	pages = {235127},
	title = {Thermoelectric transport through strongly correlated quantum dots},
	volume = {81},
	year = {2010}}

\end{document}